\documentclass[]{aa} 
\usepackage{float}
\usepackage{multirow} 
\usepackage{graphicx} 
\usepackage{arydshln} 
\usepackage{caption}
\usepackage{subcaption} 
\usepackage{graphicx} 
\usepackage{txfonts} 
\usepackage{amsfonts}
\usepackage{natbib} 
\usepackage{float} 
\usepackage{lscape} 
\usepackage{nicefrac}
\usepackage[export]{adjustbox}
\usepackage[version=4]{mhchem}
\usepackage{txfonts}
\usepackage{xcolor}

\begin{document}

\title{Signatures of UV radiation in low-mass protostars}
\subtitle{I. Origin of HCN and CN emission in the Serpens Main region}

\author{Agnieszka Mirocha\inst{1,2}, Agata Karska\inst{2}\thanks{Corresponding author: Agata Karska\newline
\email{agata.karska@umk.pl}}, Marcin Gronowski\inst{3,4}, Lars E. Kristensen\inst{5}, Łukasz Tychoniec\inst{6,7},
 Daniel Harsono\inst{6,8}, Miguel Figueira\inst{9}, Marcin Gładkowski\inst{2,10}, Michał Żółtowski\inst{2,11,12}}
\institute{
$^{1}$ Astronomical Observatory of the Jagiellonian University, ul. Orla 171, 30-244 Kraków,
Poland\\ 
$^{2}$ Institute of Astronomy, Faculty of Physics,
Astronomy and Informatics, Nicolaus Copernicus University, ul. Grudziądzka 5, 87-100 Toruń,
Poland \\
$^{3}$ Institute of Physical Chemistry Polish Academy of Sciences, ul. Kasprzaka 44/52, 01-224 Warszawa,
Poland \\
$^{4}$ Faculty of Physics, University of Warsaw, Pasteura 5, 02-093 Warsaw, Poland \\
$^{5}$ Niels Bohr Institute \& Centre for Star and Planet Formation, Copenhagen University, {\O}ster Voldgade 5--7, 1350 Copenhagen K, Denmark\\
$^{6}$ Leiden Observatory, Leiden University, P.O. Box 9513, NL-2300RA Leiden, The Netherlands\\
$^{7}$ European Southern Observatory, Karl-Schwarzschild-Str 2, D-85748 Garching near Munich, Germany\\
$^{8}$ Institute of Astronomy and Astrophysics, Academia Sinica, No. 1, Sec. 4, Roosevelt Road, Taipei 10617, Taiwan, R. O. C. \\
$^{9}$ National Centre for Nuclear Research, ul. Pasteura 7, 02-093 Warszawa, Poland \\
$^{10}$ Nicolaus Copernicus Astronomical Center, ul. Rabiańska 8, 87-100 Toruń, Poland \\
$^{11}$ University of Le Havre, Laboratoire Ondes et Milieux Complexes, UMR CNRS 6294,75 Rue Bellot, 76600 Le Havre, France \\
$^{12}$ Univ Rennes, CNRS, IPR (Institut de Physique de Rennes) – UMR 6251, F-35000 Rennes, France \\
}
\date{Received March 19, 2021; accepted September 13, 2021}
\titlerunning{Signatures of UV radiation around low-mass protostars}
\authorrunning{A.~Mirocha et al. 2021}

\abstract
{Ultraviolet radiation (UV) influences the physics and chemistry of star-forming regions,
but its properties and significance in the immediate surroundings of low-mass protostars are still poorly understood.}
{We aim to extend the use of the CN/HCN ratio, already established for high-mass protostars, to the low-mass regime to trace and characterize the UV field around low-mass protostars on $\sim 0.6\times0.6$ pc scales.}
{We present $5'\times5'$ maps of the Serpens Main Cloud encompassing 10 protostars observed with the EMIR receiver at the IRAM~30~m telescope in CN 1-0, HCN 1-0, CS 3-2, and some of their isotopologues. The radiative-transfer code RADEX and 
the chemical model Nahoon are used to determine column densities of molecules, gas temperature and density, and the UV field strength, $G_\mathrm{0}$.}
{The spatial distribution of HCN and CS are well-correlated with CO 6-5 emission that traces 
outflows. The CN emission is extended from the central protostars to their immediate
 surroundings also tracing outflows, likely as a product 
of HCN photodissociation. The ratio of CN to HCN total column densities ranges from
 $\sim$1 to 12 corresponding to G$_0$ $\approx$ $10^{1}-10^{3}$ for gas densities and
  temperatures typical for outflows of low-mass protostars.}
{UV radiation associated with protostars and their outflows is indirectly identified in a significant part of the Serpens Main low-mass star-forming region. Its strength is consistent with the values obtained from the OH and H$_2$O ratios observed with \textit{Herschel} and compared with models of UV-illuminated shocks. 
From a chemical viewpoint, the CN to HCN ratio is an excellent tracer of UV fields around low- and intermediate-mass star-forming regions.}

\keywords{astrochemistry -- stars: formation -- ISM: molecules -- ISM: individual objects:
Serpens Main -- ISM: jets and outflows --Submillimeter: ISM}

\maketitle
%
%
\section{Introduction} 
\label{section:intro}

The formation of low-mass stars is associated with many physical phenomena. The inside-out collapse of dense cores
is accompanied by the ejection of bipolar outflows and the formation of embedded disks (e.g. \citealt{Fra14,Li14}).
Ultraviolet (UV) radiation can be produced in mass accretion on the central object or bow shocks, and it irradiates the outflow cavities 
in the envelopes \citep{Spa95,vKe09a,Vis12,Dro15}. The physical conditions and chemical 
composition in star-forming regions depend on the characteristics of the above processes. 

The importance of UV radiation for star formation was initially considered only in the context 
of massive stars, where the central stars are the main source of UV photons from early stages \citep{Ces05,ZY07}. 
In dense star-forming environments, the far-UV radiation is revealed by the chemical composition of the  material,
 since it dissociates and ionizes molecules and atoms with ionization potentials below 13.6 eV \citep{Ben16}.
More energetic photons are easily absorbed in the surrounding interstellar medium \citep{Dra03}. 
Among the most useful diagnostics of UV radiation is the ratio of CN to HCN \citep{DN97,Sta05}. In the presence of 
UV photons, HCN is photodissociated to CN with a rate of $1.64\times10^{-9}$ s$^{-1}$,
which is more than one order of magnitude higher than the photodissociation of CN (\citealt{Hea17}). 
The CN/HCN ratio was used as a tracer of \mbox{UV/X-ray} radiation in several astrophysical environments
 including extragalactic photon-dominated regions (e.g. \citealt{Per07}), reflection nebulae (e.g. \citealt{Fue95}), molecular clouds (e.g. \citealt{Gre96}), 
high-mass protostars (\citealt{Sta07}) and proto-planetary disks (e.g.
\citealt{Cha12}). Recent models of the envelopes of low-mass protostars included the impact of UV radiation, but it remains hard to substantiate observationally \citep{Vis12,Dro15}.

\begin{figure} [tb]
\begin{center}
\includegraphics[width=8cm]{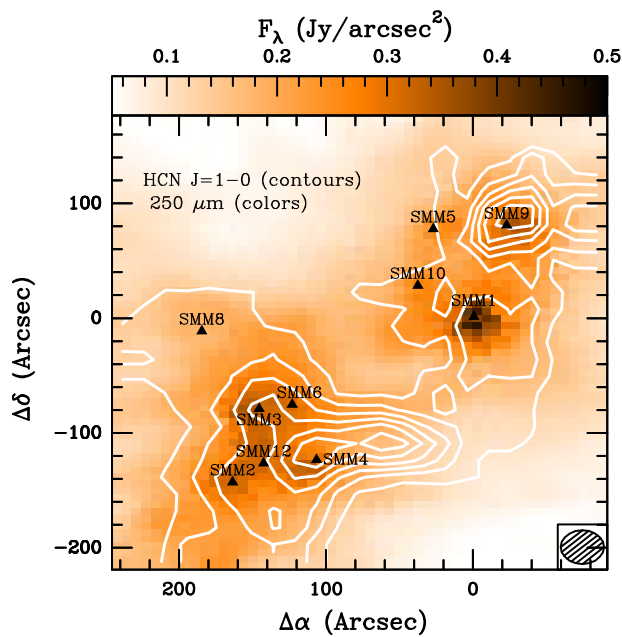} 
\caption{EMIR map of HCN 1-0 (contours) on top of the continuum emission at 250 $\mu$m
from \textit{Herschel}/SPIRE. The HCN contours start at 6~K~km s$^{-1}$ (20 $\sigma$) and are drawn in steps 
of 3~K~km~s$^{-1}$ (10 $\sigma$). The protostellar properties are shown in Table \ref{table:2}.} 
\label{fig1:seds} 
\end{center}
\end{figure}

The first detection of HCN toward a low-mass protostar NGC 1333 IRAS4B revealed a broad linewidth of 17.6~km~s$^{-1}$ in HCN~4-3  similar to those of CO~2-1, and thus the HCN line was identified
  as a possible tracer of the outflows (\citealt{Bla95}). Subsequent interferometric maps of
   Ser SMM1, SMM3, and SMM4 showed relatively compact emission on 30''$\times$30'' scales
    (\citealt{Hog99}). The line wings of HCN 4-3 were associated with the bipolar outflows
     and the line center with the cold envelope material. The HCN 1-0 emission was interpreted
      as a tracer of the outflow cavities (\citealt{Hog99}). A recent survey of protostars in
       Perseus in HCN 4-3 shows its utility as a tracer of the most energetic outflows \citep{Wal14}.
        Enhancement of HCN abundances in shocks where the gas temperatures above 200~K is
         suggested by both observations and models (\citealt{Boo01}, \citealt{Lah07}, \citealt{Pin90}). 
 
Mapping of CN emission toward the low-mass protostar, L483, shows an elongated structure
 in the outflow direction, which is narrower than that seen in the more quiescent gas and
  associated with the outflow cavity walls (\citealt{Jor04}). Based on correlations
   between CN abundances and UV field in gaseous disks around more evolved sources,
    it was suggested that the enhancement of CN in L483 is related to the UV irradiation.
     Simultaneous mapping of HCN and CN on large scales indeed confirmed the offset of
      emission peaks between the two molecules in L1157 (\citealt{Bac01}).

The presence of UV radiation around low-mass protostars was confirmed by the 
detection of warm gas traced by narrow $^{12}$CO 6-5 lines and spatially associated with
 outflow cavities in the low-mass protostar HH 46 \citep{vKe09b}. 
Firm detection of [C{\,\sc{i}}] 2-1 at the tip of the jet and the lack of CO was attributed
 to the dissociative bow shock, which itself is strong enough to produce UV photons that
  subsequently dissociate CO \citep{ND89,vDi21}. 
Their transport on $\sim$1000 AU scales is facilitated by the low densities and
scattering in the outflow cavities \citep{Spa95}. Similar signatures of UV radiation were
 observed in a dedicated APEX-CHAMP$^+$ survey of $\sim$20 protostars \citep{Yil12,Yil15}. 

Far-infrared observations with \textit{Herschel} provided access to highly-excited transitions
of abundant molecules e.g., CO and H$_2$O, and detections of new species e.g., OH$^+$ \citep{Wyr10} and H$_2$O$^{+}$ \citep{Oss10}. 
As part of the `Water in star-forming regions with \textit{Herschel}' 
\citep[WISH,][]{vDi11,vDi21} key program, a statistically significant sample of low-mass
 protostars was surveyed with the Heterodyne Instrument for the Far-Infrared \citep[HIFI,][]{dGr10} 
and the Photodetector Array Camera and Spectrometer \citep[PACS,][]{Pog10}. Among the key 
findings are: (i) ubiquitous gas at temperature of $\sim$300 K with molecular signatures
resembling shocks \citep{Her12,Goi12,Kar13,Kar18,Gre13,Man13}; (ii) a consistently low abundance of H$_2$O of $\sim10^{-6}$ \citep{Kri17}; 
(iii) high abundances of the H$_2$O photodissociation products and other hydrides, in particular OH \citep{Wam13,Ben16}; 
(iv) velocity-resolved components in H$_2$O profiles arising close to the protostar, at the 
positions of hydrides \citep{Kri13}; (v) ratios of H$_2$O/OH and CO/OH consistent with UV-irradiated shock models \citep{Kar18}.
These results indicate that UV photons affect the physical and chemical structure in the immediate surroundings of low-mass protostars. 

In this paper, we perform a ground-based follow-up study on large-scale maps of HCN and CN
 toward a low-mass star-forming region in the Serpens Main. The CN/HCN ratio is an independent
  tracer of UV photodissociation, which allows to benchmark the results from \textit{Herschel}.
 We address the following questions: What is the morphology and spatial extent of the regions affected by UV radiation? 
Do chemical models validate the utility of this ratio as a tracer of UV-irradiated gas 
in low-mass star-forming regions? What is the UV field distribution in Serpens Main?

The paper is organized as follows: Section~\ref{section:observations} describes the Serpens
 Main region and its protostars, the observations and
data reduction. Section~\ref{section:results} shows the submillimeter maps and the line
 profiles at selected positions.
Section~\ref{section:analysis} shows the determination of molecular column densities and
 their comparisons to the chemical model. Section~\ref{section:discussion} discusses the
  results in the context of complementary studies and 
Section~\ref{section:conclusions} presents the summary and conclusions. 

\section{Source sample and observations}
\label{section:observations}
\begin{table*} 
\caption{Properties of targeted protostars in the Serpens Main region} 
\label{table:2} 
\centering  
\begin{tabular}{l c c r r c l l} \hline\hline 
Source & R.A. & Dec. & $T_\mathrm{bol}$ & $L_\mathrm{bol}$ & Class & Other names\\ 
& (J2000.0) & (J2000.0) & (K) & (L$_\odot$) & &\\ 
\hline
SMM 1 & 18 29 50.0 & +01 15 20.3 & 37 & 115.2 & 0 & Ser-emb6, FIRS1, EC41, Bolo23\\
SMM 2 & 18 30 00.5 & +01 12 57.8 & 34 & 7.2 & 0 & Ser-emb4, Bolo28\\
SMM 3 & 18 29 59.6 & +01 13 59.2 & 35 & 7.1 & 0 & Ser-emb26, Bolo26\\
SMM 4 & 18 29 57.0 & +01 13 11.3 & 68 & 5.1 & I & Ser-emb22, Bolo25\\
SMM 5 & 18 29 51.4 & +01 16 38.3 & 151 & 3.7 & I & Ser-emb21, EC53, WMW24, Bolo22 \\
SMM 6 & 18 29 57.8 & +01 14 05.3 & 532 & 43.1 & I & Ser-emb30, EC90, WMW35, SVS20S, Bolo
28 \\
SMM 8 & 18 30 01.9 & +01 15 09.2 & 15 & 0.2 & 0 & Bolo30\\ 
SMM 9 & 18 29 48.3 & +01 16 42.7 & 33 & 11.0 & 0 & Ser-emb8, ISO241, WMW23, Bolo22\\
SMM 10 & 18 29 52.3 & +01 15 48.8 & 79 & 7.1 & I & Ser-emb12, WMW21, Bolo 23\\
SMM 12 & 18 29 59.1 & +01 13 14.3 & 72 & 10.0 & I & Ser-emb19, Bolo28\\
\hline 
\end{tabular}
\tablefoot{Source coordinates are taken from \cite{Sur16} except for Ser SMM8,
whose position is listed in \cite{Lee14}. The calculations of $T_\mathrm{bol}$
 and $L_\mathrm{bol}$ are presented in Appendix~\ref{app:seds}.}
\end{table*}
\begin{table*} 
\caption{Catalog of the observed molecular lines with IRAM and APEX \label{table:lines}} 
\centering 
\begin{tabular}{l c r c r c c c}
\hline \hline 
Mol. & Trans. & $E_\mathrm{u}/k_\mathrm{B}$ & \textbf{$n_\mathrm{crit}$} & Freq. & Telescope & Beam size & Efficiency \\
& & (K) &  (cm$^{-3}$) & (GHz)  & & ($\arcsec$) & $\eta_\mathrm{MB}$\\ 
\hline 
HCN & 1-0 & 4.25 & $1.7 \times 10^{5}$ ~\tablefootmark{a} & 88.631847 &  IRAM-EMIR & 28 & 0.81 \\ 
H$^{13}$CN & 1-0 & 4.14 & $1.3 \times 10^{6}$ ~\tablefootmark{a} & 86.340184 &  IRAM-EMIR & 29 & 0.81 \\ 
~ & 2-1 & 12.43 & $1.2 \times 10^{6}$ ~\tablefootmark{a} & 172.677881 &  IRAM-EMIR & 14 & 0.68 \\ 
CN & 1-0 & 5.45 & $1.1 \times 10^{5}$ ~\tablefootmark{a} & 113.490985 &  IRAM-EMIR & 22 & 0.78 \\ 
C$^{34}$S & 3-2 & 13.9 & $4.4 \times 10^{5}$ ~\tablefootmark{b} &  144.617109 & IRAM-EMIR & 16 & 0.74 \\
CS & 3-2 & 14.1 & $2.6 \times 10^{5}$ ~\tablefootmark{a} & 146.969029 &  IRAM-EMIR & 16 & 0.74 \\ 
$^{12}$CO & 6-5 & 116.2 & $1.0 \times 10^{5}$ ~\tablefootmark{c} & 691.473076 & APEX-CHAMP+ & 9 & 0.48 \\
C$^{18}$O & 6-5 & 110.6 & $2.6 \times 10^{5}$ ~\tablefootmark{b} & 658.553278 & APEX-CHAMP+ & 10 & 0.48\\
\hline 
\end{tabular} 
\begin{flushleft}
\tablefoot{Molecular data adopted from the Leiden Atomic and Molecular Database (LAMDA, \citealt{Sch05})
and the JPL database \citep{Pic98}.}
\tablefoottext{a}{\citet{Shi15}, assuming optically thin transition lines for an excitation temperature of 50 K.}
\tablefoottext{b}{Calculated based on JPL database \citep{Pic98} assuming an excitation temperature of 50 K.}
\tablefoottext{c}{\citet{Yil12}}
\end{flushleft}
\end{table*}

\subsection{Serpens as a low-mass star-formation site}
\label{subsection:serpens}

Serpens Main is a well-studied low-mass star-forming region located at a distance
 of 436$\pm$9 pc \citep{Ort17}. The identification and classification of young 
stellar objects (YSOs) was done in the region as part of the \textit{Spitzer} ‘From Molecular Cores to
Planet-forming Disks’ (c2d) survey \citep{Har07,Eno09,Eva09,Dun15}. 
Submillimeter sources were studied using continuum observations at 12, 25, 60,
 100 $\mu$m \citep{Hur96}, 800, 1100, 1300, 2000 $\mu$m \citep{Cas93} and 3 mm \citep{Tes98}.
  The outflow activity in Serpens Main was characterized using CO 2-1 \citep{Dav99}, and 
more recently with CO 3-2 and CO 6-5 for a subsample of sources (\citealt{Gra10},
 \citealt{Dio10}, \citealt{Yil15}). Ser SMM1, SMM3 and 
SMM4 were observed with \textit{Herschel} as part of the WISH and `Dust, Ice, and Gas in Time' 
\citep[DIGIT,][]{Gre13,Gre16,YL18} programs. 

Figure \ref{fig1:seds} shows the continuum map at 250 $\mu$m corresponding to the region we observed with the
IRAM 30 m. The map was obtained with the \textit{Herschel} Spectral and Photometric Imaging 
REceiver (SPIRE; \citealt{Gri10}) as part of the ‘\textit{Herschel} Gould Belt Survey project' (\citealt{And10}). 

For the purpose of our analysis, we re-calculated the spectral energy distributions (SEDs) for 
all protostars in the region using the new continuum measurements at 70, 160, 250, 350 and 500 $\mu$m 
from PACS and SPIRE covering the peak of the SEDs and following the procedures outlined in \citet{And10}, \citet{Kir13}, \citet{Kon15}.
Figure~\ref{seds} and Table~\ref{SED_data} show the SEDs of the protostars and provide
 the flux density at each wavelength. Table~\ref{table:2} shows the bolometric luminosities,
  temperatures and the classification of the sources based on \citet{Eva09}. We note that Ser SMM1 has a bolometric luminosity consistent with intermediate-mass protostars. However, it is known to consist of 5 protostars contributing to the total luminosity, and we regard it as a boarderline low-mass protostar (\citealt{Hul17}).

\subsection{Observations and data reduction}
\label{subsection:data}
The observations at the IRAM 30 m telescope were performed between 14-Jul-2009 and 17-Jul-2009
 as part of the project ‘HCN/CN as UV-tracers in YSO envelope-outflow interfaces’
  (PI: L. Kristensen). The Eight MIxer Receiver (EMIR) bands E090 and E150 were used to observe HCN 1-0, CN 1-0
    and CS 3-2, and provided also additional detections 
of C$^{34}$S 3-2, H$^{13}$CN~\mbox{1-0} and H$^{13}$CN 2-1. The frequency range covers $\pm$ 53.59 MHz from the central line. The
backend was the Versatile SPectrometer Array (VESPA) autocorrelator and the 1 MHz filterbank
reaching a spectral resolution of 39 kHz (E150 band) and 78 kHz (E090 band).
Table~\ref{table:lines} shows the full list of molecular transitions observed with EMIR with the
respective frequency-dependent beam sizes and main beam efficiencies, $\eta_{\mathrm{MB}}$ from 0.68 to 0.81, used to convert antenna temperatures to main beam temperatures ($T_{\mathrm{MB}}$). 

\begin{figure*}[tb]
  \begin{minipage}[t]{.5\textwidth}
  \begin{center}  
    \includegraphics[angle=0,height=9cm]{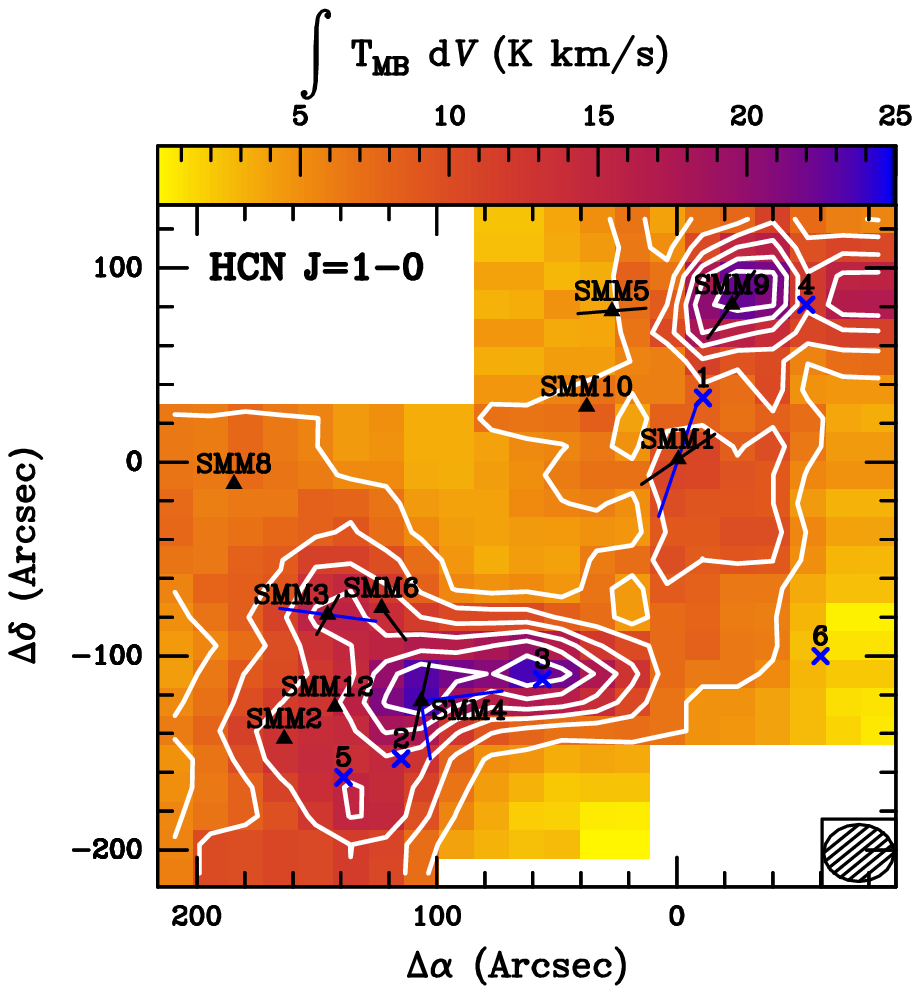}
     \includegraphics[angle=0,height=9cm]{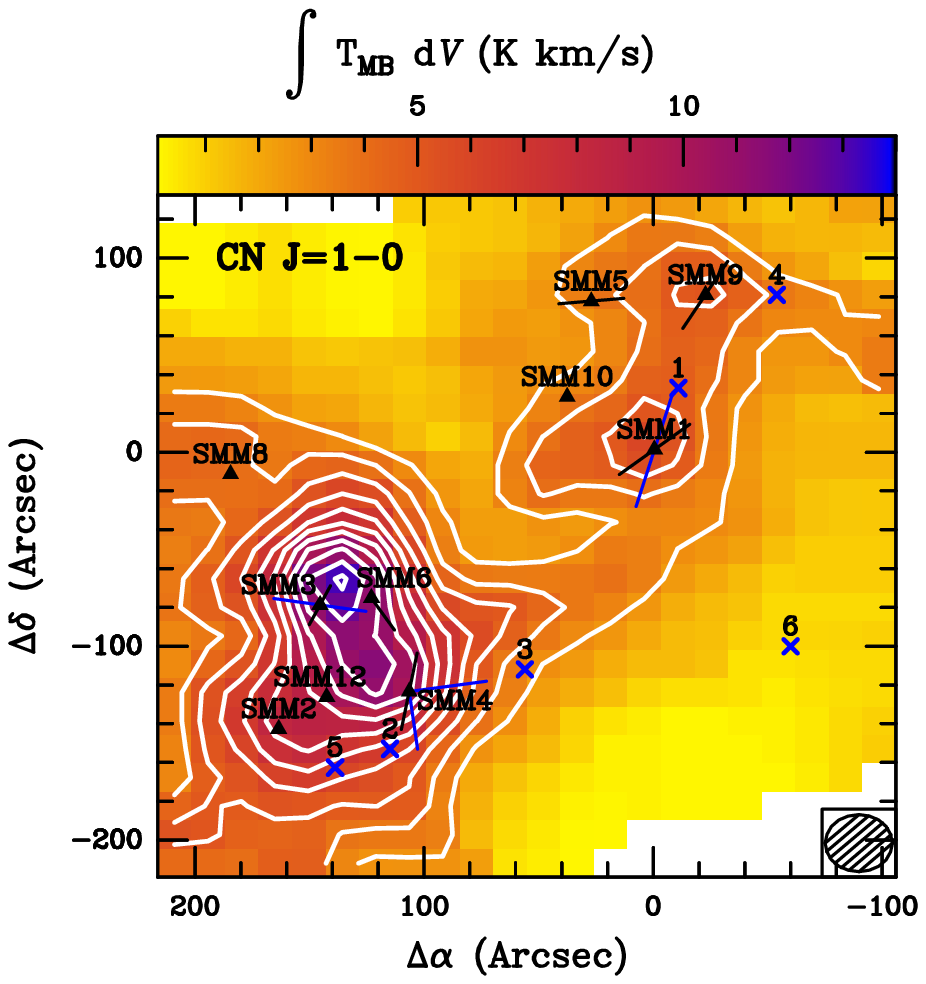}
    \end{center}
  \end{minipage}
  \hfill
  \begin{minipage}[t]{.5\textwidth}
  \begin{center}  
    \includegraphics[angle=0,height=9cm]{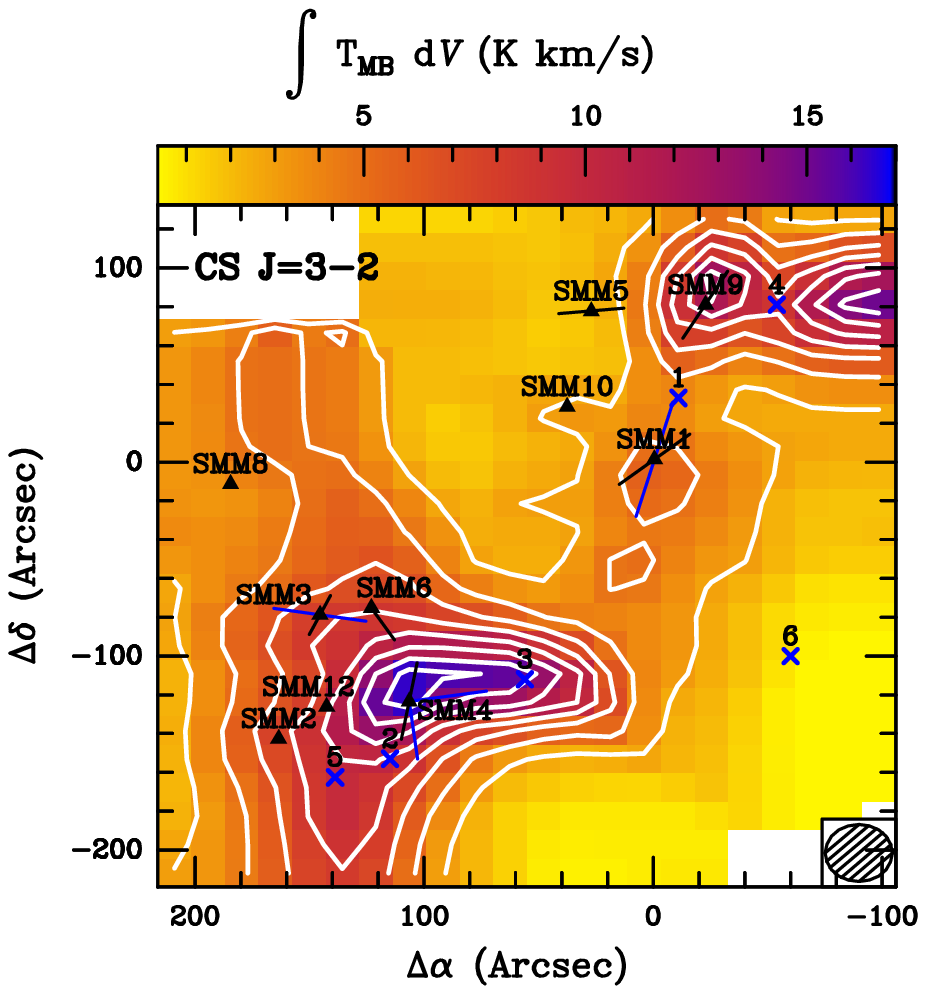}
     \includegraphics[angle=0,height=9cm]{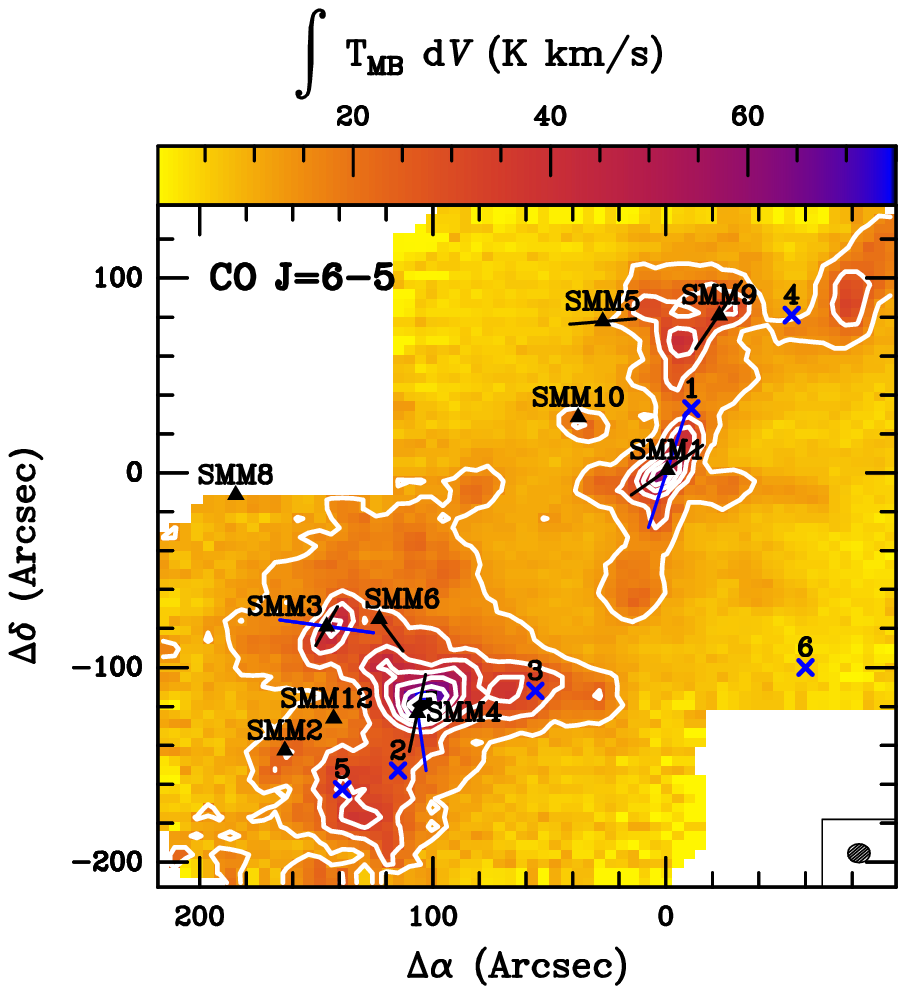}
    \end{center}
  \end{minipage}
    \caption{\label{iram_maps} Integrated intensity maps of HCN 1-0 (upper left panel), CS
3-2 (upper right panel), CN 1-0 (bottom left panel) and CO 6-5 (bottom right panel) toward
the Serpens Main star-forming region shown in contours and colors. Black triangles show the positions of protostars
(see Table 1) and blue crosses show the outflow positions selected for analysis (see Table 3). 
Solid lines show outflow directions from CO 6-5 (black; \citealt{Yil15}) and
CO 3-2 (blue; \citealt{Dio10}). The center of the maps (0,0) corresponds 
to ($\alpha$, $\delta$) = 18$^{\mathrm{h}}$29$^{\mathrm{m}}$49$\fs$6,
 +01$\degr$15$\arcmin$20$\farcs$5. Contour levels start at 20 $\sigma$ (6~K~km s$^{-1}$), 10 $\sigma$ (3~K~km s$^{-1}$),
 7 $\sigma$ (3.5~K~km s$^{-1}$) and 70 $\sigma$ (14~K~km s$^{-1}$) with steps of 10 $\sigma$, 3 $\sigma$, 3.5 $\sigma$ and 50 $\sigma$, for HCN, CN, CS and CO respectively.}
\end{figure*}
\begin{figure*} 
\includegraphics[width=12cm, center]{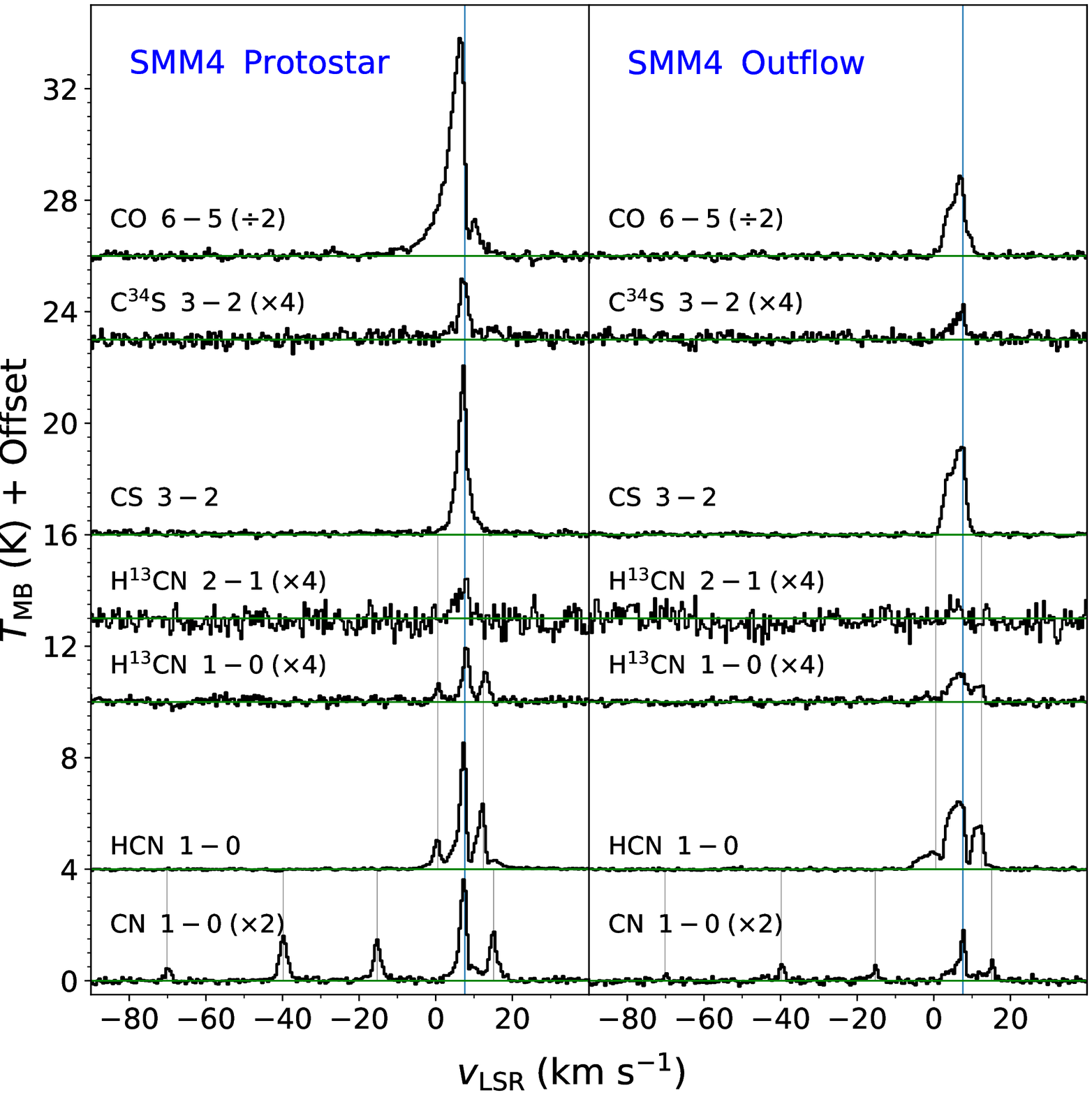} 
\caption{ \label{SMM4spectra} Line profiles of CO 6-5, C$^{34}$S 3-2, CS 3-2, H$^{13}$CN 2-1,
 H$^{13}$CN 1-0, HCN 1-0 and CN 1-0 lines at the protostellar position Ser~SMM4 (left) and
  the associated outflow position no. 3 (right). 
  Blue vertical lines show the main line component, and grey
     lines show the hyperfine splitting components. The maps in Figure 2 show the sum of all
      hyperfine components detected in the spectra of HCN and CN. } 
\end{figure*}

The on-the-fly (OTF) mapping mode was used in cross-directions to obtain two 340$^{\prime\prime}\times 340^{\prime\prime}$
 maps centered at ($\alpha_\mathrm{J2000}$,$\delta_\mathrm{J2000}$)=18$^{\mathrm{h}}$29$^{\mathrm{m}}$49$\fs$6,
+01$\degr$15$\arcmin$20$\farcs$5, and 18$^{\mathrm{h}}$29$^{\mathrm{m}}$56$\fs$6, +01$\degr$14$\arcmin$00$\farcs$3. 
The merging and data reduction were carried out with the CLASS package within GILDAS\footnote{See
http://www.iram.fr/IRAMFR/GILDAS}. For the sake of analysis, the EMIR spectra were
baseline-corrected and resampled to a resolution of 0.5 km s$^{-1}$, which is optimal for the observed linewidths in the range from $\sim$2~km~s$^{-1}$ to $\sim$21~km~s$^{-1}$ (Section 3.2). We fit a zeroth order
 baseline to all of the spectra in order to remove the continuum. The rms of extracted spectra varies from 0.024 K to 0.125 K in 0.5 km s$^{-1}$ bins (see Table~\ref{table:fluxes}). Figure \ref{iram_maps} shows the
   size and extent of the maps after the merging of two datasets and beam convolution.
    The CN map was convoluted to a the beam size of HCN.

The CHAMP+ dual-beam heterodyne receiver on the Atacama Pathfinder Experiment (APEX)
 telescope used for CO 6-5 observations at 691.5 GHz was originally presented in \cite{Yil15}.
  The observations were performed on June 16th 2009 using position-switching in the OTF
   mode resulting in maps covering 340$^{\prime\prime}\times 340^{\prime\prime}$. The
    Fast Fourier Transform Spectrometer (FFTS) was used as the backend with a resolution
     of 0.079 km s$^{-1}$ \citep{Kle06}. The rms varies from 0.12 to 0.23 K in 0.5 km s$^{-1}$ bins (Table ~\ref{table:fluxes}). 
Similarly, observations of the Ser SMM1 protostar in C$^{18}$O 6-5 were obtained with
 the APEX/CHAMP+ on October 23th 2009. The data reduction and analysis were performed
  in a similar way to the IRAM observations using the CLASS software.

\section{Results} 
\label{section:results}

In the following sections, we present IRAM 30~m maps and line profiles 
providing complementary information about the emission from 
protostars and their outflows, and large-scale cloud emission.
Differences in spatial extent reflect the range of gas and dust 
distributions, and associated physical conditions and processes. 
We calculate ratios of various transitions to indicate species tracing similar physical components, the gas temperature,
and line opacities. 

\subsection{Spatial extent of line emission}
\label{subsection:emission}

Figure \ref{iram_maps} shows the line emission associated with protostars and their outflows
 in HCN, CN and CS lines obtained with IRAM 30~m and CO with APEX. The integrated line intensities in HCN and CN are the sum of all hyperfine components detected in the spectra 
(see Figure \ref{SMM4spectra}). Maps in H$^{13}$CN 1-0 and C$^{34}$S 3-2 are presented in Appendix B.

HCN 1-0 emission is associated with clusters of protostars 
in the south-east and north-west parts of the map where dust emission at 250 $\mu$m is also detected (Figure 1). 
The emission peaks are the strongest at the positions of Ser SMM4 and SMM9 protostars and their outflows, and significantly weaker at Ser SMM1 and SMM3 - in part due to self-absorption in their line profiles (Figure A.3). Qualitatively, the pattern of emission in
HCN is similar to the CO 6-5 - a well-established outflow tracer (Figure 2; bottom right panel). The enhancement of HCN emission along the outflow is in agreement with previous surveys of low-mass protostars \citep{Bac01,Wal14}.
Any differences between HCN and CO likely stem from the higher critical density of the 
\mbox{HCN 1-0}, its lower upper energy level (see Table 2), and its 
significantly easier photodissociation due to UV photons than CO (see Section~\ref{section:discussion}). 

CS 3-2 emission shows a very similar spatial distribution compared to HCN 1-0 and CO 6-5, with most 
prominent structures associated with Ser SMM4. Some differences are seen, mostly in the 
surroundings of Ser SMM9, where CS emission is substantially weaker, in contrast to HCN which
 shows similar line strengths towards both protostars. 
Additionally, CS shows a relatively strong emission to the west of Ser SMM9, which coincides 
with the outflow from that protostar. The differences may result from CS 3-2 tracing
 higher-density gas than HCN 1-0 \citep{Shi15}.
 
In contrast to HCN and CS, CN 1-0 emission is more closely associated with the positions of protostars
and the continuum peaks at 250 $\mu$m, but not with the outflows (Figures 1 and 2). The emission is the strongest towards
Ser SMM4, SMM3, and SMM6, and the region in between these three protostars.
 The differences in spatial extent between CN and the outflow tracers suggest
  a different physical origin (see Section~\ref{subsection:extent}).

\subsection{Detection rates and line profiles}
\label{subsection:lines}
To explore the gas properties in the Serpens Main, we select 10 protostellar positions,
 5 outflow positions, and 1 position not associated with any known protostar. The selection
  benefits from previous observations of this star-forming region in other tracers and detections of line emission with IRAM
(Figure~\ref{iram_maps}). Figure~\ref{SMM4spectra} shows the line profiles of targeted species and transitions toward Ser SMM4
  and one of its outflow positions. Appendix~C shows the line profiles in the remaining positions. 

\begin{figure*}
\centering 
\begin{subfigure}{.45\textwidth} 
\centering
\includegraphics[width=1.\linewidth]{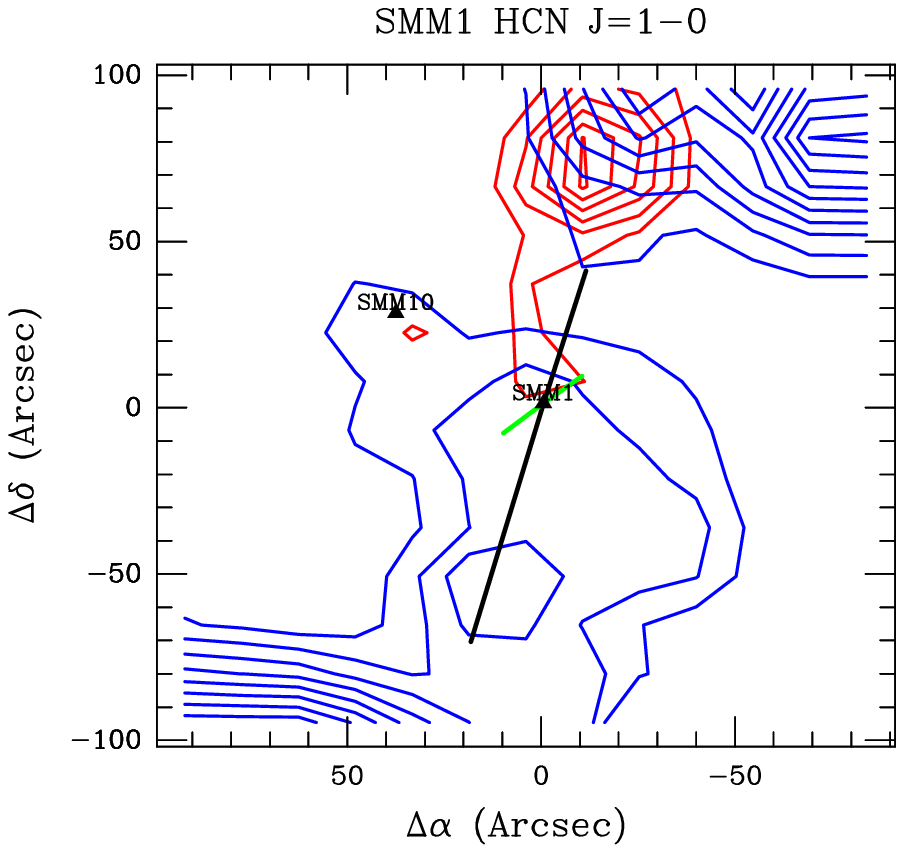} 
\end{subfigure}
\begin{subfigure}{.45\textwidth} 
\centering
\includegraphics[width=1.\linewidth]{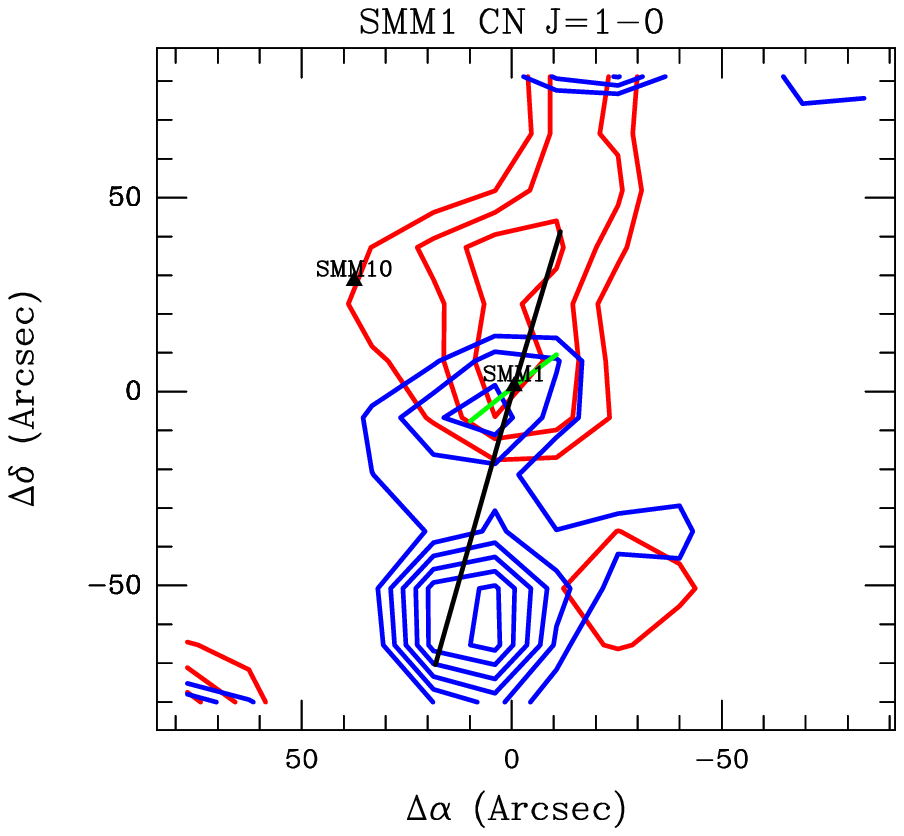} 
\end{subfigure}
\begin{subfigure}{.45\textwidth} 
\centering
\includegraphics[width=1.\linewidth]{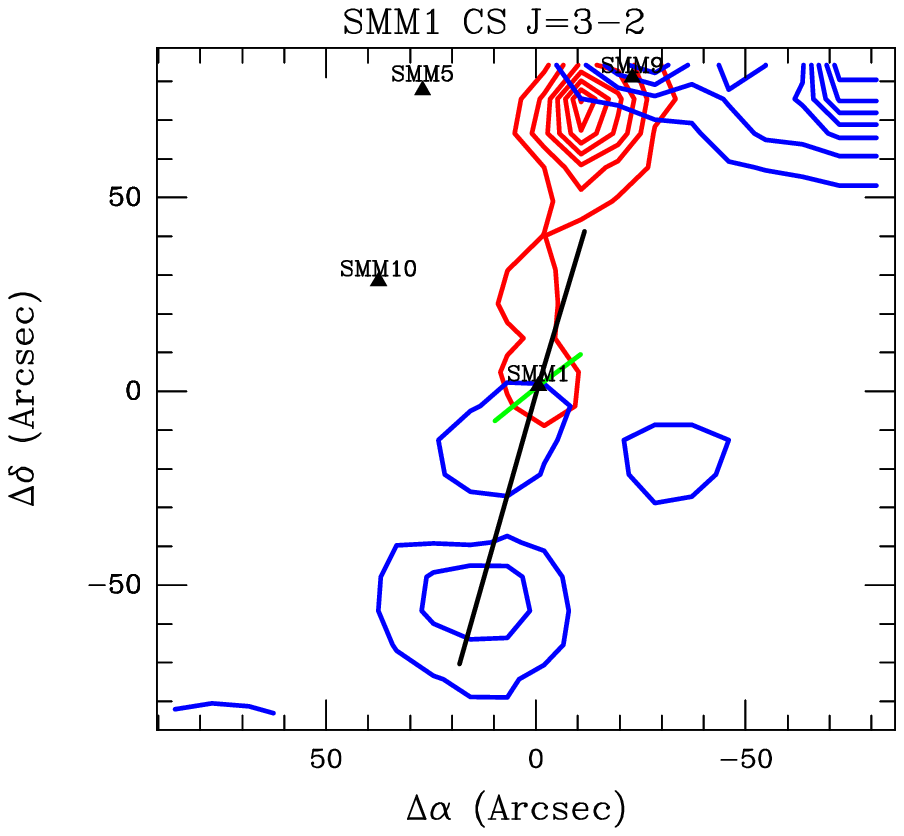} 
\end{subfigure}
\begin{subfigure}{.45\textwidth} 
\centering
\includegraphics[width=1.\linewidth]{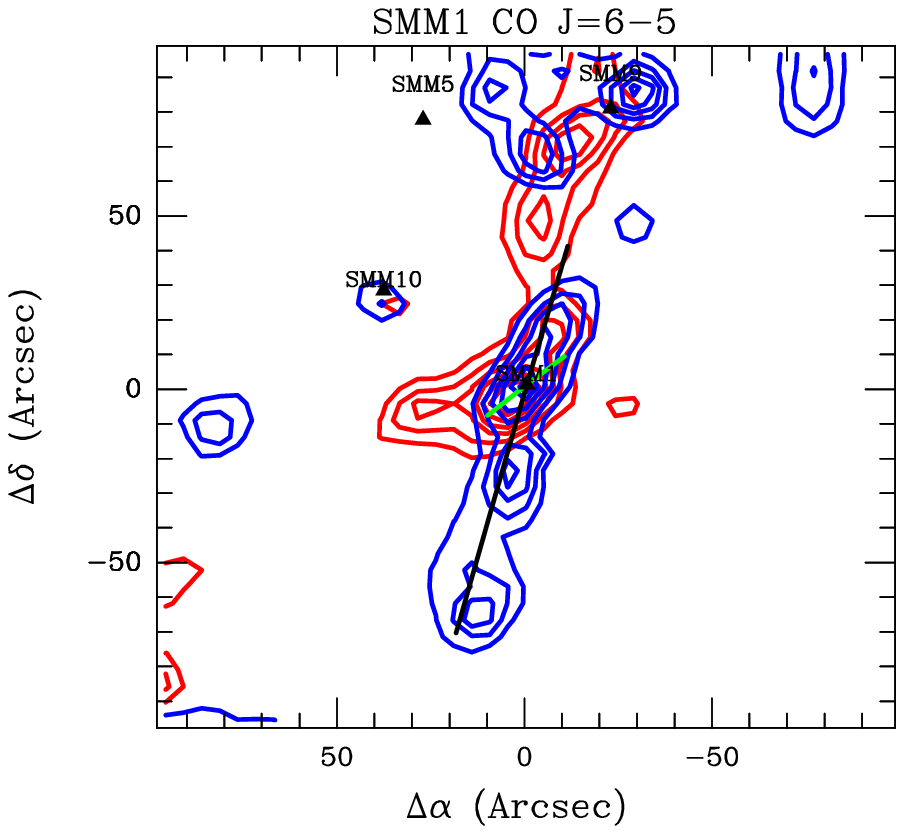} 
\end{subfigure}
\caption{Emission in the line wings of HCN 1-0, CN 1-0, CS 3-2 and CO 6-5 toward Ser SMM1 (blue and red contours). The velocity ranges used to calculate the integrated intensities are provided in Appendix D. Black and green
 lines correspond to outflow directions in CO 3-2 and CO 6-5 (the same as in Figure 2).}
  \label{smm1_red_blue}
\end{figure*}

All targeted species and their isotopologues are detected at the selected protostar positions
 in the Serpens Main region, except for H$^{13}$CN 2-1 which is detected only towards Ser SMM4
  and Ser SMM9. Similarly, emission in the outflow 
positions is always detected in HCN, CS, CN and CO. Considering the targeted rare isotopologue lines,
 detections are seen in C$^{34}$S 3-2 , H$^{13}$CN 1-0, but not in H$^{13}$CN 2-1.
In the case of HCN and CN, up to 3 and 5 components are detected as the result of the hyperfine splitting, respectively. 

The line shapes at both the protostar and outflow positions show relatively broad profiles, 
with outflow wings in CS extending to 13.6 km s$^{-1}$ and 9.8 km s$^{-1}$ at the protostellar
 and outflow positions, respectively (Figure \ref{SMM4spectra}). Clearly, the beam sizes of 
IRAM encompass a substantial amount of outflow emission even at the protostellar position. 

In order to quantify the emission in the line wings, we use the line profile of C$^{18}$O 6-5
 towards Ser SMM1, which is a well-established envelope tracer based on high S/N observations and modeling (\citealt{Kri10}, \citealt{Yil13}). The velocity range for the C$^{18}$O
  line sets the limits for the inner velocity ranges for the line wings in the observed
   transitions (Figure \ref{wing_ranges} in Appendix ~\ref{app:fluxes}). The outer velocity ranges are calculated 
separately for each line based on the signal detected above 2$\sigma$, selected by the visual inspection of the line profiles. 

Figure \ref{smm1_red_blue} shows the subset of the IRAM 30m large-scale maps toward Ser
 SMM1 where emission is integrated solely in the line wings velocity range. Emission in
  HCN, CN, and CS resembles that of warm gas in the outflows traced by CO 6-5, but some
   differences in morphology are clearly present. For example, HCN emission in line wings in the close vicinity of Ser SMM1 is much weaker than that of CN, which cannot be assigned to the self-absorption of material in the envelope. Low HCN emission might be rather related
     to the presence of large cavities around Ser SMM1, where ionizing radiation is capable of photodissociating HCN (\citealt{Hul16}, \citealt{Hul17}). A full analysis of the outflow properties will be presented in a forthcoming paper (Karska et al., in preparation).

In the case of Ser SMM4 (Figure \ref{SMM4spectra}), the emission in the line wings
 of CS 3-2 is 61\% of the total profile at the protostellar position and 72\% at
  the outflow position. Similar characteristics is 
seen in HCN line wings (48\% and 68\%, respectively). In contrast, CN 1-0 is detected 
mostly at source velocity, with 40\% of emission in the line wings. Among all 15
positions, exceptionally broad line wings in HCN and CS are detected at 
the Ser SMM1, Ser SMM9 and Ser SMM10 protostellar positions and outflow positions 
nr 1, 4, and 5, exceeding 70\% of the total profile in HCN and 58\% in CS. In fact, these high fractions of line emission in the wings are upper limits, because the opacity effects decrease the emission primarily in the source positions. In case of the weaker lines like CN, the fraction of wing versus on-source emission might be slightly lowered due to a limited signal-to-noise ratio of the spectra, which might not recover the wings at full. Higher signal-to-noise observations of rare isotopologue lines as well as CN would be needed to fully recover the on-source and wing emission.

The shapes of line profiles clearly indicate that the emission in various species
 is detected in different physical components of the protostellar systems. For the
  forthcoming analysis, we will consider the emission in the line wings alone and
   in the fully integrated profiles separately.

\subsection{Line ratios}
\label{subsection:ratios}
In this section, we calculate molecular line ratios at protostar and selected positions in the Serpens Main region separately for the fully integrated profiles and for the line wings. We discuss (i) the ratios of different isotopologues,
 constraining the line opacities, and (ii) the ratios of different species, reflecting 
their differences in spatial extent and relative abundances.

\subsubsection{Ratios of different isotopologues}
\label{subsection:isotopologues}


The line ratio of the same transition of two isotopologues is a tracer of line opacity of
 the more abundant species, assuming that the emission in the other isotopologue is optically
  thin. We calculate line opacities of HCN 1-0 and \mbox{CS 3-2} lines both at the
   protostellar and outflow positions using \citep{Gol84,Ma15}:
\begin{equation} 
\label{eq1} \frac{T_{\mathrm{HCN}}}{T_{\mathrm{H^{13}CN}}} = \left(\frac{\nu_{\mathrm{HCN}}}{\nu_{\mathrm{H^{13}CN}}}\right)^2
\frac{X[\mathrm{HCN}]}{X[\mathrm{H^{13}CN}]} \frac
{1-\exp(-\tau_{\mathrm{HCN}})}{\tau_{\mathrm{HCN}}} ,
\end{equation} 

\noindent where $X_{\mathrm{i}}$ and $T_{\mathrm{i}}$ refer to the abundance and antenna temperature
 of the isotopologue, and the same excitation for both isotopologues is assumed. We adopt an abundance ratio of HCN and H$^{13}$CN following the standard interstellar ratio of $^{12}$C/$^{13}$C of 68 \citep{Mil05}, and 20 for the CS and C$^{34}$S abundance ratio \citep{Ter10}, and assume that both isotopologues
   arise from the same physical region, described by the excitation temperature, $T_\mathrm{exc}$,
    and in local thermodynamic equilibrium (LTE). 

Optical depths at the protostellar positions range from $\sim$2.3 to 12.8 for HCN and from
 $\sim$0.3 to 4.4 for CS when the emission from the entire profile is considered (see Table
  \ref{table:taus}). At the outflow positions, the optical depths are $\sim$4.9-8.0 for HCN
   and $\sim$0.6-4.9 for CS. These values are in fact lower limits, because line profiles
    of HCN and CN show self-absorption toward many positions (see Figures \ref{Spectra_protostars}
     and \ref{Spectra_outflows}). To avoid this effect, we calculate optical depths in the line wings alone. 

HCN emission is optically thin ($\tau<1$) in the line wings towards the Ser SMM3, SMM5, and
 SMM10 protostellar positions (which also include emission from the outflows) and pure outflow
  positions 1 and 2. CS emission is optically thin in Ser SMM3, SMM6, and SMM10, and outflow
   position 3. In the remaining positions, $\tau$ ranges from 1.3 to 2.7 (HCN) and from 1.2
    to 5.6 (CS), indicating optically thick emission. 
    
    The optical depth of the CN emission is calculated using the ratios of its hyperfine-splitted
 components using entire line profiles. The ratio of the 
F=3/2$\rightarrow$1/2, F=5/2$\rightarrow$3/2, and F=1/2$\rightarrow$1/2 components is 0.1235:0.3333:0.0988
 in the optically thin limit \citep{Ska83,cdms}. We find that CN lines are generally optically thin toward
  the selected positions, with the exception of Ser SMM1, SMM5, and SMM10 where $\tau$ is
   $\sim1.1-2.1$ for the two strongest components (see Table \ref{table:taus}). 

 Similar method is used to verify the assumption that the H$^{13}$CN emission is optically thin. We use the ratios of the HCN hyperfine-splitted components as a proxy for H$^{13}$CN following \cite{Lou12}. The ratio of the F=2$\rightarrow$1 to F=1$\rightarrow$1 lines, which is a recommended tracer by \cite{Mul16}, indicates optical depths in the range from 0.77 to 2.76, and a median of 1.29 (see Table E.4). Thus, the line is optically thick toward some positions, and may not provide a reliable benchmark for HCN. The optical depths of HCN determined using hyperfine-splitted components show $\tau_\mathrm{HCN}$ in the range from 1.48 to 4.12, with a median of 2.74. These values are generally lower than those determined using H$^{13}$CN, but do not change the conclusion that HCN is optically thick.

\begin{figure} 
\centering 
\includegraphics[width=10cm]{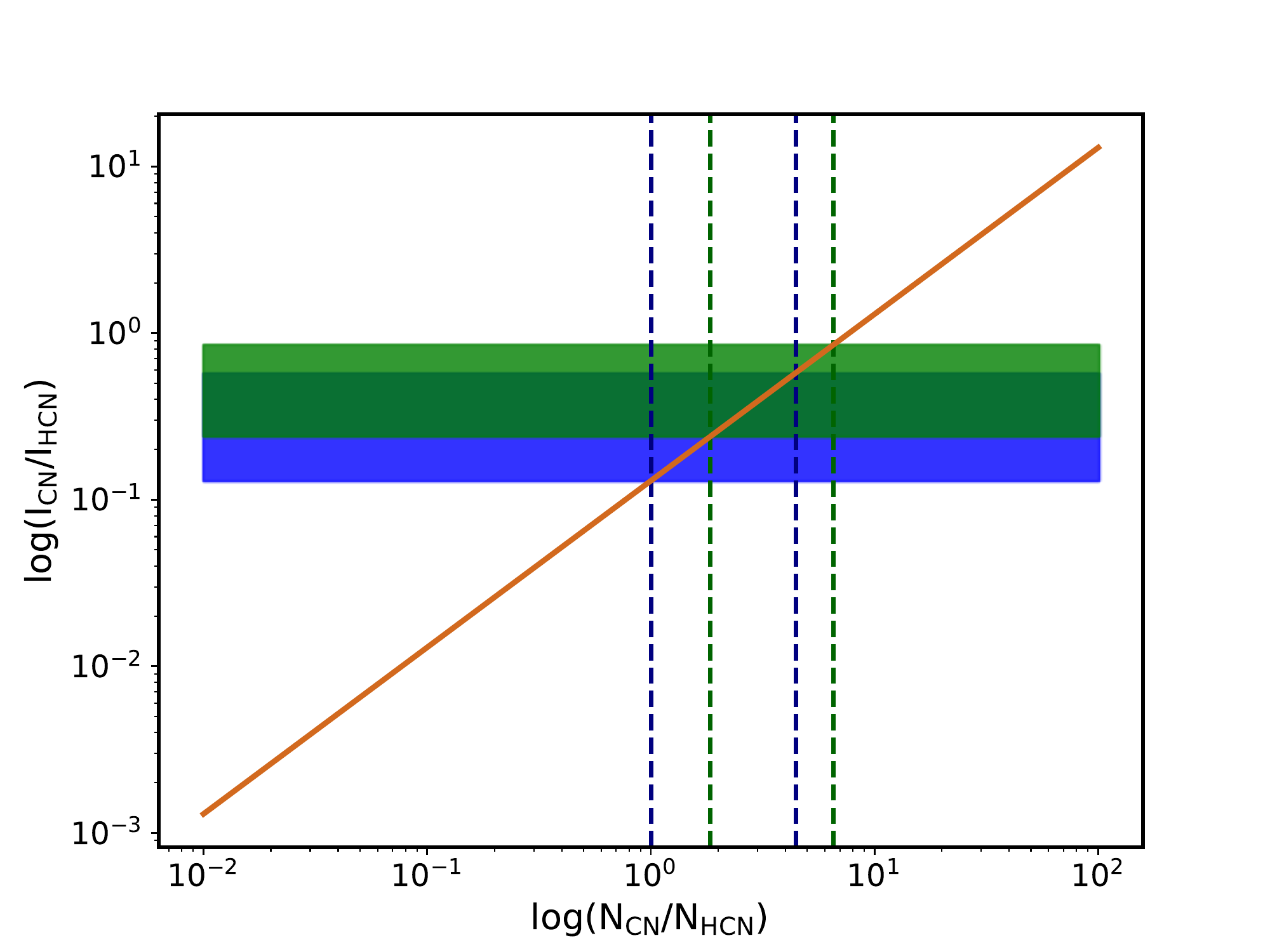}
\caption{The ratio of CN to HCN column densities obtained with RADEX for 
$n_\mathrm{H} = 10^5$ cm$^{-3}$ and $T_\mathrm{kin} = 50$ K for a range of line
 intensity ratios (orange line). The observed line intensity ratio is shown as a
  rectangle: light green color corresponding to the values observed at the 
protostar position, blue color for the outflow positions and dark green color
 for both regions. Column density ranges are marked with black and navy
  emdashed lines for protostars and outflow positions, respectively.} 
\label{model} 
\end{figure}
\begin{table} \caption{Model-dependent column density ratios of CN to HCN 
at protostar and off-source positions} 
\centering 
\label{RADEX_Ns} 
\begin{tabular}{c c c c} 
\hline\hline $n_\mathrm{H_2}$ & $T_\mathrm{kin}$ &
$N_\mathrm{CN}$/$N_\mathrm{HCN}$ &
$N_\mathrm{CN}$/$N_\mathrm{HCN}$\\ 
$($cm$^{-3})$ & (K) & Protostars & Off-source \\ 
\hline 10$^{4}$ & 50 & 1.57-5.55 & 0.85-3.79 \\
10$^{4}$ & 100 & 1.40-4.98 & 0.76-3.40 \\
10$^{4}$ & 200 & 1.33-4.70 & 0.72-3.21 \\ 
10$^{4}$ & 300 & 1.32-4.66 & 0.71-3.18 \\ 
10$^{4}$ & 400 & 1.45-5.13 & 0.78-3.50 \\ 
10$^{4}$ & 500 & 1.56-5.53 & 0.85-3.77 \\ 
\hline
10$^{5}$ & 50 & 1.85-6.55 & 1.00-4.67 \\
10$^{5}$ & 100 & 1.87-6.23 & 1.01-4.52 \\
10$^{5}$ & 200 & 2.01-7.12 & 1.09-4.86 \\ 
10$^{5}$ & 300 & 2.14-7.60 & 1.16-5.18 \\ 
10$^{5}$ & 400 & 2.22-7.87 & 1.20-5.37 \\ 
10$^{5}$ & 500 & 2.27-8.02 & 1.23-5.47 \\ 
\hline
10$^{6}$ & 50 & 2.52-8.93 & 1.37-6.10 \\ 
10$^{6}$ & 100 & 2.84-10.04 & 1.54-6.85 \\ 
10$^{6}$ & 200 & 3.17-11.22 & 1.72-7.65\\
10$^{6}$ & 300 & 3.33-11.80 & 1.80-8.05\\
10$^{6}$ & 400 & 3.01-10.67 & 1.63-7.28\\
10$^{6}$ & 500 & 2.75-9.73 & 1.49-6.64\\
\hline 
\end{tabular} 
\end{table}

Qualitatively, the optical depths determined for HCN, CS, and CN are in agreement with simple
 calculations for total line profiles using the 1D non-LTE radiative-transfer code RADEX \citep{vdT07}.
  Adopting a hydrogen density of $10^5$ cm$^{-3}$ and a gas kinetic temperature of 50 K typical
   for low-mass protostars \citep{Mot14}, we find optical
depth of CS 3-2 line as 2.1, assuming column densities of $5 \times 10^{13}$ cm$^{-2}$. Using a revised collisional rates for HCN adopted from \citealt{Mul16}, the optical depth calculated for HCN 1-0 is 2.9, assuming column densities of $3 \times 10^{14}$ cm$^{-2}$. This value is in agreement with the optical depth calculated from F=2$\rightarrow$1 to F=1$\rightarrow$1 ratio.
 Similar optical depths (within a 
 factor of 2) are found for gas temperatures of 300 K and number densities of $10^4$ cm$^{-3}$.

In summary, HCN 1-0 and H$^{13}$CN 1-0 are optically thick, and CN 1-0 is typically optically thin. Thus, the opacity effects influence the relative spatial distribution of CN and HCN, and their ratio. 

\subsubsection{Ratios of different species}
\label{subsection:statistics}
Line ratios of various species show differences related to their local abundances and
 excitation conditions. The CN/HCN intensity ratio is expected to trace regions affected by UV radiation. 

Protostars with the highest CN/HCN intensity ratio, where the impact of UV might be the strongest,
 are located in the south-east subcluster. Here, the peak of CN 1-0 emission is detected close
  to Ser SMM3 and SMM6 (Figure 2). Both protostars drive outflows detected in CO 6-5, and less
   clearly in HCN (see also Figure C.1). In fact, the HCN/CO ratio at Ser SMM3 and SMM6 is on
    the low side (0.27 and 0.44 respectively) compared to the average of 0.41$\pm$0.22 for
     protostar positions and 0.59$\pm$0.23 for off-sources positions (Table 3). 

The lowest CN/HCN intensity ratio is seen toward Ser SMM9, and to a lesser extent toward
 Ser SMM8 and SMM10. The former source drives a large outflow, where bright emission in
  HCN likely results from its high abundance in the warm gas (similarly to off-source
   position 3). The latter sources are characterized by a very weak molecular emission
    in general (Figure 2). Ser SMM8 is a low luminosity sources ($L_\mathrm{bol}$ of 0.2
     $L_{\odot}$, see Table 1), which is located on the map edge in CO 6-5. Ser SMM10 is
      associated with CO 6-5 peak, but is clearly very weak in both CN and HCN.

The median value of the CN to HCN intensity ratio at the protostellar positions is
 $0.63 \pm 0.19$. Taking into account the emission in the line wings, the ratio is
  slightly lower and equals $0.40 \pm 0.18$. It is generally consistent with the
   ratios measured for each source using the full profile. The ratio calculated for
    the selected off-source positions equals $0.43 \pm 0.18$ and $0.20 \pm 0.18$,
     for the full profile and wings, respectively.
  
The CN/HCN intensity ratio for the Class 0 protostars is $0.61 \pm 0.23$ and for
 the Class I protostars is $0.67 \pm 0.17$, indicating that the evolutionary stage
  does not strongly affect the measured ratios. A slightly higher CN/HCN ratio for
   Class I protostars is consistent with increasing opening angles of the outflows
    for more evolved sources and larger areas affected by UV radiation (see Discussion). The opacity effects overestimate the ratio of CN and HCN, and could have a slightly  stronger impact on the more massive envelopes of Class 0 sources (Section 3.3.1).

\section{Analysis}
\label{section:analysis}
In this section, we calculate column densities of the molecules with the radiative-code
 RADEX \citep{vdT07} and compare them to the results from the chemical-code Nahoon, run for a set of gas
  temperatures, densities and UV radiation fields. 
\subsection{Column densities of molecules from observations}
\label{subsection:column_densities}


In case of optically thin lines and LTE conditions, the column density of the 
upper level of a given molecule, $N_\mathrm{u}$, can be calculated using Equation~\ref{eq3},
 where $\beta = 8\pi k / h c^3$ is a constant equal to 1937 cm$^{-2}$, $W$ is the integrated
intensity of the emission line ($\int{T_{mb} \, dV}$), $A$ is the Einstein coefficient and $\nu$ is the transition 
frequency between the upper and lower level expressed in GHz \citep[e.g., ][]{Yil15}. 
\begin{equation} 
\label{eq3} N_\mathrm{u} = \beta \, \frac{\nu \,W}{A} 
\end{equation} 

In order to calculate the total column density of a given molecule, 
we use the following expression and adopt a gas temperature of 50 K \citep{Yil12}. 
\begin{equation} 
\label{eq4} N_\mathrm{tot} = Q(T_\mathrm{exc}) \, \exp(\frac{E_\mathrm{u}}{k_\mathrm{B} \, T_\mathrm{exc}}) \frac{N_\mathrm{u} }{g_\mathrm{u} } ,
\end{equation} 
where $Q$($T$) refers to the 
temperature-dependent partition function for the given molecule, $E_\mathrm{u}$ is
the upper energy level, $g_\mathrm{u}$ is level degeneracy, and 
$k_\mathrm{B}$ is the Boltzmann constant. Appendix~\ref{app:fluxes} shows the column densities
 obtained for all molecules, both at the protostellar and the outflow positions.

\begin{figure} 
\centering 
\includegraphics[width=10cm]{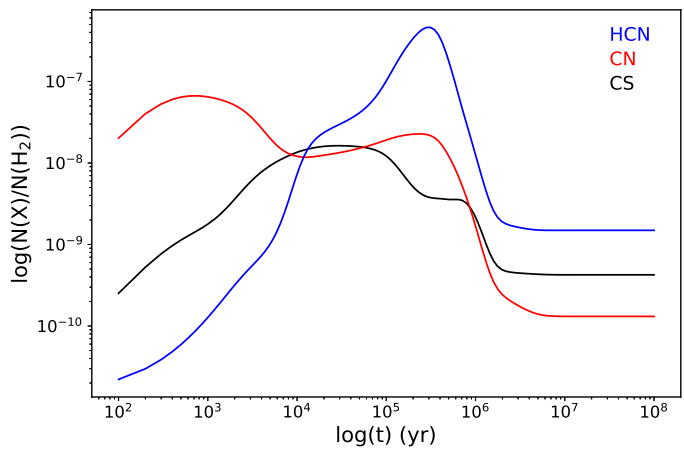} 
\caption{Time evolution of CN
(red line), HCN (blue line) and CS (black line) abundances obtained with Nahoon astrochemical code
with initial parameters of $n_\mathrm{H} = 10^4$ cm$^{-3}$, $T = 10$ K, $A_\mathrm{V}$ =
5 mag. The initial abundances of CN, HCN and CS equal zero at $t = 0$. The assumed
 cosmic-ray ionization rate is $1.3\times10^{17}$ s$^{-1}$, dust to gas mass ratio
is 0.01, dust grain radius is $10^{-5}$ cm and grain density is 3 g cm$^{-3}$.} 
\label{starless}
\end{figure}
\begin{figure} 
\centering 
\includegraphics[width=10cm]{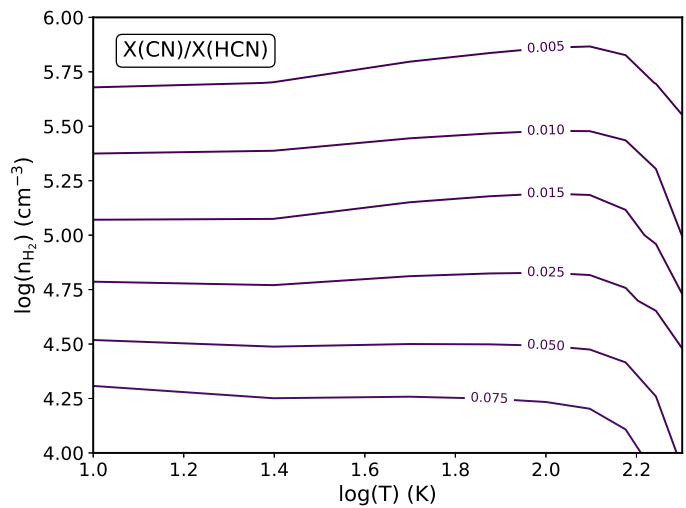} 
\caption{Contour plot of Nahoon sets
of models of CN/HCN abundances ratio with fixed visual extinction $A_\mathrm{V}$ = 5 mag
 and 10$^{7}$ yr after star formation began in the cloud.} 
\label{AV5} 
\end{figure}

Because of the optical thickness of HCN (Section~\ref{subsection:isotopologues}),
 we also employ the non-LTE radiative transfer code RADEX 
to obtain independent determinations of the column densities. In order to mimic
 the optically thin case, for the calculations we adopt a HCN column density of
  10$^8$ cm$^{-2}$. We vary the column density of CN from 10$^6$ to 10$^{10}$ cm$^{-2}$.
   The calculations are computed assuming a linewidth of 1.0 km s$^{-1}$ for typical
    physical conditions of the gas in low-mass star forming regions - the number
     densities, $n_\mathrm{H}$ of the order of $10^{4}$-$10^{6}$ cm$^{-3}$ 
and the kinetic temperatures, $T_\mathrm{kin}$, of 50-300 K \citep{Yil13,Mot14}. The models include hyperfine splitting, but no internal radiation field.

Figure \ref{model} shows an example model calculated for a density of $10^{5}$ cm$^{-3}$
 and kinetic temperature of 50 K, and its comparison to observations. The preimage of the
  observed line intensity ratios is equal to the ratios of column densities. Table~\ref{RADEX_Ns}
   shows calculations for various sets of $n_\mathrm{H}$ and $T_\mathrm{kin}$. The column
    densities ratios of CN to HCN increase both as a function of density and temperature,
     but their resulting range is relatively narrow. For the protostellar positions, the
      CN/HCN column density ratios are in the range from 1.3 to 11.8 ($1\sigma=3.2$).
       In case of off-source positions, which likely represent a larger variety of environments,
        the CN/HCN column densities range from 0.7 to 8.1 ($1\sigma=2.3$). Thus,
         the observed line ratios correspond to column densities ratios of the order of
          $\sim$1-12 irrespective of the gas parameters.

\subsection{Chemical model}
\label{subsection:nahoon}
\begin{table*} 
\caption{Dominant processes in CN and HCN chemistry at 50 K} 
\centering %
\label{reactions_50} 

\begin{tabular}{c c c c} 
\hline\hline 
Molecule & Weak UV fields & Medium UV fields & Strong UV fields \\ 
& (G$_0$ = $10^{-3} - 10^{-1})$ & (G$_0$ = $10^{-1} - 10^{1})$ & (G$_0$ = $10^{1} - 10^{6})$ \\ 
\hline 
\multirow{8}{*}{CN} & \multicolumn{3}{c}{\textbf{Destruction}}\\ 
& O + CN $\rightarrow$ N + CO & CN + h$\nu$ $\rightarrow$ C + N & CN + h$\nu$ $\rightarrow$ C + N\\ 
& CN + N $\rightarrow$ C + N$_2$ & O + CN $\rightarrow$ N + CO & \\
\vspace{2.5 pt} &\multicolumn{3}{c}{\textbf{Production}}\\ 
&N + CH $\rightarrow$ H + CN & N + C$_2$ $\rightarrow$ C + CN & HCN$^+$ + e$^-$ $\rightarrow$ H + CN\\
&CNC$^+$ + e$^-$ $\rightarrow$ C + CN & H + CN$^+$ $\rightarrow$ CN + H$^+$ & N + CH $\rightarrow$ H + CN\\
&N + C$_2$ $\rightarrow$ C + CN & & H + CN$^+$ $\rightarrow$ CN + H$^+$\\ & & & N + C$_2$ $\rightarrow$ C + CN\\
\hline
\multirow{9}{*}{HCN} & \multicolumn{3}{c}{\textbf{Destruction}}\\ 
&HCN + C$^+$ $\rightarrow$ H + CNC$^+$ & HCN + C$^+$ $\rightarrow$ H + CNC$^+$ & HCN + h$\nu$ $\rightarrow$ H + CN\\
&HCN + H$^+$ $\rightarrow$ H + HNC$^+$ & & \\
&HCN + h$\nu$ $\rightarrow$ H + CN & & \\
\vspace{2.5 pt} 
&\multicolumn{3}{c}{\textbf{Production}}\\ 
&N + CH$_2$ $\rightarrow$ H + HCN & H + CCN $\rightarrow$ C + HCN & HCNH$^+$ + e$^-$ $\rightarrow$ H + HCN\\
&H + CCN $\rightarrow$ C + HCN & HCNH$^+$ + e$^-$ $\rightarrow$ H + HCN & H + CCN $\rightarrow$ C + HCN\\
&HCNH$^+$ + e$^-$ $\rightarrow$ H + HCN & & \\ 
\hline \end{tabular} 
\end{table*}
The Nahoon chemical code is used to calculate theoretical abundances of 
molecules for a set of physical conditions and UV field strengths\footnote{We used the latest version of Nahoon code -- Nahoon\_kida.uva.2014}. It is a well-known almost purely
 gas-phase chemical code for astronomical applications (\citealt{Wak15}). The Nahoon solver
  computes the chemical evolution in time including 489 species and 7509 gas-phase and
   selected gas-grain reactions based on rate coefficients from the Kinetic Database
    for Astrochemistry (KIDA) database\footnote{http://kida.obs.u-bordeaux1.fr/}. 

The UV radiation in Nahoon is described through the relation
between the visual extinction $A_\mathrm{V}$ and the photodissociation rate coefficient $k$ as follows: 
\begin{equation} \label{eq5} 
k = \alpha e^{-\gamma A_\mathrm{V}} ,
\end{equation} 
where, $\mathrm{\alpha}$ and $\mathrm{\gamma}$ are the coefficients
of photodissociation for HCN, equal to $1.64 \times 10^{-9}$ s$^{-1}$ and $3.12$ s$^{-1}$, respectively \citep{Hea17}.

\begin{figure} 

\centering \includegraphics[width=9cm]{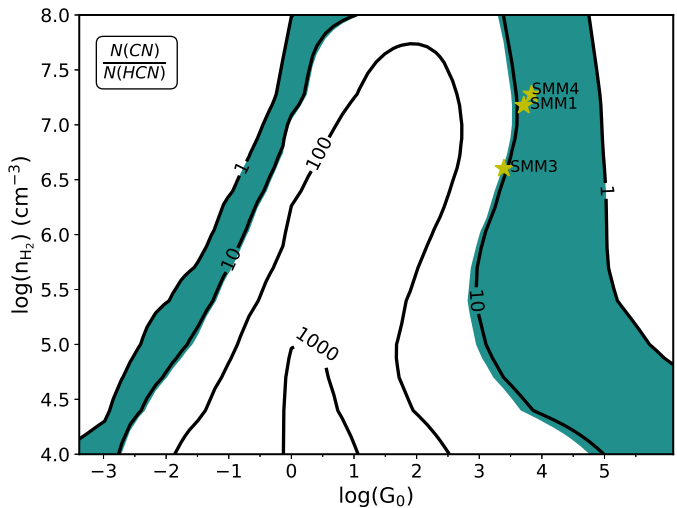} 
\caption{The column density ratio of CN to HCN from Nahoon for a 
range of hydrogen densities and UV field strengths assuming $T = 50$ K. 
The corresponding range of ratios from observations and radiative transfer models are shown in green (see Table 4). Yellow stars show the positions of Ser SMM1, SMM3, and SMM4 assuming their gas number densities at 1000 AU from \cite{Kri12}.} 
\label{G0_50} 
\end{figure}
\begin{figure*}
\centering 
\begin{subfigure}{.45\textwidth} 
\centering
\includegraphics[width=1.\linewidth]{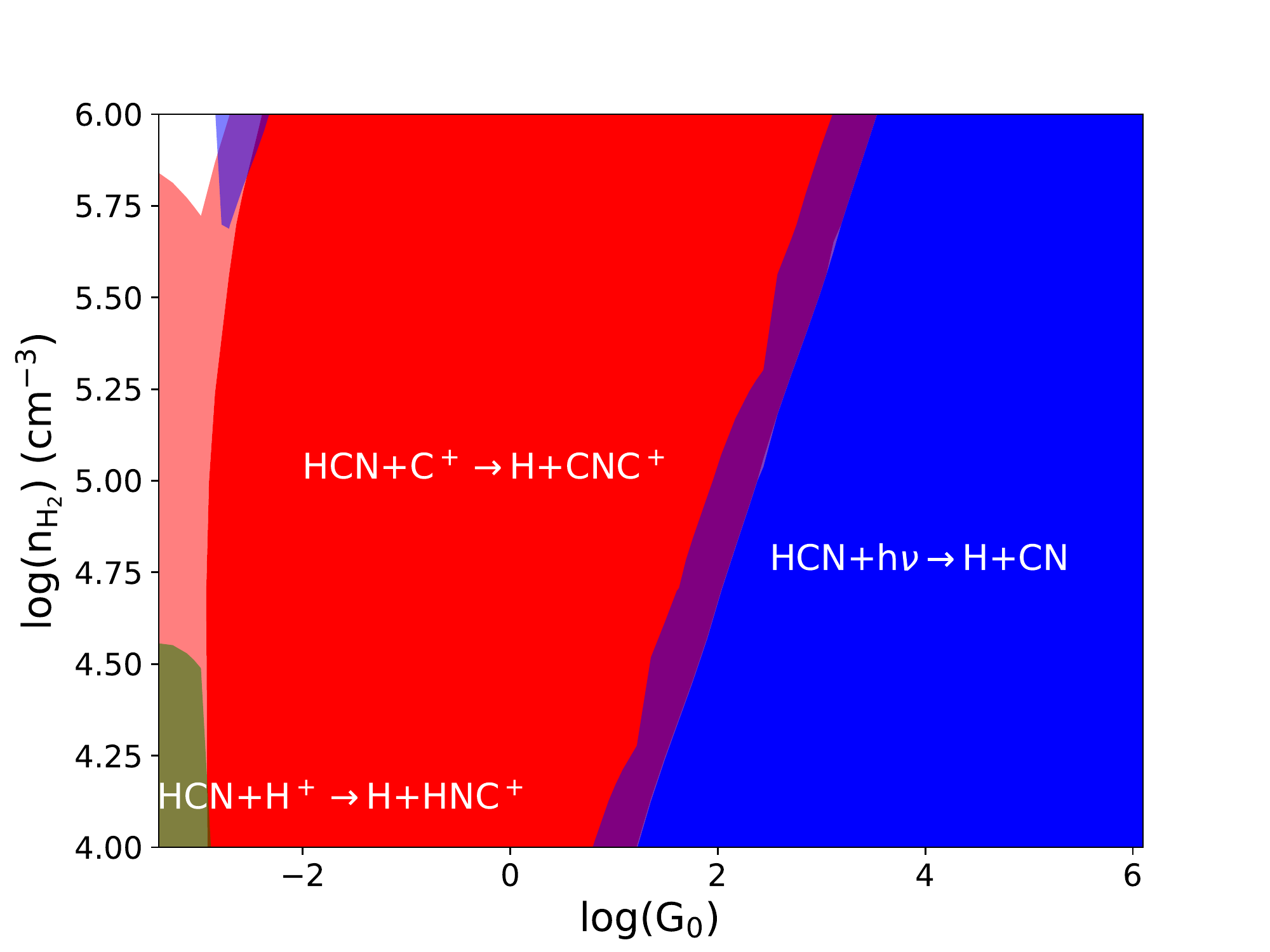} 
\caption{} 
\end{subfigure}
\begin{subfigure}{.45\textwidth} 
\centering
\includegraphics[width=1.\linewidth]{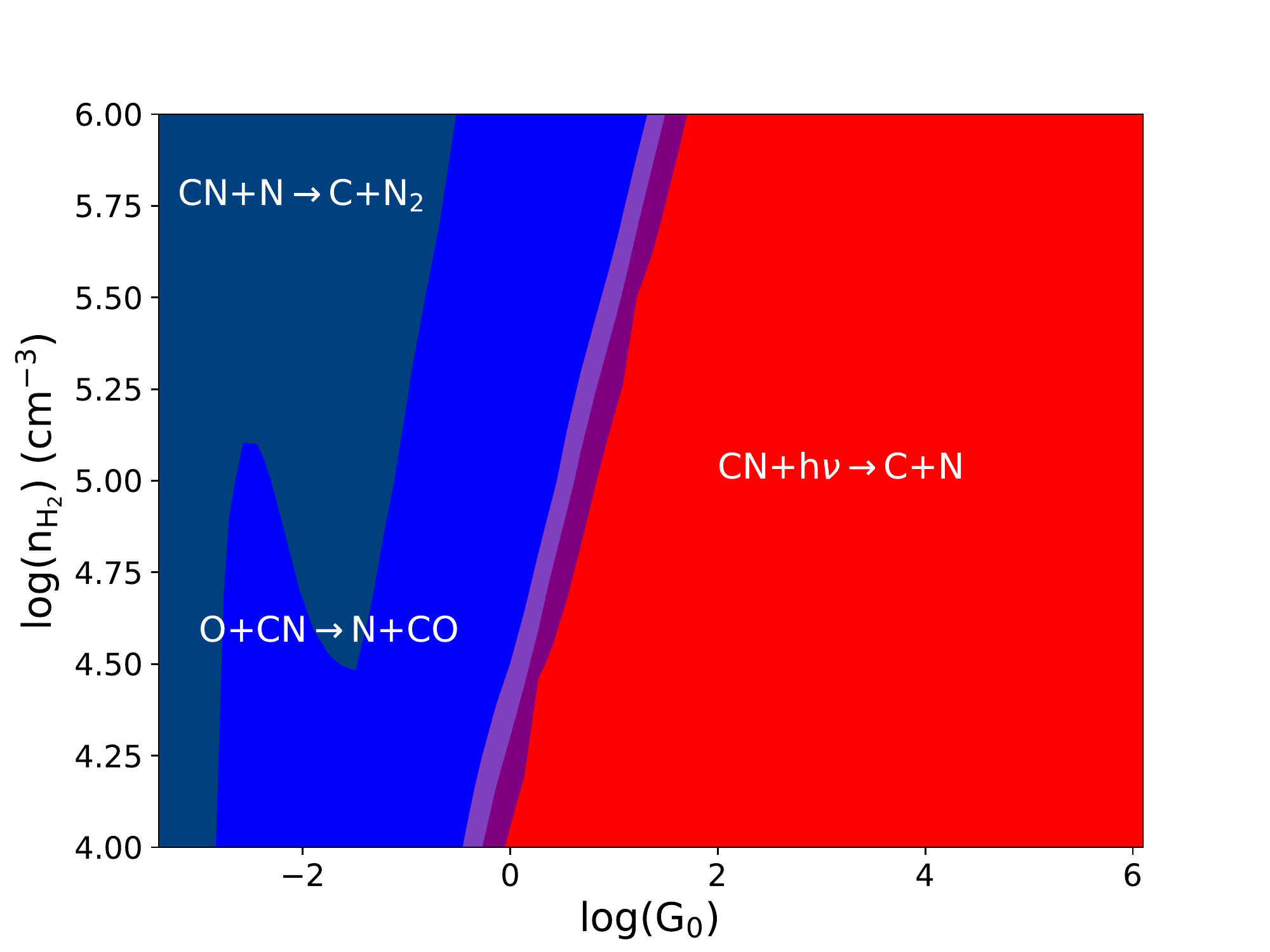} 
\caption{} \end{subfigure}
\caption{Dominant reactions
of HCN destruction (left) and CN destruction (right) assuming $T~=~50$ K. Reactions
 contributing at least to 50$\%$ of the total flux are marked with full colors.
  Transparent colors at the borders of different regimes correspond to 30$\%$-50$\%$ contribution. 
}
 \label{dest_50}
\end{figure*}

The evolution of the chemical network starts at the time of dense cloud formation.
 Figure~\ref{starless} shows a model corresponding to a typical dense cloud with
  a temperature of 10 K and hydrogen total density of $n_\mathrm{H} = 10^4$ cm$^{-3}$.
   The chemical composition of the CN, HCN and CS molecules becomes stable after
    $10^{7}$ yr; the HCN abundance is higher
than that of CN. Assuming that star-formation begins at $t = 10^6$~yr in a dense cloud,
 we use model abundances for all 489 species at this time as an input data for the forthcoming set of models.

The closest neighborhood of low-mass protostars is simulated based on the initial
 abundances of all species from the modeling of pre-stellar cores. We adopt a cosmic-ray
  ionization rate of $1.3\times 10^{17}$ s$^{-1}$ \citep{Cra78}. The sets of models
   are run for a temperature range between 10 and 200 K and a total hydrogen
    densities from $10^4$ cm$^{-3}$ to $10^6$ cm$^{-3}$. 

Figure \ref{AV5} shows the model results assuming a visual extinction of 5 mag which
 corresponds to a lack of UV radiation. In that case, HCN is more abundant than CN
  by about 2-3 orders of magnitude. The column density ratio of CN to HCN weakly
   depends on the gas temperature until $\gtrsim100$ K, so we fix the gas temperature
    at 50 K. 

Figure \ref{G0_50} shows the abundance ratio of CN to HCN as a function of
$n_\mathrm{H}$ and G$_0$ (at $T=50$ K), where G$_0$ is the far-ultraviolet radiation field (6 eV$<h\nu<13.6$ eV) in the units of the Habing Field, $1.6\times10^{-3}$ ergs cm$^{-2}$ s$^{-1}$ \citep{Hab68,kau99}. The models show a weak impact of hydrogen
 number densities on the $N_\mathrm{CN}$/$N_\mathrm{HCN}$ ratio, 
which mostly depends on the value of G$_0$. 
The ratio increases with G$_0$ until it equals $\sim1$. Subsequently, the
 $N_\mathrm{CN}$/$N_\mathrm{HCN}$
 decreases for higher G$_0$. Similar calculations performed at 300 K show a very
  similar pattern (See Fig.~\ref{G0_300}).  

The $N_\mathrm{CN}$/$N_\mathrm{HCN}$ ratios obtained from observations and radiative transfer
 models are $\sim1-12$ (see Section 4.1). 
Chemical models are consistent with the observed column density ratio of CN to HCN for
 UV fields that are 10$^3$ larger than the average interstellar value (Figure 8). 
The observations are also consistent with models with significantly lower 
UV fields, of 10$^{-3}$-10$^{-2}$ times the average interstellar UV field. For three sources observed with \textit{Herschel}, gas densities at 1000 AU and independent signatures of UV fields were found \citep{Kri12,Kri17,Kar18}. Using Figure 8, we estimate $G_\mathrm{0}$ toward Ser SMM1, SMM3 and SMM4 of $10^3-10^4$ the interstellar value. We discuss these values and compare them to other measurements in Section 5.3.

In order to understand the chemistry leading to the CN/HCN$>1$ in the presence of UV fields, both production and destruction of each molecule need to be investigated. Table \ref{reactions_50} shows the dominant reactions at 50 K for three UV field regimes
 - weak (G$_0$ = $10^{-3} - 10^{-1}$), medium (G$_0$ = $10^{-1} - 10^{1})$, and strong
  (G$_0$ = $10^{1} - 10^{6})$. These reactions account for more than 80$\%$ of accumulated
   total flux of all reactions, where the reaction flux is defined as reactants' abundances
    multiplied by the reaction rate coefficient.  
Additional reactions, where CN or HCN production or destruction is greater than 30$\%$,
 are listed in Appendix F. 

In the regime of strong UV fields (G$_{0}>10^{1}$), the dominant destruction route of both
 HCN and CN is photodissociation by UV photons. Similarly, in medium UV fields CN
  is also directly destroyed by UV photons, and HCN is removed by the reaction with C$^{+}$,
   which production also requires UV radiation. At a gas temperature of 300 K, corresponding
    to the bulk of gas in the outflows from low-mass protostars \citep{Kar18}, the impact of
     photodissociation of HCN becomes dominant already in the medium UV fields (see Appendix F).
      For low UV fields, the main destruction route of CN by collisions with H$_2$ leads to
       the production of HCN, and subsequent decrease in the CN/HCN ratio.

Figure \ref{dest_50} shows which reactions dominate the molecular destruction for a range
 of UV fields and also gas densities. HCN and CN photodissociation is more efficient at
  lower densities, and stronger UV fields are required to maintain photodissociation in
   denser regions. Clearly, outflow cavities carved by jets and winds in the envelopes of
    protostars facilitate the irradiation of the bulk of gas \citep[see also][]{Spa95,vKe09a}.

To summarize, the astrochemical models calculated using the Nahoon code reproduce
 the observed CN/HCN ratios in the Serpens Main for UV fields in the range from $10^3$ to $10^{5.5}$ times the interstellar radiation field, assuming gas densities of 10$^5$ cm$^{-3}$. Independent measurements of gas densities in individual protostars are needed to narrow down the determination of G$_\mathrm{0}$ for specific objects. Clearly, the strength of the UV field has an important effect on the chemical reactions at play and the resulting abundance ratios of molecules.

\section{Discussion}
\label{section:discussion}
\subsection{Spatial extent of HCN and CN in low-mass protostars}
\label{subsection:extent}

The immediate environment of protostars is subject to multiple physical and chemical processes
 which can be traced using a broad range of molecules and their transitions. Here,
  we propose to use the CN to HCN ratio as a tracer of UV irradiation associated with
   low-mass star formation on cluster scales ($\sim$1 pc). 

Large-scale emission from both CN and HCN is clearly detected in the Serpens Main
 (Figure 10, see also Section 3.1). The HCN emission shows a more extended 
 pattern, but overall the emission peaks of both species are connected with the positions of protostars and their outflows. On scales of individual objects, the morphology of CN 1-0 and HCN 1-0
    emission resembles that of CO 6-5 and CS 3-2 (Figure 4), which points to the outflow
     origin and presence in dense gas \citep{Yil15}. Yet, CN is visually more compact than
      HCN for some sources, which is also evidenced by the lower median CN/HCN intensity
       ratio toward outflow positions with respect to the protostellar positions. Thus,
        the impact of UV radiation, and the photodissociation of HCN to CN, is the
         strongest in the close neighborhood of the protostars.

The spatial offset between CN and HCN was also detected in the single prior mapping of a
 $\sim$pc-size outflow from a low-mass protostar L1157 in both tracers \citep{Bac01}.
  The HCN 1-0 emission in L1157 is basically co-spatial with CO \mbox{2-1}, whereas the
   CN 1-0 emission only reaches about half of the outflow extent seen in HCN. The
    compactness of the CN emission with respect to other outflow tracers was also seen
     in L483 \citep{Jor04}. The CN emission was interpreted as the tracer of outflow cavity
      walls by comparison to more evolved young stellar objects, where the CN emission
       is proportional to the strength of the UV field \citep{Jor04}. 
\begin{figure} 
\centering 
\includegraphics[width=9cm]{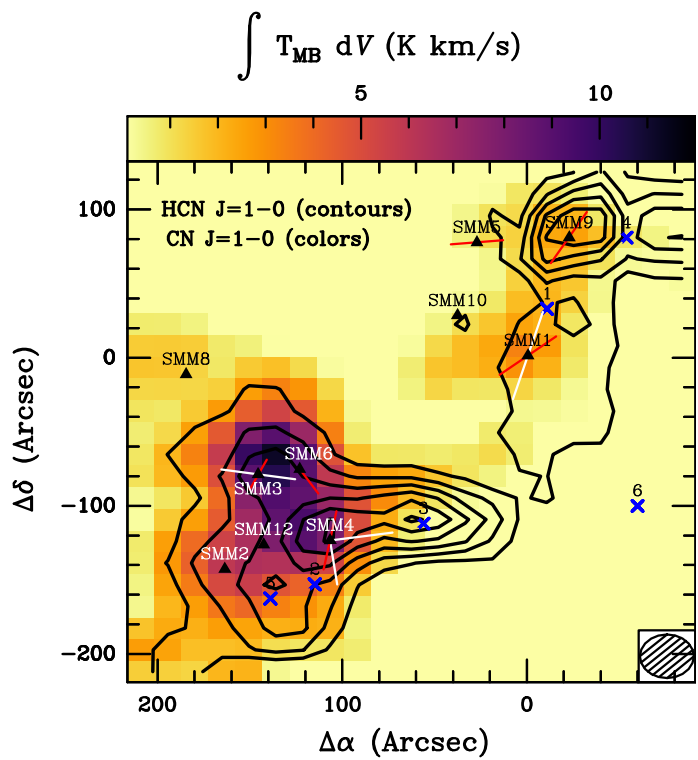}
\caption{Map of CN 1-0 (colors) and \mbox{HCN 1-0}
(contours) in the Serpens Main. The labels are the same as in Figure~\ref{iram_maps}.
 Contour levels start at 8~K~km s$^{-1}$ with the steps of 3~K~km s$^{-1}$.
The CN emission has been resampled to the beam size of HCN to compare the same emitting regions.} 
\label{cn10_and_hcn10} 
\end{figure}
\citet{Yil15} used $^{13}$CO 6-5 emission as an alternative way to quantify the spatial
 extent of UV radiation for a sample of $\sim30$ Class 0/I protostars. They isolated a
  narrow component of the line profiles attributed to the UV-heated gas, in excess of
   envelope emission. The gas affected by UV was found to be aligned with the outflow
    direction on spatial scales of $\sim$1000-2000 AU, consistent with the extent of
     the CN/HCN enhancement. The elongated pattern of $^{13}$CO 6-5 was more clearly
      detected in Class 0 sources, but the total amount of gas affected by the UV radiation was not found to be correlated with the evolutionary stage. 
      
\begin{figure} 
\includegraphics[width=0.55\textwidth]{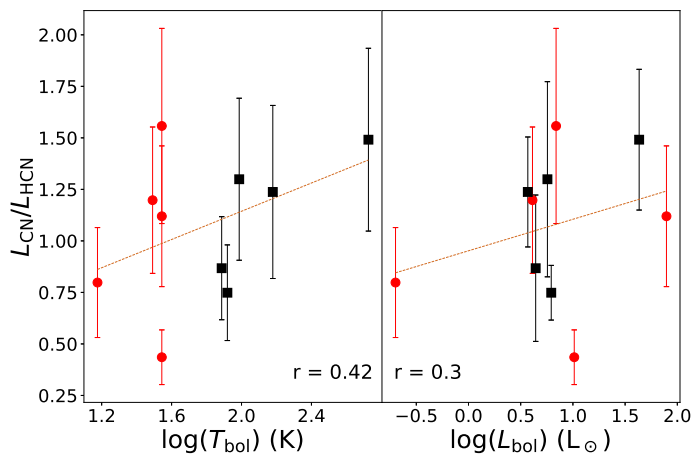} 
\caption{Correlations of line luminosity ratios with bolometric temperature and
 luminosity of protostars. Class 0 protostars are marked with red dots, while
  Class I with black squares. The Pearson coefficient of the correlation (r) is shown. }
\label{Tbol_Lbol_ratio} 
\end{figure}

      Figure~\ref{Tbol_Lbol_ratio} shows the CN/HCN luminosity ratio as a function of $T_\mathrm{bol}$ and $L_\mathrm{bol}$
         for all protostars observed in the Serpens Main. The strength of the correlations is quantified using the Pearson coefficient, $r$, where for the sample of 10 protostars $|r|$ of 0.33, 0.67, and 1 corresponds to a 1$\sigma$, 2$\sigma$, and $3\sigma$ correlation, respectively. Clearly, the Pearson coefficients of 0.3-0.4 indicate a lack of any correlation that would suggest evolutionary changes in the CN/HCN emission during the Class 0/I phases. It is consistent with a lack of significant changes in the [O I] line emission from Class 0 to Class I in a sample of about 90 protostars observed with \textit{Herschel}/PACS, which is also connected with the amount of UV radiation \citep{Kar18}.

The spatial extent of CN and HCN emission implies that UV photons are wide-spread in the
 Serpens Main region on cluster scales. The enhancement of the CN/HCN ratio is associated
  with the parts of outflows mostly affected by UV from protostars and does not depend on
   the evolutionary stage. The changes in the spatial patterns of CN and HCN with evolution
    on small scales would require higher spatial resolution observations.

\subsection{Chemical effects}
\label{subsection:chem}

In the presence of strong UV radiation, HCN photodissociates into CN and H, while CN
 requires even more energetic photons (>12.4~eV) to be dissociated (\citealt{vDi87}). This
  should lead to a higher abundance of CN molecules and an increase of the CN/HCN column
   density ratio. Despite many limitations and approximations, the chemical model presented
    in Section 4 demonstrates that the CN/HCN ratio is sensitive to the UV field. The strength
     of UV radiation distinguishes the dominant reactions. The density and temperature
      have less effect on chemistry, at least below a few hundred K. Three regimes can
       be defined: weak (G$_0$ < $10^{-1}$), intermediate (G$_0 \sim 10^{-1}-10^1$) and
            strong (G$_0$ > $10^{1}$) UV field. They correspond to regimes where the bulk of carbon is either in the molecular (CO), atomic (C), or ionized (C$^+$) form, which has a significant impact on chemical networks.

In starless, dark, non-turbulent clouds, HCN is more abundant than CN by about an
 order of magnitude (\citealt{Pra97}). This is in agreement with our model of a
  starless cloud after $10^5-10^6$ yr of cloud evolution (Figure~\ref{starless}).
   With the further chemical evolution of a starless cloud, the CN/HCN ratio decreases.
    In a weak UV field, the chemistry is similar to the star-less, dark, non-turbulent 
    interstellar cloud. Dominant formation channels of CN and HCN are reactions of nitrogen with hydrocarbons
(CH and CH$_2$, respectively), while destruction of CN is mostly dominated by the
reactions with neutral oxygen or nitrogen. 
HCN is less reactive than CN, and reactions with neutral atoms are not very efficient in this case.
 Many reaction channels drive the destruction of HCN, but reactants are less abundant than neutral
  atoms. The simultaneous impact of a few different reactions is not as effective as the
   reaction ruling the CN destruction. Those factors lead to the higher HCN abundances
    compared to CN. 
    
\begin{figure} 
\includegraphics[width=9cm]{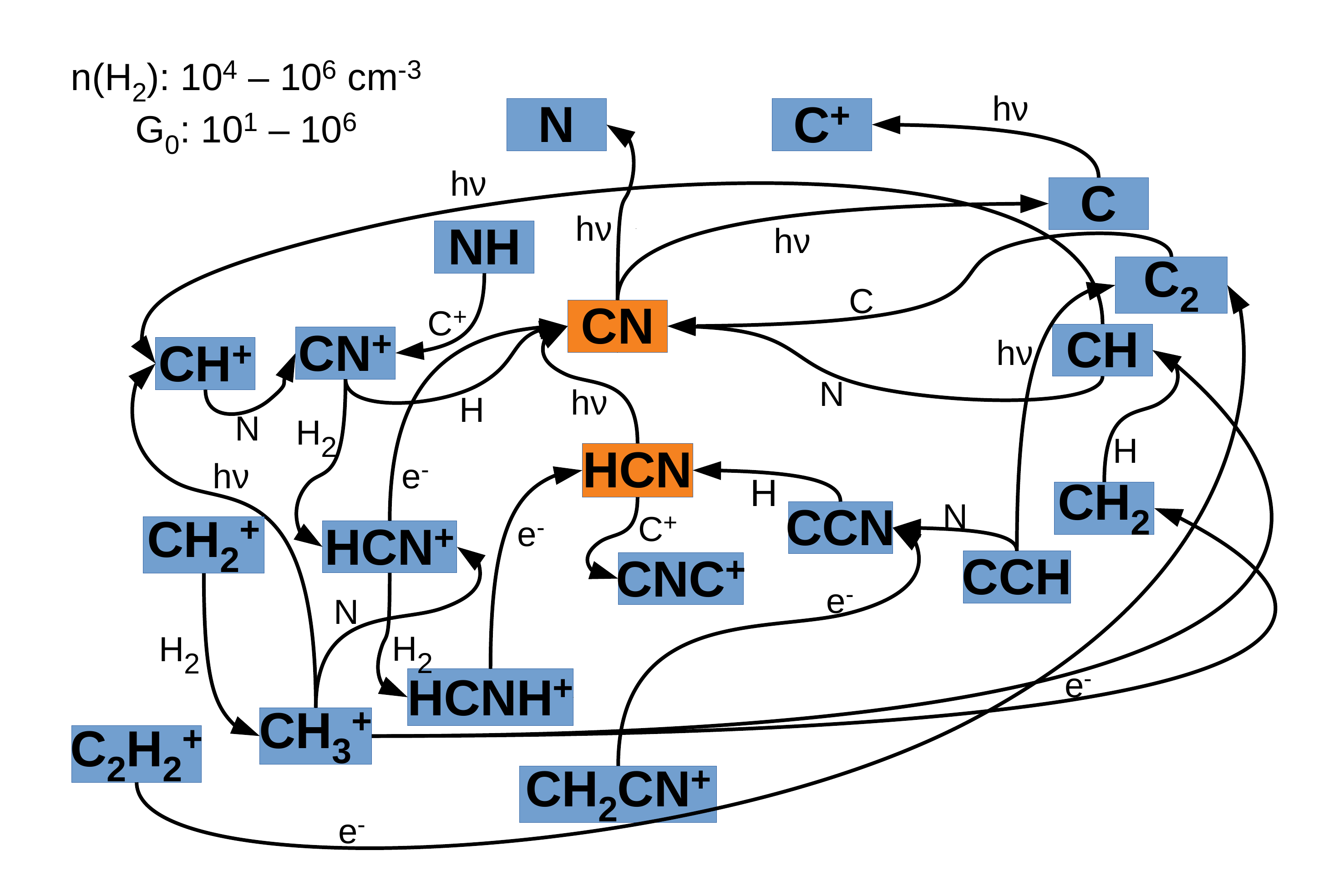} 
\caption{Reaction network predicted by the Nahoon model assuming the UV field strength is higher than 10 G$_0$. The dominant reactions are listed in Table ~\ref{reactions_50}}.
\label{reactions_largeG0} 
\end{figure}

As the intensity of UV radiation increases, photodissociation and photoionization become
 more important. For strong radiation field, the destruction of CN is dominated by reaction:
\ce{CN ->[UV photons] C +N}.\\
However, at intermediate UV field strength, photodissociation of CN is not efficient
 and the reaction with neutral oxygen is important:\\
\ce{CN + O -> N + CO}.\\
For intermediate and strong radiation fields, \ce{C+} is significantly more abundant than \ce{C}.
 It opens a very effective channel of HCN destruction:\\
\ce{C ->[UV photons] C+ + e-}\\
\ce{HCN + C+ -> H + CNC+}\\
\ce{CNC+ + e- -> CN + C}\\
The presence of this channel leads to a significantly lower concentration of HCN than CN
 for intermediate UV fields. The HCN molecule can be destroyed directly by UV photons:\\
\ce{HCN ->[UV photons] H + CN}.\\
This reaction is dominant for the strong UV fields. Nevertheless, the exact balance between
 the two ways of destroying HCN depends on temperature and density.

The CN/HCN ratio can also be shaped by several factors neglected in the chemical model,
 including grain chemistry, evaporation, turbulence, shocks, the spectrum of UV emission
  of a particular object. Ice chemistry, as well as grain sublimation, can significantly
   alter the results. Although not yet detected in the ISM \citep{Boo15}, HCN is postulated to exist in the
    interstellar ice with the abundance of $10^{-8}$ relative to hydrogen \citep{Kal15}. Thus, HCN can sublimate under the influence of increasing temperature
     or UV light, leading to a decrease in the CN/HCN ratio. CN may be also produced by the
      photodegradation and photoevaporation of CN-bearing complex organic molecules trapped
       in ice. Even though a high CN reactivity causes it to be present at a low concentration
        in interstellar ice and dust, thermal degradation of CN-bearing dust may be an
         important CN source, as postulated for comets (\citealt{Hanni2020,Lippi2013}). 

Finally, European Space Agency’s Rosetta spacecraft, measured a higher concentration of HCN
 than CN in the comet 67P/Churyumov–Gerasimenko (\citealt{Hanni2020}). Considering that
  the chemistry of N-bearing species in 67P/C-G is similar to interstellar ice chemistry
   \citep{Jor20}, we expect that the CN/HCN ratio predicted by the presented chemical model
    is slightly overestimated.

\subsection{UV field strengths in the Serpens Main}
\label{subsection:UV}

Observations of CN and HCN elucidate the presence of UV fields in the immediate surroundings of low- and intermediate-mass protostars. To determine the strengths of the UV fields, we calculated chemical models to reproduce the observed column density ratios. 

The measured CN/HCN ratios are consistent with the chemical models that include UV
 radiation, and for a broad range of physical conditions typical for low-mass star-forming
  regions. The UV field strengths, G$_0$, in excess of 10$^3$ best reproduce the
   observed CN/HCN ratios assuming gas temperature of 50 K (Section 3.2). At a gas temperatures
    of 300 K, the CN/HCN ratio is consistent with UV fields of 0.1-10 (Fig.~\ref{G0_300}). In both regimes,
     the dominant destruction reaction of HCN is indeed the photodissociation by UV photons. 

An alternative measure of the UV fields around protostars was recently provided by
 the observations of H$_2$O as part of the WISH program on \textit{Herschel} \citep{vDi21}.
  Far-infrared molecular spectra indicate that the bulk of H$_2$O forms in non-dissociative
   shocks in outflows at gas temperature $\sim$300 K \citep{Her12,Gre16,Kar18}. Some H$_2$O
    emission arises in hotter, $T\gtrsim700$ K regions with enhanced emission from hydrides,
     e.g. OH$^+$, where dissociative shocks are at play \citep{Kri13}. The fraction of H$_2$O
      in the outflows is clearly photodissociated, as evidenced by low absolute abundances
       and unexpectedly bright emission of OH and O \citep{Wam13,Kri17}. The impact of UV
        photons on non-dissociative shocks was modeled by \citet{Mel15}. The models reproduced
         the observed ratios of H$_2$O/OH and CO/OH in low-mass protostars for UV fields
          with G$_0$ of 0.1-10 \citep{Kar18}. The same ratios showed a few orders of magnitude
           disagreement with fully-shielded shock models, providing further evidence for
            H$_2$O photodissociation (\citealt{Kar14}). 
\begin{figure}
\centering 
\includegraphics[width=\linewidth]{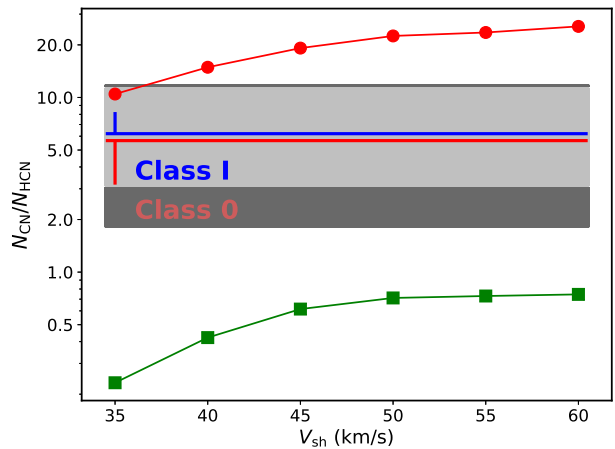} 
\caption{\label{shock_models} The observed CN/HCN column density ratio compared with
shock models with UV for a preshock density of $10^4$ cm$^{-3}$ \citep{Leh20}.
  The column densities are shown as a function of shock velocity for the precursor
   and post-shock regions (in red circles), and  the postshock region alone (in green squares).
    Column densities ratios obtained for Class 0 and Class I protostars are shown with dark grey
     and light grey rectangles, respectively. The horizontal lines show median
     CN/HCN ratios, and the vertical lines - their standard deviations for Class 0 (red) and Class I (blue) sources, respectively}. 
\end{figure}

The spectrum of the UV radiation adopted in \citet{Mel15} is based on average interstellar
 measurements, and is best suited for modeling the external irradiation. The spatial extent
  of the enhanced CN/HCN ratio and $^{13}$CO 6-5 emission suggest that UV photons in most
   protostars are likely produced in-situ, in the immediate surrounding of the outflow shocks
    \citep[see Section 5.1, ][]{Yil15}. Recent models by \citet{Leh20} provide predictions
     for molecular abundances arising in shocks where UV emission originates from the shock
      itself, i.e. the self-generated UV radiation. The calculations are done for a single
       gas pre-shock density of 10$^4$ cm$^{-3}$, which corresponds to (post-shock) gas
        densities of $\sim$10$^6$ cm$^{-3}$, assuming a compression factor of 100 typical
         for dissociative shocks. The UV fields generated by shocks with velocities in
          the range from 35 to 60 km s$^{-1}$ are $\sim10-400$ larger than the interstellar
           radiation field. In this regime, CN/HCN$>1$ when considering the entire shock
            region; CN/HCN$<1$ for lower-velocity shocks and the postshock region alone
             (see Figure~\ref{shock_models}). 

Observations of CN and HCN in the Serpens Main are consistent with the \citet{Leh20}
 shock models with velocities of 35 km s$^{-1}$, but only for the regions with the highest
  CN/HCN ratios (Figure~\ref{shock_models}). These shocks produce UV fields $G_{\mathrm{eff}}\sim25$, 
  where $G_{\mathrm{eff}}$ is the flux of UV photons normalized to the average 
  interstellar UV field \citep{Leh20}, and have a relatively short lifetime of 
  $\sim10^3$~yrs. The UV fields are lower than G$_{0}$ of $\sim10^3$ predicted by the 
  chemical model with UV (Section 4), which could be due to many different factors 
  and assumptions. In particular, neither of the models reproduces the physical 
  structure of the protostellar envelope with outflows.  

The models of the CN/HCN ratio in the dense envelopes of high-mass protostars were pioneered by
\citet{Sta07}, who considered both the effect of UV and X-ray radiation fields. 
The 1D radiative-transfer models accounted for the temperature and density gradients
 in the envelope, and the abundance profiles of the two molecules. The observed CN/HCN
  ratios in the envelopes were well-reproduced with $G_{0}<100$ at 
  $T_{\mathrm{gas}}\gtrsim200$ K, and densities of 10$^6$ cm$^{-3}$. 
  For lower gas temperatures, radiation from X-ray photons was invoked to explain 
  the observations. 
  
\cite{Bru09a} extended the 1D models by carving out a low-density outflow region in the envelope, which allowed a more efficient irradiation. The introduction of outflow cavity walls facilitated a significant increase in the volume of FUV irradiated gas. An example 2D model of AFGL 2591 successfully reproduced the CO$^{+}$ emission at $G_0>10^2$ \citep{Bru09b}.

Recent observations of additional hydrides provide further estimates of the radiation 
fields in high-mass protostars. \citet{Ben16} used new envelope models including outflow 
cavities to reproduce the abundances of e.g. CH$^+$, OH$^+$ and H$_2$O$^+$. The UV fields of 20-600 times the average interstellar radiation field were invoked to reproduce the observations of high-mass protostars. Lower UV fields, 2-400 times the interstellar radiation field, were found for low-mass protostars. The impact of X-ray photons on the chemistry was not observed in the envelopes.
For gas densities of 10$^6$ cm$^{-3}$, the ratios of $\sim1-12$ observed in the 
Serpens Main are consistent with the models of \citet{Sta07} with $G_0\lesssim10^2$
 and gas temperatures $\gtrsim300$ K. Inclusion of the lower-density outflow cavities
  similar to \citet{Ben16}, would better reproduce the CN/HCN ratios due to easier
   propagation of UV photons at lower densities. Such a modeling, however, is outside
    of the scope of this paper.

In summary, the CN/HCN ratio provides information about UV field distribution on cluster scales. Independent measurements of gas densities allow the estimate of the strengths of the UV fields. Their values in the surrounding of low-mass protostars in Serpens are of the order of $10^{3-4}$ the interstellar radiation field, qualitatively in agreement with observations and modeling of other species. While opacity effects influence the results of calculations, UV fields clearly play a significant role even around Solar-type protostars and affect the chemistry and physical properties of material in the forming disks (\citealt{Vis09}, \citealt{Dro16}). 

\section{Conclusions}
\label{section:conclusions}
IRAM 30~m / EMIR observations of CN and HCN emission pin-point the location 
of the impact of UV radiation on the chemistry of low-mass star forming region in the Serpens Main. 
A combination of simple models using the radiative transfer code RADEX and the chemical code Nahoon 
allows us to determine physical conditions of the gas, column densities of molecular species,
and estimate the UV radiation field strength. The main conclusions of our study are the 
following: 
\begin{itemize} 
\item The large-scale spatial extent of HCN and CN emission show differences, with 
HCN resembling more the CO emission tracing outflows and CN concentrated closer
 to the individual protostar positions. Yet, on spatial scales on individual sources,
  both tracers are associated with outflows.
  
 \item {Analysis of the ratios of hyperfine-splitted components shows that HCN 1-0 and H$^{13}$CN 1-0 are optically thick, and CN 1-0 is typically optically thin. Thus, the opacity effects influence the spatial distribution of CN/HCN enhancements by underestimating HCN emission and overestimating the ratio of CN and HCN in dense regions. Column densities of HCN corrected for optical depth and determined from a simple scaling of H$^{13}$CN agree within a factor of 2.}
  
 \item {For typical physical conditions of the gas in low-mass star forming regions column density ratio of CN and HCN is in the range of 1--12. For gas temperatures below $\sim200$ K, derived UV field strengths are weakly dependent on the assumed gas densities in the models. At higher gas temperatures, the CN/HCN ratios are also consistent with lower UV fields, in line with measurements obtained using hydrides with \textit{Herschel} \citep{Ben16}.}
 
\item {Enhancements of the CN/HCN ratios trace UV field strengths in excess of 10$^3$ times the interstellar UV field toward the analyzed protostars and off-source positions in Serpens Main. Adopting measurements of gas densities at 1000 AU in the low-mass envelopes from \cite{Kri12}, G$_0$ of $\sim10^3-10^4$ is inferred toward Ser SMM1, SMM3, and SMM4.}
 
\item {The chemical network for nitrogen-bearing species is sensitive to UV photons. The CN/HCN ratio is primarily driven by the destruction of HCN by UV radiation field for G$_{0}>10^2$ in the immediate surroundings of low-mass protostars.}
 
  \item {The luminosity ratio of CN and HCN toward low- and intermediate protostars in Serpens is not correlated with bolometric temperature and luminosity, suggesting a similar amount of UV photons affecting the gas. This results is consistent with far-infrared observations of [O I] toward a larger sample of sources, where no significant changes have been seen between Class 0 and Class I low-mass protostars \citep{Kar18}.}
  
\end{itemize}

The CN/HCN ratio as a tracer of the UV field can now be - with all the limitations - extended to the low-mass regime. Higher resolution observations are needed to fully exploit 
the impact of UV radiation on the low-mass star 
forming regions on scales of individual objects. Detailed 3D modeling of protostellar
 envelopes with outflow cavities is necessary to fully constrain the strengths of the
  UV fields in physical components of young stellar objects.

\begin{acknowledgements} We thank the anonymous referee for a careful reading of the manuscript and many constructive comments. AM, AK, MG, and MŻ acknowledge support from the Polish National Science
Center grant 2016/21/D/ST9/01098. AK also acknowledges support from the First TEAM grant of the Foundation for Polish Science No. POIR.04.04.00-00-5D21/18-00 and the
 hospitality of the StarPlan group in the University of Copenhagen during
  the manuscript preparation. The research of LEK is supported by a research grant (19127) from VILLUM FONDEN. DH acknowledges support from the EACOA Fellowship from the East Asian Core Observatories Association. MŻ acknowledges financial support from the European Research Council (Consolidator Grant COLLEXISM, grant agreement 811363),
 the Institut Universitaire de France, and the Programme National ‘Physique
  et Chimie du Milieu Interstellaire’ (PCMI) of CNRS/INSU with INC/INP cofunded
   by CEA and CNES. MF acknowledge support from the Polish National Science Center
    grant UMO-2018/30/M/ST9/00757. This article has been supported by the Polish National
Agency for Academic Exchange under Grant No. PPI/APM/2018/1/00036/U/001. The research
 has made use of data from the \textit{Herschel} Gould Belt survey (HGBS) project
  (http://gouldbelt-herschel.cea.fr). The HGBS is a \textit{Herschel} Key Programme
   jointly carried out by SPIRE Specialist Astronomy Group 3 (SAG 3), scientists
    of several institutes in the PACS Consortium
(CEA Saclay, INAF-IFSI Rome and INAF-Arcetri, KU Leuven, MPIA Heidelberg),
 and scientists of the \textit{Herschel} Science Center (HSC). 
\end{acknowledgements}

\bibliographystyle{aa} 
\bibliography{amirocha} 

\begin{appendix} 

\section{Spectral Energy Distributions}
\label{app:seds}

Broad-band observations are needed in order to determine physical properties of a protostar. \cite{Dun15} studied protostars in the Serpens molecular cloud using 2MASS
(\citealt{Skr06}) and Spitzer IRAC/MIPS (\citealt{Eva09}) observations covering the range
1.25–70 $\mu$m, photometry from Wide-field Infrared Survey Explorer at 12 and 22 $\mu$m (WISE;
\citealt{Wri10}), SHARC-II 350 $\mu$m (\citealt{Sur16}), the SCUBA Legacy Catalog 450 and 850 $\mu$m
(\citealt{dFr08}) and 1.1 mm observations from Bolocam dust survey (\citealt{Eno07}). The Serpens
Main region was also observed in the \textit{Herschel} Gould Belt survey project (\citealt{And10}) at 70, 160, 250, 350 and 500 $\mu$m.
SPIRE/PACS photometry in the Serpens molecular cloud is discussed in \citet{Fio21}. The flux densities used in the analysis are shown in Table~\ref{SED_data}.

\begin{table*} 
\caption{Comparison of $T_\mathrm{bol}$ and $L_\mathrm{bol}$ with literature values}
\label{table:seds-comp} 
\centering  
\begin{tabular}{l | c c | c c | c c  } \hline\hline 
Source & $T_\mathrm{bol}$ & $T_\mathrm{bol}$ (reference) & $L_\mathrm{bol}$ & $L_\mathrm{bol}$ (reference) & \multicolumn{2}{c}{Difference} \\ 
& (K) & (K) & (L$_\odot$) & (L$_\odot$) & \multicolumn{2}{c}{\%} \\ 
\hline
SMM 1 & 37 & 39 (1) & 115.2 & 108.7 (1)& -5 & +6\\
SMM 2 & 34 & 28 (2) & 7.2 & 8.0 (2) & +21 & -10\\
SMM 3 & 35 & 37 (1) & 7.1 & 27.5 (1) & -5 & -74 \\
SMM 4 & 68 & 28 (1) & 5.1 & 13.6 (1) & +143 & -63 \\
SMM 5 & 151 & 130 (2) & 3.7 & 4.8 (2) & +16 & -23 \\
SMM 6 & 532 & 530 (2) & 43.1 & 42.0 (2) & +0.4 & +3 \\
SMM 8 & 15 & \dots & 0.2 & \dots & -- & -- \\ 
SMM 9 & 33 & 29 (2) & 11.0 & 14.0 (2) & +14 & -21 \\
SMM 10 & 79 & 62 (2) & 7.1 & 7.6 (2) & +27 & -7 \\
SMM 12 & 72 & 100 (2) & 10.0 & 6.2 (2) & -28 & +61 \\
\hline 
\end{tabular}
\tablefoot{References: (1) \citealt{Kar18}, (2) \citealt{Dun15}. No previous determination of  $T_\mathrm{bol}$ and $L_\mathrm{bol}$ for SMM 8 was found in the literature.}
\end{table*}

\begin{table*} 
\caption{Continuum fluxes in Jy}      
\centering       %
\label{SED_data}    

\begin{tabular}{l c c c c c c c c c} 
\hline 
$\lambda$ & SMM1  & SMM2  & SMM3 & SMM4 & SMM5  & SMM6 & SMM9 & SMM10 & SMM12 \\ 
($\mu$m) &  &  &  & & & & & & \\ 
\hline 
1.25 &- &- &- & 6.0$\times$10$^{-4}$ & 3.0$\times$10$^{-4}$ & 2.1$\times$10$^{-2}$ &- &- &-\\
1.65 &- &- &- & 3.2$\times$10$^{-3}$ & 9.0$\times$10$^{-4}$ & 2.1$\times$10$^{-1}$ &- &- &-\\
2.17 & -& -& -& 1.0$\times$10$^{-2}$ & 4.3$\times$10$^{-3}$ & 1.0$\times$10$^{0}$ & -&- &-\\
3.6 & 9.0$\times$10$^{-4}$ & 4.0$\times$10$^{-4}$ & 2.8$\times$10$^{-3}$ & 2.4$\times$10$^{-2}$ & 3.1$\times$10$^{-2}$ & 2.5$\times$10$^{0}$ & 2.0$\times$10$^{-3}$ & 7.4$\times$10$^{-3}$ & 2.8$\times$10$^{-3}$\\ 
4.5 & 2.6$\times$10$^{-3}$ & 1.2$\times$10$^{-3}$ & 5.8$\times$10$^{-3}$ & 3.3$\times$10$^{-2}$ & 7.3$\times$10$^{-2}$ & 3.0$\times$10$^{0}$ & 7.0$\times$10$^{-3}$ & 3.3$\times$10$^{-2}$ & 3.0$\times$10$^{-2}$\\ 
5.8 & 2.3$\times$10$^{-3}$ & 2.1$\times$10$^{-3}$ & 7.8$\times$10$^{-3}$ & 4.1$\times$10$^{-2}$ & 1.4$\times$10$^{-1}$ & 5.1$\times$10$^{0}$ & 1.2$\times$10$^{-2}$ & 4.1$\times$10$^{-2}$ & 1.0$\times$10$^{-1}$\\
8 & 3.5$\times$10$^{-3}$ & 3.5$\times$10$^{-3}$ & 3.6$\times$10$^{-2}$ & 5.6$\times$10$^{-2}$ & 2.1$\times$10$^{-1}$ & 5.4$\times$10$^{0}$ & 1.7$\times$10$^{-2}$ & - & 2.0$\times$10$^{-1}$\\ 
12 & 6.2$\times$10$^{-3}$ & - & - &- & 1.8$\times$10$^{-1}$ & 7.6$\times$10$^{0}$ & 1.2$\times$10$^{-2}$ & 3.8$\times$10$^{-2}$ & 2.2$\times$10$^{-1}$\\ 
22 & 1.0$\times$10$^{0}$ & - & - & -& 9.5$\times$10$^{-1}$ & 1.0$\times$10$^{1}$ & 2.1$\times$10$^{-1}$ & 8.0$\times$10$^{-1}$ & 3.1$\times$10$^{0}$\\ 
24 & 1.2$\times$10$^{0}$ & 1.0$\times$10$^{-1}$ & 1.1$\times$10$^{-1}$ & 2.8$\times$10$^{-1}$ & 9.9$\times$10$^{-1}$ &- & 2.2$\times$10$^{-1}$ & 1.6$\times$10$^{0}$ & 2.6$\times$10$^{0}$\\
70 & 2.6$\times$10$^{2}$ & 1.1$\times$10$^{1}$ & 7.7$\times$10$^{0}$ & 1.9$\times$10$^{0}$ & 8.5$\times$10$^{0}$ & 2.6$\times$10$^{1}$ & 1.5$\times$10$^{1}$ & 8.8$\times$10$^{0}$ & 1.2$\times$10$^{1}$\\
160 & 5.8$\times$10$^{2}$ & 3.8$\times$10$^{1}$ & 4.2$\times$10$^{1}$ & 2.6$\times$10$^{1}$ & 1.8$\times$10$^{0}$ & 3.1$\times$10$^{1}$ & 6.1$\times$10$^{1}$ & 1.4$\times$10$^{1}$ & 4.3$\times$10$^{1}$\\
250 &  2.4$\times$10$^{2}$ & 3.1$\times$10$^{1}$ & 4.2$\times$10$^{1}$ & 5.9$\times$10$^{1}$ & 4.0$\times$10$^{0}$ & 2.9$\times$10$^{1}$ & 5.9$\times$10$^{1}$ & 6.9$\times$10$^{0}$ & 5.3$\times$10$^{1}$\\
350 & 1.2$\times$10$^{2}$ & 3.1$\times$10$^{1}$ & 3.5$\times$10$^{1}$ & 4.7$\times$10$^{1}$ & 1.6$\times$10$^{1}$ & 1.6$\times$10$^{1}$ & 4.3$\times$10$^{1}$ & 6.5$\times$10$^{0}$ & 2.7$\times$10$^{1}$\\
450 & 1.0$\times$10$^{2}$ &-  & 3.1$\times$10$^{1}$ &- &- &- &- &- &-\\
500 & 5.5$\times$10$^{1}$ & 1.8$\times$10$^{1}$ & 1.7$\times$10$^{1}$ & 2.9$\times$10$^{1}$ & 1.2$\times$10$^{1}$ & 9.5$\times$10$^{0}$ & 2.3$\times$10$^{1}$ & 5.1$\times$10$^{0}$ & 1.8$\times$10$^{1}$\\
850 & 1.5$\times$10$^{1}$ & 8.9$\times$10$^{0}$ & 8.9$\times$10$^{0}$ & - & 3.9$\times$10$^{0}$ & - & 1.3$\times$10$^{1}$ & 3.0$\times$10$^{0}$ &\\
1100 & 6.5$\times$10$^{0}$ & 2.1$\times$10$^{0}$ & 2.1$\times$10$^{0}$ & - & 1.5$\times$10$^{0}$ & 1.6$\times$10$^{0}$ & 3.0$\times$10$^{0}$ & 1.0$\times$10$^{0}$ & 2.1$\times$10$^{0}$\\
\hline \end{tabular} 
\end{table*}

\begin{figure} 
\includegraphics[width=8cm]{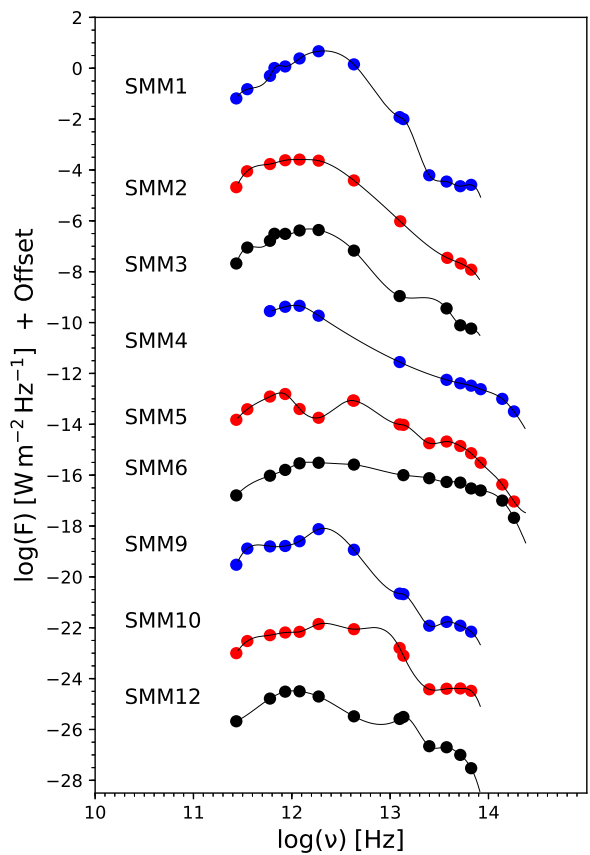} 
\caption{SED of protostars in the Serpens Main region. The offset between each plot is 3 orders of magnitude.} 
\label{seds} 
\end{figure}

Based on Spectral Energy Distributions (SEDs), the bolometric temperature and luminosity can be calculated for each of the observed
protostars. The bolometric luminosity is determined by integrating the SEDs over frequency:
\begin{equation} \label{eq6} L_{bol} = 4 \pi \, d^2 \, \int F_\nu d\nu \end{equation} where $d$ is the
cloud distance of $436 \pm 9.2$ pc (\citealt{Ort17}). The bolometric temperature is calculated as
described in \cite{Mye93}: \begin{equation} \label{eq7} T_{bol} = 1.25 \, 10^{-11} \, \bar{\nu}
\end{equation} where $\bar{\nu}$ is the mean frequency given by: \begin{equation} \label{eq8}
\bar{\nu} = \frac{\int \nu F_\nu d\nu}{ \int F_\nu d\nu} \end{equation}

Using Scipy \textit{splrep} and \textit{splev} functions, cubic smooth spline interpolation of the
photometric data is obtained while calculating the protostars parameters. Integration along the
resulting axis is obtained with the composite trapezoidal rule (\textit{Scipy} package\footnote{https://www.scipy.org/}). The
photometric data allow us to perform the integration along a wide range of wavelengths (Tab.~\ref{SED_data}) with the exception
of SMM 8. For this source we have only 4 photometric points from the 
\textit{Herschel} Gould Belt survey so the calculated bolometric temperature and luminosity may be underestimated.

Table~\ref{table:seds-comp} shows bolometric temperatures and luminosities, and their comparison to the literature values from \cite{Kar18} and \cite{Dun15}. The differences are lower than 30\% for most of the sources, with the exception of Ser SMM3, SMM4, and SMM 12 where $L_\mathrm{bol}$ differs by a factor of $\sim2-4$. The differences are mostly driven by the far-infrared measurements: \cite{Kar18} use PACS $5\times5$ single footprint maps covering a field of view of 47''$\times$47'' and the range from 55 to 190 $\mu$m, and \cite{Dun15} use significantly less sensitive \textit{Spitzer}/MIPS photometry. Here, we use photometric bands at 70 $\mu$m, 100 $\mu$m and 160 $\mu$m from the \textit{Herschel} Gould Belt survey, with the beam sizes of 5.6'', 6.8'', and 10.7'', respectively\footnote{Quick-Start Guide to Herschel–PACS the Photometer https://www.cosmos.esa.int/}. For Ser SMM3, there is a factor of 5-6 difference in the flux density at 70 $\mu$m (7 Jy vs. 36 Jy) and 160 $\mu$m (44 Jy vs. 182 Jy) between the PACS photometry and the 47''$\times$47'' from spectroscopy. Similarly, for Ser SMM4 the 160 $\mu$m flux differs by a factor of $\sim$6, suggesting that the large field-of-view of the spectroscopy mode might include some additional emission from background or nearby sources.

\section{Maps in additional tracers}
\label{app:maps}

\begin{figure} \includegraphics[width=8cm]{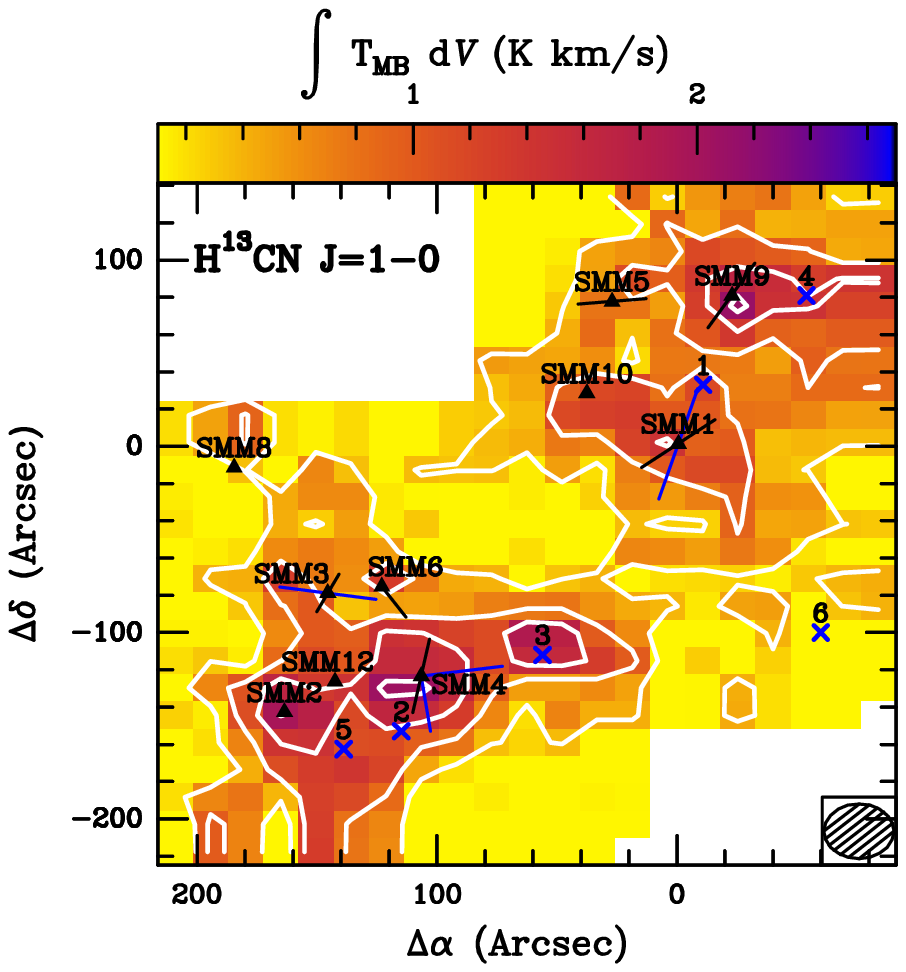} \caption{Similar to Figure~\ref{iram_maps}
but for H$^{13}$CN $1-0$. The first contour is at 3 $\sigma$ (0.5~K~km s$^{-1}$) with steps
of 3 $\sigma$}. \label{h13cn10} \end{figure}

\begin{figure} \includegraphics[width=8cm]{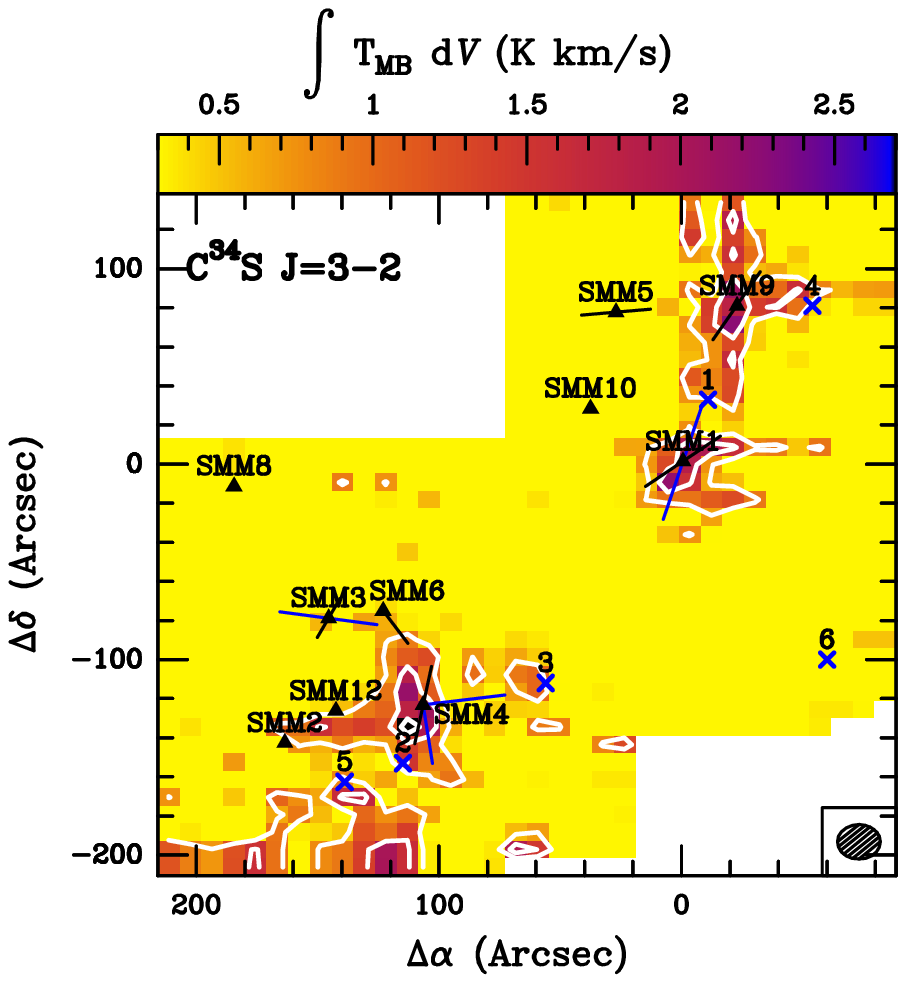} \caption{Similar to Figure~\ref{iram_maps}
but for C$^{34}$S $3-2$. The first contour is at 3 $\sigma$ (1.2~K~km s$^{-1}$), with steps of 1.5 $\sigma$.} \label{c34s32} \end{figure}

IRAM 30m spectral line maps in HCN and CS isotopologues are presented in 
Figure~\ref{h13cn10} and Figure~\ref{c34s32}. The H$^{13}$CN 1-0 emission
 is a sum of all hyperfine splitting components. The H$^{13}$CN 1-0 is 
 spatially consistent with \mbox{HCN 1-0}, although a few times weaker.
  It peaks around Ser SMM2, Ser SMM4 and Ser SMM9, as well as at the outflow
   position no.~3. Except for the Ser SMM2 protostar, the peaks in H$^{13}$CN
    1-0 are co-spatial with the HCN emission peaks. The emission of H$^{13}$CN
     1-0 line is similarly strong in the south-east and north-west regions of the map.

The C$^{34}$S 3-2 line does not show such extended emission, although it is detected in
 all outflow positions. The emission is concentrated mostly near Ser SMM1, Ser SMM4
  and Ser SMM9 sources, unlike the CS 3-2 which is not that significant around Ser SMM1.
   The isotopologues trace the same gas as their regular counterparts.

\section{Line profiles across the maps}
\label{app:lines}

\begin{figure*} 
\centering
\includegraphics[width=15cm]{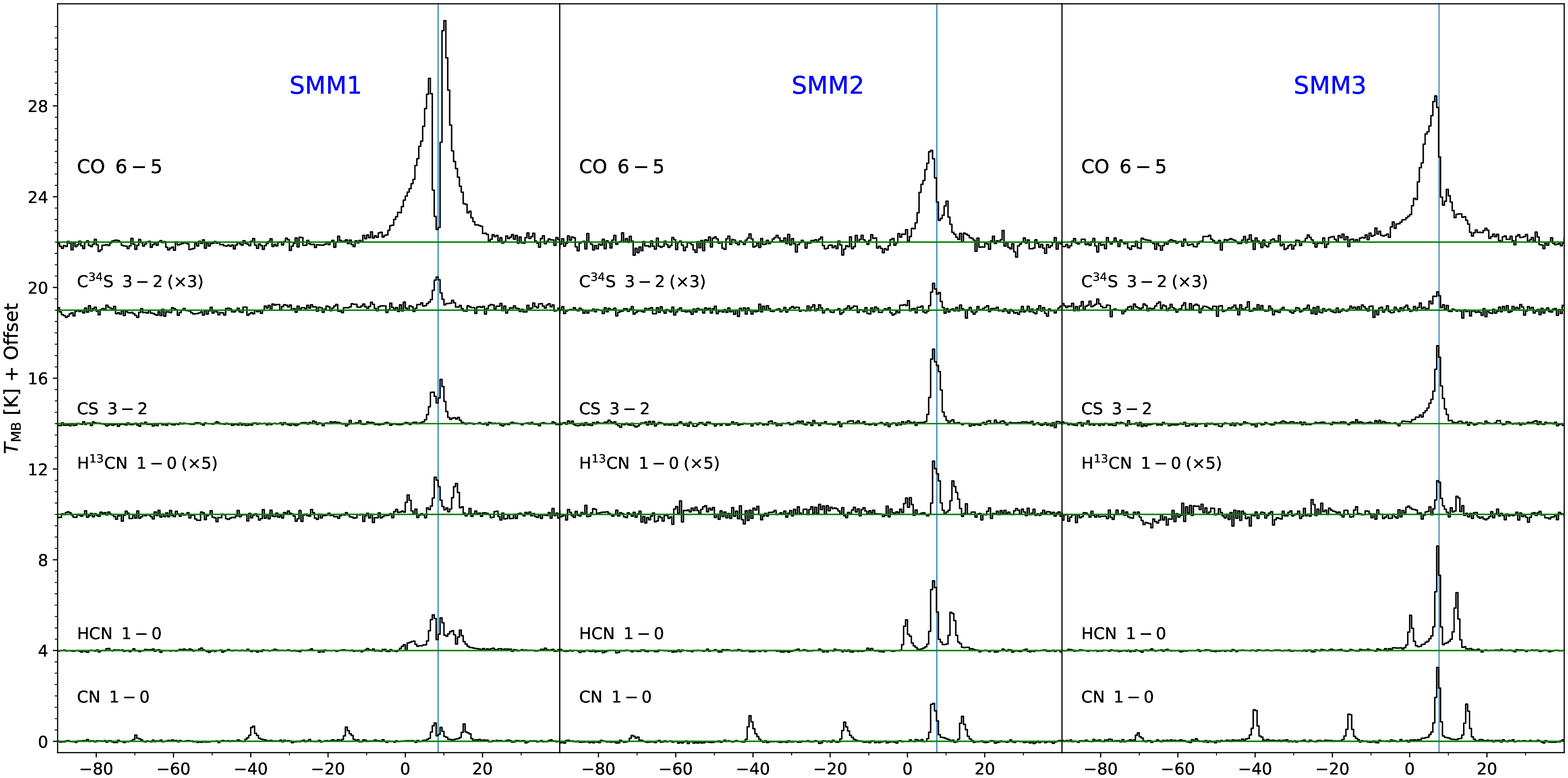}
\includegraphics[width=15cm]{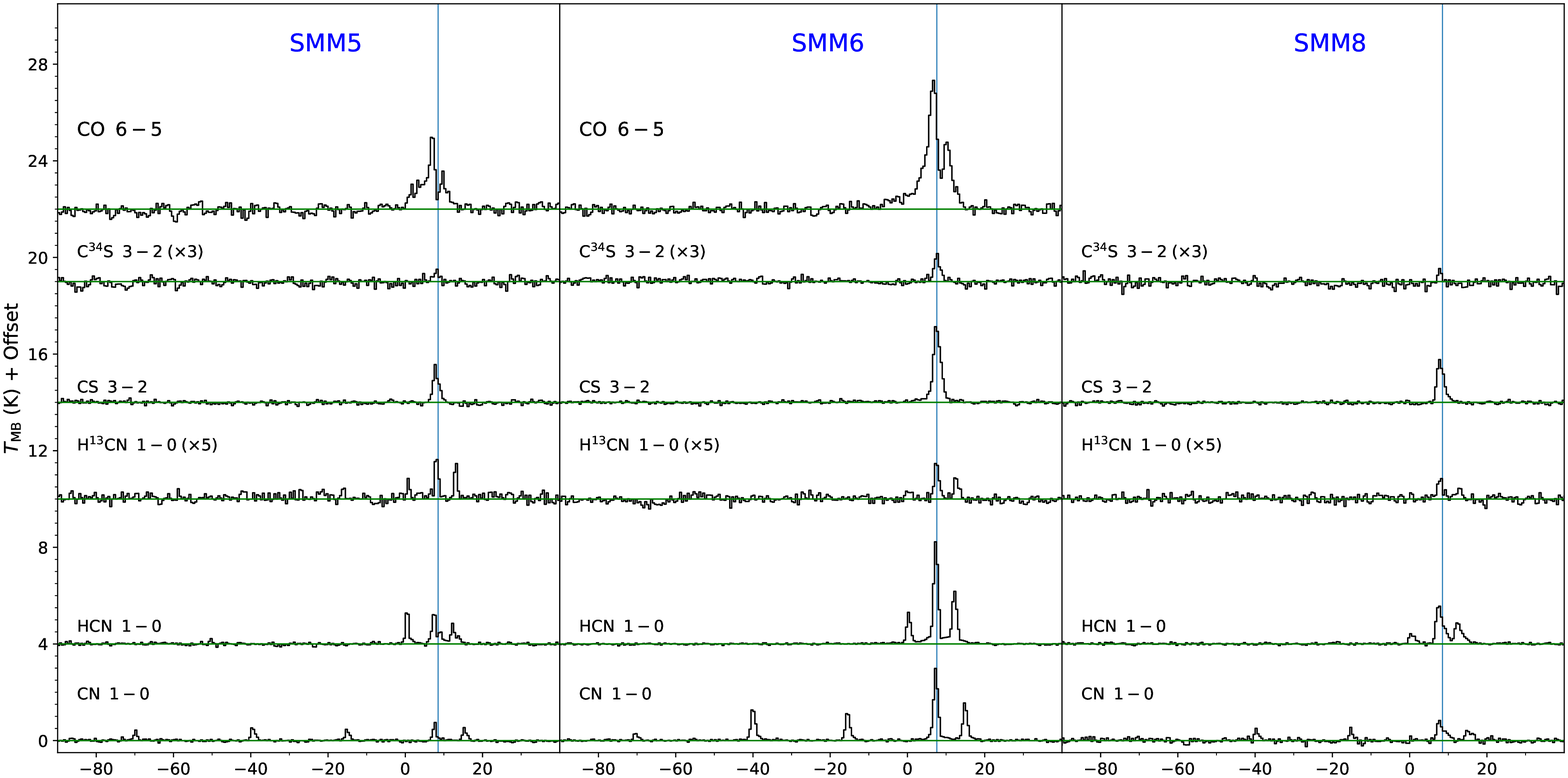} 
\includegraphics[width=15cm]{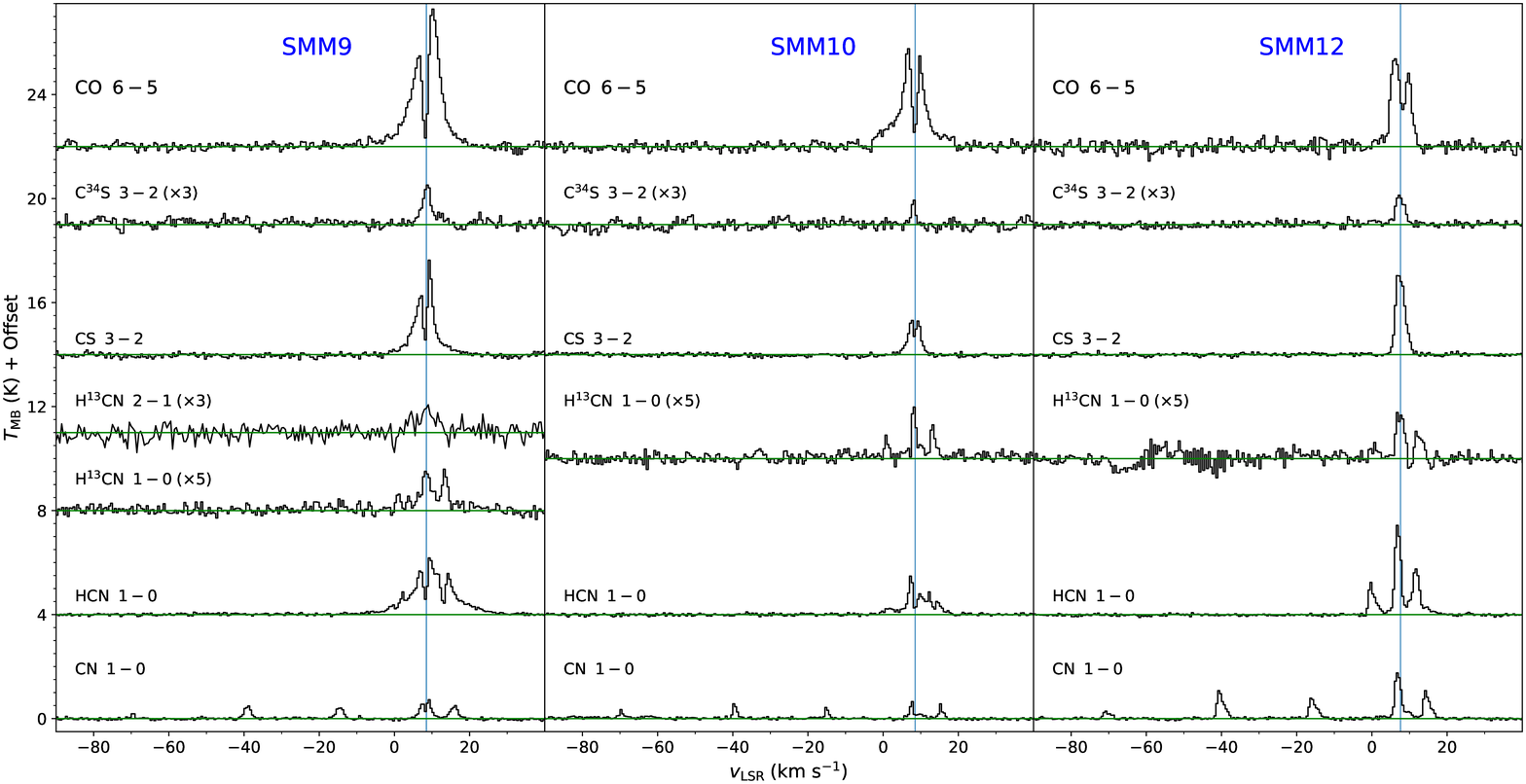}
\caption{Line profiles of CO 6-5, C$^{34}$S 3-2, CS 3-2 , H$^{13}$CN 2-1, H$^{13}$CN 1-0,
 HCN 1-0 and CN 1-0 in the protostar positions.}
\label{Spectra_protostars}
\end{figure*}

\begin{figure*} 
\centering
\includegraphics[width=15cm]{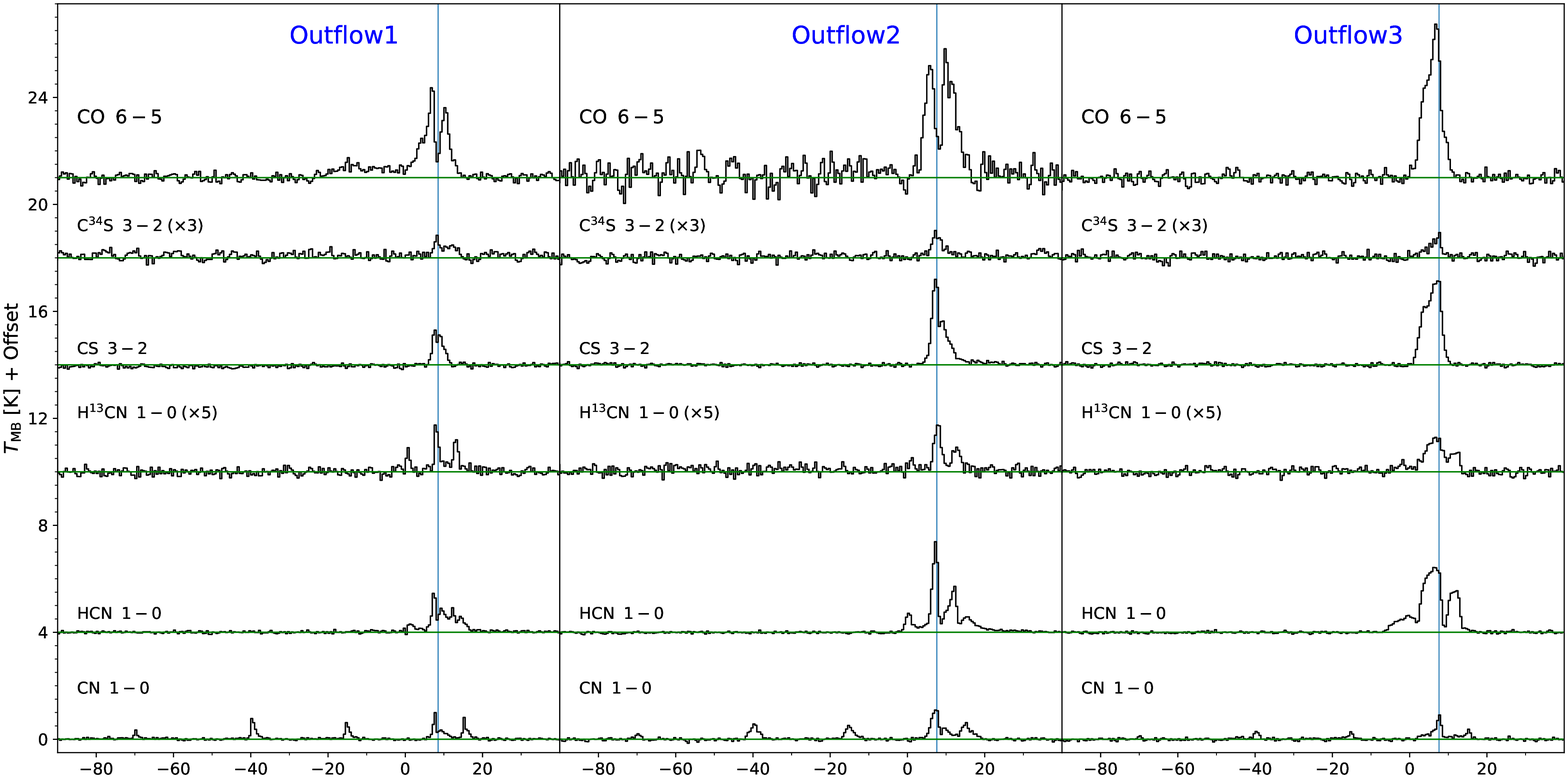}
\includegraphics[width=15cm]{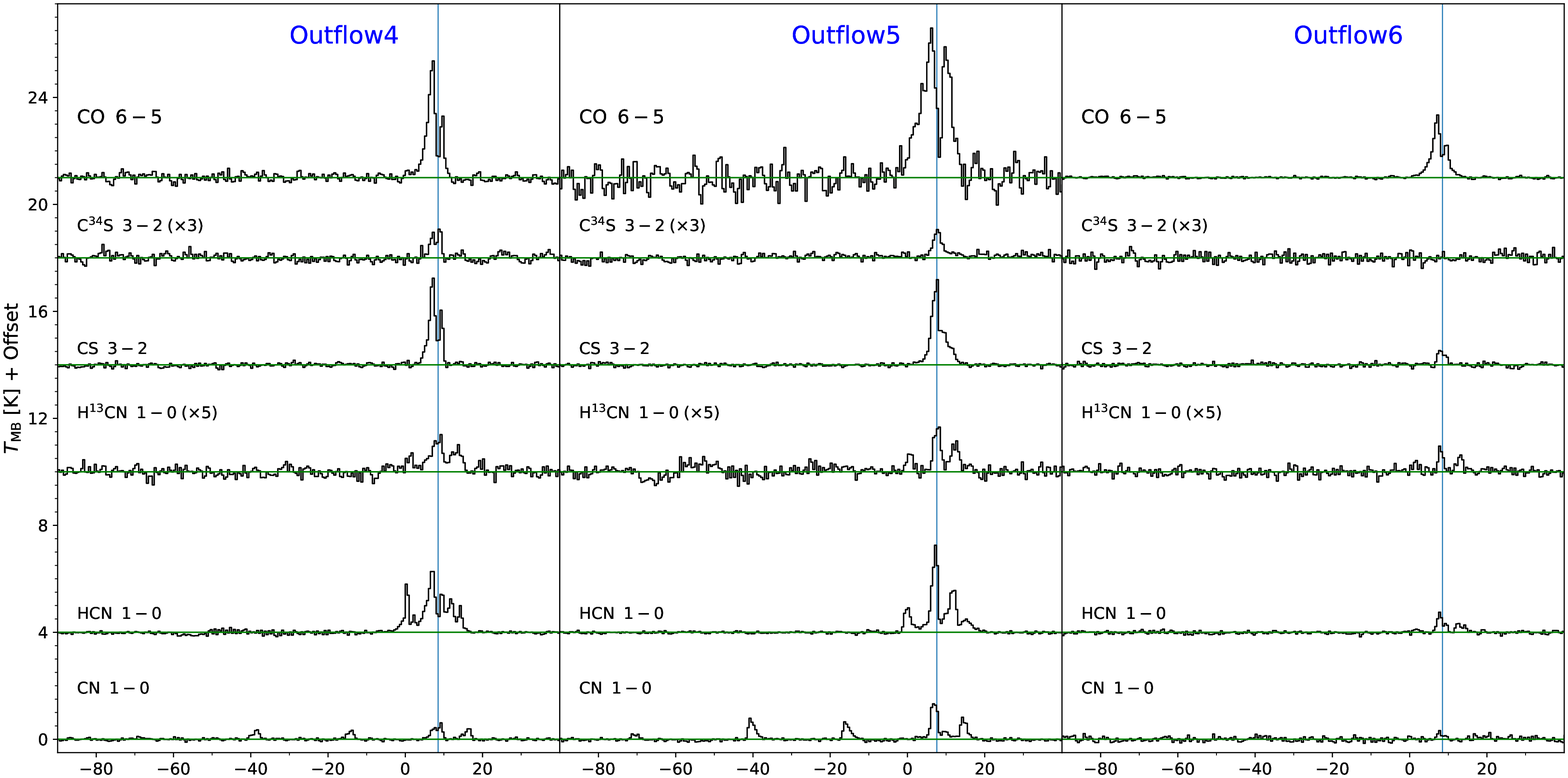}
\caption{Line profiles of CO 6-5, C$^{34}$S 3-2, CS 3-2 , H$^{13}$CN 2-1, H$^{13}$CN 1-0,
 HCN 1-0 and CN 1-0 in the outflow positions.}
\label{Spectra_outflows}
\end{figure*}

A majority of targeted lines are detected at protostar and outflow positions. 
Figures~\ref{Spectra_protostars} and \ref{Spectra_outflows} show line profiles in CO 6-5
 observed with APEX-CHAMP$^+$ and C$^{34}$S 3-2, CS 3-2, H$^{13}$CN 1-0, HCN 1-0 and
  CN 1-0 lines obtained with IRAM 30m. Note that the line profiles of Ser SMM4 and its
   outflow position are shown in Figure~\ref{SMM4spectra}. The H$^{13}$CN 2-1 line is
    detected only toward Ser~SMM4 and Ser SMM9. The Ser SMM8 protostar is outside of
     the mapping area in CO 6-5.


\section{Line fluxes and observed column densities}
\label{app:fluxes}
\begin{figure*} 
\begin{subfigure}{.5\textwidth} 
\centering
\includegraphics[width=1.\linewidth]{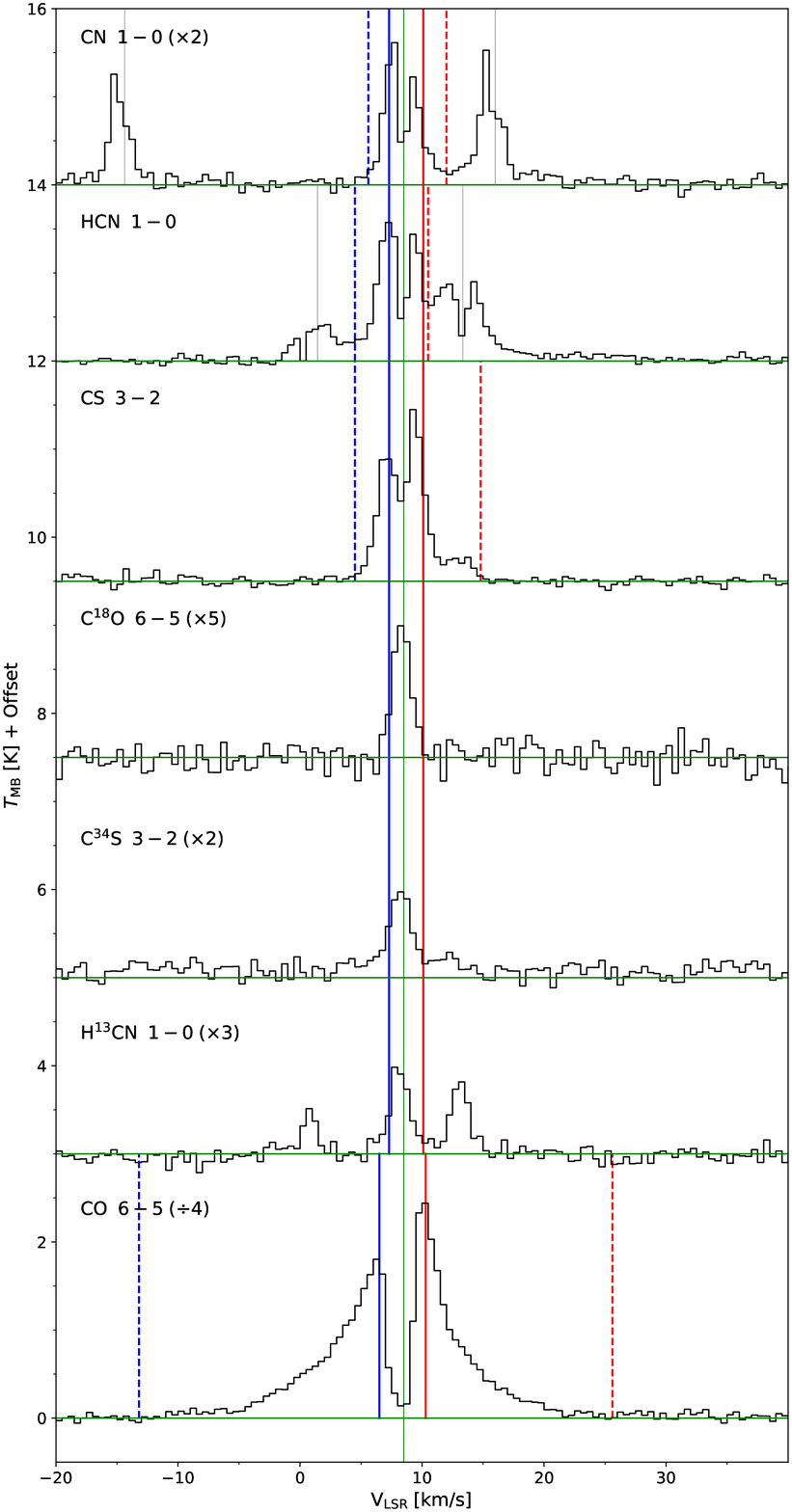} 
\caption{} 
\end{subfigure}
\begin{subfigure}{.5\textwidth} 
\label{3map} 
\centering
\includegraphics[width=1.\linewidth]{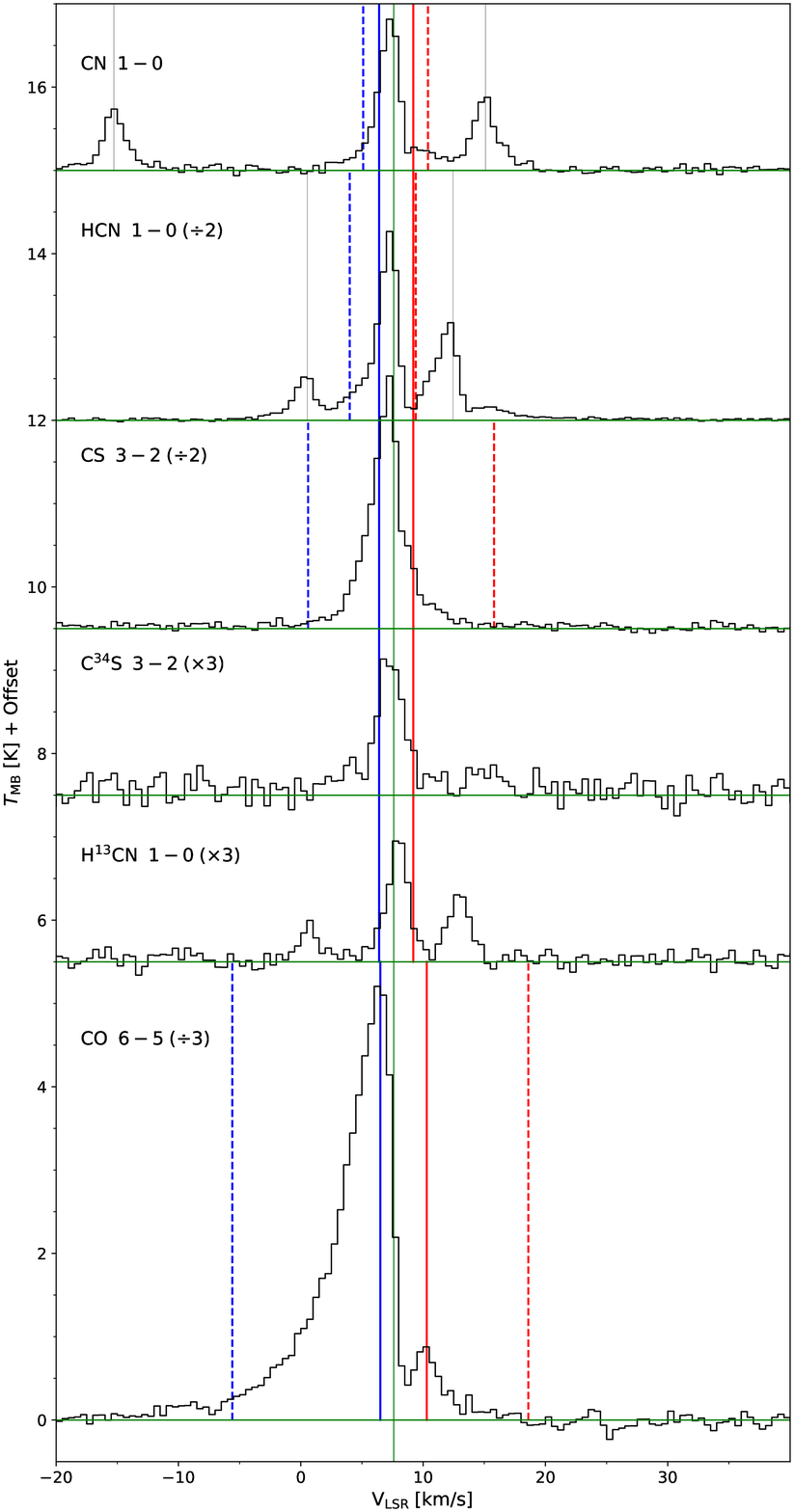} 
\caption{} \end{subfigure}
\caption{\label{wing_ranges}  Line emission for Ser SMM1 (a) and SMM4 (b) with ranges
 for calculations of blue and red line wing emission. The C$^{18}$O 6-5 line profile
  for Ser SMM1 is used as a proxy for the velocity range corresponding to the envelope 
  emission \citep{Kri10}.}
\end{figure*}

\begin{table*}
\centering
\caption{Velocity ranges (in km s$^{-1})$ used to calculate emission in the line wings}\label{Tab.wing_ranges}

\begin{tabular}{l c c c c }
 \hline \hline
Line & \multicolumn{2}{c}{SMM 1} &\multicolumn{2}{c}{SMM 4} \\ 
& blue & red & blue & red \\
\hline \hline
\multirow{5}{*}{CN 1-0} & (-71.0,-70.0) & (-68.4,-68.1) & (-70.9, -70.9) & (-69.3, -68.7) \\
& (-42.2, -39.6) & (-38.0, -36.2) & (-41.6, -40.5) & (-38.9, -37.6) \\
& (-16.7, -15.1) & (-13.5, -13.0 & (-19.2, -16.0) & (-14.4,-11.6) \\
& (5.6, 7.8) & (9.4, 12.0) & (5.1, 6.9) & (8.5, 10.4) \\
& (12.6, 15.3) & (16.9, 17.6) & (13.5, 14.4) & (16, 17.5) \\ \hline
\multirow{3}{*}{HCN 1-0} & (-1.3, 0.735) & (2.335, 3.5) &  (-3.3, -0.2) & (1.4, 4.0) \\
& (4.5, 7.8) & (9.4, 10.5) & (4.0, 6.9) & (8.5, 9.4) \\
& (10.5, 12.64) & (14.24, 16) & (9.4, 11.7) & (13.3, 18.9) \\ \hline
CS 3-4 & (4.5, 7.3) & (9.4, 14.8) & (0.6, 6.9) & (8.5, 15.8)\\ \hline
CO 6-5 & (-11.3, 7.8) & (9.4, 21.7) & (-13.2, 6.9) & (8.5, 17.3) \\
\hline
\end{tabular}
\tablefoot{The velocity ranges for CN 1-0 and HCN 1-0 are obtained separately for each hyperfine component.}
\end{table*}

Figure \ref{wing_ranges} shows line profiles in Ser SMM1 and SMM4 from IRAM 30m
 compared to the C$^{18}$O line profile from APEX in order to select the velocity
  ranges for calculating emission in the line wings.
  
Table~\ref{Tab.wing_ranges} shows the velocity ranges used to calculate integrated intensities in the line wings for each of the species. The extent of the line wings, $V_\mathrm{out}$, corresponds to the velocity range where the emission exceeds 2$\sigma$. The inner boundaries of the line wings are calculated using the C$^{18}$O 6-5 line, with $V_\mathrm{in}$ corresponding to the velocity where the C$^{18}$O emission drops below the 
3$\sigma$ level in Ser SMM1 (Figure \ref{wing_ranges}). The adopted ranges have been visually inspected to contain the entire wing emission and omit absorption at the line centers.

Table~\ref{table:fluxes} shows line fluxes and level-specific column densities
 calculated at protostar positions. The integrated intensity is measured at 3$\sigma$
  level for each spectrum separately. The peak temperature $T_\mathrm{peak}$ corresponds
   to the line maximum or to the maximum of the strongest hyperfine component. The uncertainties in $T_\mathrm{peak}$ are rms values calculated using the baseline in the vicinity of each line. Column
    densities at the upper lever $N_\mathrm{up}$ are calculated as described in
     Section~\ref{subsection:column_densities} (Equation~\ref{eq3}).

Table~\ref{table:columns} shows total column densities $N_\mathrm{tot}$ calculated
 for total line profiles and wings separately (Equation~\ref{eq4}) assuming $T_{\mathrm{exc}}=50$ K. The values are
  corrected using the optical depth correction factor $C_\mathrm{\tau}$ (\citealt{Gol99}) defined as:
\begin{equation} \label{eq9}
C_\mathrm{\tau} = \frac{\tau}{1-\exp^{-\tau}} 
\end{equation}
where $\tau$ is the optical depth. 

Table~\ref{table:HCNs} shows a comparison of HCN column densities calculated using corrections for optical depth and using scaled H$^{13}$CN 1-0 column densities. Optical depths are calculated using the ratio of the hyperfine-splitted components, the F=2$\rightarrow1\div$1$\rightarrow1$ ($N$(HCN)$_{\tau_1}$), and using Equation 1 (see Section 3.3.1) for the full line profile of HCN ($N$(HCN)$_{\tau_2}$). The H$^{13}$CN 1-0 column density is multiplied by a factor of 68 corresponding to the standard interstellar $^{12}$C/$^{13}$C ratio (\citealt{Mil05}). The mean ratios of column densities determined using H$^{13}$CN 1-0 scaling and HCN corrected for optical depth, $N$(HCN)$_{\tau_1}$ and $N$(HCN)$_{\tau_2}$, is 0.7 and 2.1, respectively. Thus, the HCN column densities calculated using these methods agree within a factor of $\sim$2.

\begin{sidewaystable*}[ht!]
\caption{Integrated fluxes and level-specific column densities at the positions of protostars}\label{table:fluxes}
\begin{tabular}{p{1.3cm} p{1.6cm} p{1.8cm} p{1.8cm} p{1.8cm} p{1.8cm} p{1.6cm} p{1.8cm} p{1.6cm} p{1.8cm} p{1.6cm} p{1.8cm}} 
\hline\hline 
Line & & SMM1 & SMM2 & SMM3 & SMM4 & SMM5 & SMM6 & SMM8 & SMM9 & SMM10 & SMM12 \\
\hline 
\multirow{4}{*}{CN 1-0} & $\int{T_{\mathrm{mb}} \, dV}$& 6.3 $\pm$ 0.9 & 8.5 $\pm$ 1.3 & 12.2 $\pm$ 1.8 & 10.2 $\pm$ 1.5 & 2.7 $\pm$ 0.4 & 10.6 $\pm$ 1.6 & 3.0 $\pm$ 0.5 & 4.9 $\pm$ 0.8 & 3.0 $\pm$ 0.5 & 10.1 $\pm$ 1.5\\
& (K km s$^{-1}$) & & & & & & & & & &  \\
& $T_\mathrm{peak}$ (K) & 0.89 $\pm$ 0.03 & 1.92 $\pm$ 0.04 & 3.42 $\pm$ 0.03& 1.90 $\pm$ 0.03 & 0.94 $\pm$ 0.04 & 3.17 $\pm$ 0.03& 0.94 $\pm$ 0.13 & 0.84 $\pm$ 0.04& 0.78 $\pm$ 0.03 & 1.85 $\pm$ 0.03 \\
& $N_\mathrm{up}$ (cm$^{-2}$) & 1.30($\pm$0.07) $\times$10$^{13}$ & 1.74($\pm$0.09) $\times$10$^{13}$ & 2.50($\pm$0.13) $\times$10$^{13}$ & 2.11($\pm$0.11) $\times$10$^{13}$ & 5.17($\pm$0.03) $\times$10$^{12}$ & 2.21($\pm$0.11) $\times$10$^{13}$ & 6.22($\pm$0.04) $\times$10$^{12}$ & 1.00($\pm$0.05) $\times$10$^{13}$ & 6.25($\pm$0.03) $\times$10$^{12}$ & 2.08($\pm$0.10 )$\times$10$^{13}$ \\
\hline

\multirow{4}{*}{HCN 1-0} & $\int{T_{\mathrm{mb}} \, dV}$ & 10.3 $\pm$ 1.6 & 13.1 $\pm$ 1.9 & 14.4 $\pm$ 2.2 & 21.7 $\pm$ 3.0 & 4.0 $\pm$ 0.8 & 13.1 $\pm$ 1.9 & 6.8 $\pm$ 1.1 & 20.7 $\pm$ 3.1 & 7.3 $\pm$ 1.1 & 14.2 $\pm$ 2.2\\
& (K km s$^{-1}$) & & & & & & & & & &  \\
& $T_\mathrm{peak}$ (K) & 1.76 $\pm$ 0.04 & 3.48 $\pm$ 0.03 & 4.95 $\pm$ 0.03& 4.69 $\pm$ 0.03 & 1.72 $\pm$ 0.04 & 5.15 $\pm$ 0.03 & 1.72 $\pm$ 0.04 & 2.40 $\pm$ 0.05 & 1.75 $\pm$ 0.03 & 3.50 $\pm$ 0.04 \\
& $N_\mathrm{up}$ (cm$^{-2}$) & 6.49($\pm$0.33) $\times$10$^{12}$ & 8.26($\pm$0.40) $\times$10$^{12}$ & 9.17($\pm$0.47) $\times$10$^{12}$ & 1.35($\pm$0.06) $\times$10$^{13}$ & 2.58($\pm$0.16) $\times$10$^{12}$ & 8.29($\pm$0.40) $\times$10$^{12}$ & 4.33($\pm$0.23) $\times$10$^{12}$ & 1.28($\pm$0.05) $\times$10$^{13}$ & 4.58($\pm$0.23) $\times$10$^{12}$ & 9.07($\pm$0.46) $\times$10$^{12}$ \\
\hline

\multirow{4}{*}{CS 3-2} & $\int{T_{\mathrm{mb}} \, dV}$ & 5.9 $\pm$ 1.1 & 8.6 $\pm$ 1.4 & 8.3 $\pm$ 1.6 & 14.8 $\pm$ 3.1 & 2.6 $\pm$ 0.5 & 8.5 $\pm$ 1.3 & 4.5 $\pm$ 0.7  & 12.4 $\pm$ 2.5 & 3.6 $\pm$ 0.8 & 9.9 $\pm$ 1.5 \\
& (K km s$^{-1}$) & & & & & & & & & &  \\
& $T_\mathrm{peak}$ (K) & 1.57 $\pm$ 0.05 & 3.10 $\pm$ 0.06 & 2.90 $\pm$ 0.06 & 4.30 $\pm$ 0.06 & 1.48 $\pm$ 0.06 & 3.2 $\pm$ 0.05 & 2.08 $\pm$ 0.05 & 2.70 $\pm$ 0.07 & 1.43 $\pm$ 0.05 & 3.26 $\pm$ 0.04 \\
& $N_\mathrm{up}$ (cm$^{-2}$) & 4.06($\pm$0.11) $\times$10$^{12}$ & 5.88($\pm$0.14) $\times$10$^{12}$ & 5.72($\pm$0.15) $\times$10$^{12}$ & 1.17($\pm$0.03) $\times$10$^{13}$ & 1.87($\pm$0.05) $\times$10$^{12}$ & 5.86($\pm$0.13) $\times$10$^{12}$ & 3.07($\pm$0.07) $\times$10$^{12}$ & 8.59($\pm$0.24) $\times$10$^{12}$ & 2.47($\pm$0.08) $\times$10$^{12}$ & 6.83($\pm$0.15) $\times$10$^{12}$ \\
\hline

\multirow{4}{*}{C$^{34}$S 3-2} & $\int{T_{\mathrm{mb}} \, dV}$ & 1.3 $\pm$ 0.3 & 0.9 $\pm$ 0.2 & 0.5 $\pm$ 0.1 & 1.5 $\pm$ 0.3 & 0.3 $\pm$ 0.1 & 0.8 $\pm$ 0.2 & 0.3 $\pm$ 0.1 & 1.8 $\pm$ 0.3 & 0.4 $\pm$ 0.1 & 0.6 $\pm$ 0.1 \\
& (K km s$^{-1}$) & & & & & & & & & &  \\
& $T_\mathrm{peak}$ (K) & 0.56 $\pm$ 0.05 & 0.47 $\pm$ 0.04 & 0.36 $\pm$ 0.04 & 0.64 $\pm$ 0.05 & 0.42 $\pm$ 0.05 & 0.56 $\pm$ 0.03 & 0.35 $\pm$ 0.05 & 0.62 $\pm$ 0.04 & 0.40 $\pm$ 0.04 & 0.42 $\pm$ 0.05 \\
& $N_\mathrm{up}$ (cm$^{-2}$) & 7.03($\pm$0.20) $\times$10$^{11}$ & 4.75($\pm$0.14) $\times$10$^{11}$ & 2.73($\pm$0.10) $\times$10$^{11}$ & 8.29($\pm$0.22) $\times$10$^{11}$ & 1.83($\pm$0.08) $\times$10$^{11}$ & 4.23($\pm$0.12) $\times$10$^{11}$ & 1.41($\pm$0.07) $\times$10$^{11}$ & 1.02($\pm$0.03) $\times$10$^{12}$ & 2.16($\pm$0.10) $\times$10$^{11}$ & 3.17($\pm$0.10) $\times$10$^{11}$ \\
\hline

\multirow{4}{*}{H$^{13}$CN 1-0} & $\int{T_{\mathrm{mb}} \, dV}$ & 1.6 $\pm$ 0.2 & 1.5 $\pm$ 0.2 & 0.6 $\pm$ 0.1 & 1.6 $\pm$ 0.3 & 0.8 $\pm$ 0.1 & 0.9 $\pm$ 0.1 & 0.1 $\pm$ 0.1 & 2.4 $\pm$ 0.3 & 1.0 $\pm$ 0.2 & 1.6 $\pm$ 0.3 \\
& (K km s$^{-1}$) & & & & & & & & & &  \\
& $T_\mathrm{peak}$ (K) & 0.45 $\pm$ 0.02 & 0.64 $\pm$ 0.02 & 0.41 $\pm$ 0.03 & 0.61 $\pm$ 0.03 & 0.42 $\pm$ 0.04 & 0.40 $\pm$ 0.02 & 0.21 $\pm$ 0.02 & 0.38 $\pm$ 0.03 & 0.45 $\pm$ 0.04 & 0.46 $\pm$ 0.03 \\
& $N_\mathrm{up}$ (cm$^{-2}$) & 3.32($\pm$0.17) $\times$10$^{11}$ & 3.08($\pm$0.15) $\times$10$^{11}$ & 1.21($\pm$0.08) $\times$10$^{11}$ & 3.41($\pm$0.18) $\times$10$^{11}$ & 1.63($\pm$0.10) $\times$10$^{11}$ & 1.82($\pm$0.08) $\times$10$^{11}$ & 2.93($\pm$0.37) $\times$10$^{10}$ & 5.01($\pm$0.22) $\times$10$^{11}$ & 2.17($\pm$0.13) $\times$10$^{11}$ & 3.43($\pm$0.18) $\times$10$^{11}$ \\
\hline

\multirow{4}{*}{CO 6-5} & $\int{T_{\mathrm{mb}} \, dV}$ & 75.6 $\pm$ 11.6 & 22.7 $\pm$ 3.7 & 52.5 $\pm$ 8.1 & 94.0 $\pm$ 14.3 & 13.1 $\pm$ 2.2 & 29.6 $\pm$ 4.6 & - & 36.2 $\pm$ 5.6 & 24.3 $\pm$ 3.9 & 17.3 $\pm$ 2.8 \\
& (K km s$^{-1}$) & & & & & & & & & &  \\
& $T_\mathrm{peak}$ (K) & 9.76 $\pm$ 0.18 & 4.04 $\pm$ 0.23 & 6.44 $\pm$ 0.17 & 15.61 $\pm$ 0.21 & 2.97 $\pm$ 0.17 & 5.33 $\pm$ 0.15 & - & 5.29 $\pm$ 0.12 & 3.76 $\pm$ 0.17 & 3.73 $\pm$ 0.18 \\
& $N_\mathrm{up}$ (cm$^{-2}$) & 3.28($\pm$0.04) $\times$10$^{15}$ & 9.83($\pm$0.12) $\times$10$^{14}$ & 2.27($\pm$0.03) $\times$10$^{15}$ & 4.07($\pm$0.05) $\times$10$^{15}$ & 5.69($\pm$0.07) $\times$10$^{14}$ & 1.28($\pm$0.02) $\times$10$^{15}$ & - & 1.57($\pm$0.02) $\times$10$^{15}$ & 1.07($\pm$0.01) $\times$10$^{15}$ & 7.49($\pm$0.09) $\times$10$^{15}$ \\

\hline\hline
\end{tabular}
\begin{flushleft}
\tablefoot{Calibration uncertainty of 15\% of the flux is included.}
\end{flushleft}
\end{sidewaystable*}

\begin{sidewaystable*}
\caption{Column densities and corrected column densities calculated for the total flux and the wings}\label{table:columns}

\begin{tabular}{p{1.3cm} p{1.3cm} p{1.8cm} p{1.8cm} p{1.8cm} p{1.8cm} p{1.6cm} p{1.8cm} p{1.6cm} p{1.8cm} p{1.6cm} p{1.8cm}}
\hline\hline 
Line & & SMM1 & SMM2 & SMM3 & SMM4 & SMM5 & SMM6 & SMM8 & SMM9 & SMM10 & SMM12 \\
\hline \multirow{4}{*}{CN 1-0} & $N_\mathrm{tot}$ (cm$^{-2}$) & 1.38($\pm$0.17) $\times$10$^{15}$ & 1.14($\pm$0.22) $\times$10$^{15}$ & 1.47($\pm$0.32) $\times$10$^{15}$  & 1.34($\pm$0.27) $\times$10$^{15}$ & 6.28($\pm$0.72) $\times$10$^{14}$ & 1.33($\pm$0.28) $\times$10$^{15}$ & 3.74($\pm$0.92) $\times$10$^{14}$ & 6.18($\pm$1.30) $\times$10$^{14}$ & 7.52($\pm$0.81) $\times$10$^{14}$ & 1.28($\pm$0.26) $\times$10$^{15}$ \\ 
& $N_\mathrm{tot}^\mathrm{corr}$ (cm$^{-2}$) & 1.14($\pm$0.17) $\times$10$^{15}$ & 1.54($\pm$0.22) $\times$10$^{15}$ & 2.17($\pm$0.32) $\times$10$^{15}$ & 1.84($\pm$0.27) $\times$10$^{15}$ & 4.68($\pm$0.72) $\times$10$^{14}$ & 1.83($\pm$0.28) $\times$10$^{15}$ & 5.12($\pm$0.92) $\times$10$^{14}$ & 8.54($\pm$1.30) $\times$10$^{14}$ & 5.05($\pm$0.81) $\times$10$^{14}$ & 1.68($\pm$0.26) $\times$10$^{15}$\\ 
& $N_\mathrm{wings}$ (cm$^{-2}$) & 4.93($\pm$0.11) $\times$10$^{14}$ & 4.53($\pm$0.13) $\times$10$^{14}$ & 2.94($\pm$0.12) $\times$10$^{14}$ & 5.71($\pm$0.09) $\times$10$^{14}$ & 5.90($\pm$1.40) $\times$10$^{13}$ & 1.96($\pm$0.11) $\times$10$^{14}$ & 1.09($\pm$0.24) $\times$10$^{14}$ & 3.33($\pm$0.13) $\times$10$^{14}$ & 1.02($\pm$0.12) $\times$10$^{14}$ & 5.13($\pm$0.12) $\times$10$^{14}$ \\ 
& $N_\mathrm{wings}^\mathrm{corr}$ (cm$^{-2}$) & 1.25($\pm$0.01) $\times$10$^{15}$ & 4.53($\pm$0.13) $\times$10$^{14}$  & 2.94($\pm$0.12) $\times$10$^{14}$ & 1.21($\pm$0.01) $\times$10$^{15}$ & 5.90($\pm$1.40) $\times$10$^{13}$ & 3.56($\pm$0.11) $\times$10$^{14}$ & 1.09($\pm$0.24) $\times$10$^{14}$ & 6.45($\pm$0.13) $\times$10$^{14}$ & 1.02($\pm$0.12) $\times$10$^{14}$ & 1.50($\pm$0.01) $\times$10$^{15}$ \\  \hline

\multirow{4}{*}{HCN 1-0} & $N_\mathrm{tot}$ (cm$^{-2}$) & 1.16($\pm$0.03) $\times$10$^{15}$ & 1.14($\pm$0.03) $\times$10$^{15}$ & 3.38($\pm$0.36) $\times$10$^{14}$ & 1.24($\pm$0.05) $\times$10$^{15}$ & 5.75($\pm$0.12) $\times$10$^{14}$ & 6.45($\pm$0.31) $\times$10$^{14}$ & 1.62($\pm$0.18) $\times$10$^{14}$ & 1.81($\pm$0.04) $\times$10$^{15}$  & 7.66($\pm$0.18) $\times$10$^{14}$ & 1.15($\pm$0.04) $\times$10$^{15}$ \\  
& $N_\mathrm{tot}^\mathrm{corr}$ (cm$^{-2}$) & 1.74($\pm$0.26) $\times$10$^{14}$ & 2.10($\pm$0.31) $\times$10$^{14}$ & 2.38($\pm$0.836) $\times$10$^{14}$ & 3.56($\pm$0.49) $\times$10$^{14}$ & 6.63($\pm$1.20) $\times$10$^{13}$ & 2.24($\pm$0.31) $\times$10$^{14}$ & 1.08($\pm$0.18) $\times$10$^{14}$ & 3.40($\pm$0.51) $\times$10$^{14}$ & 1.24($\pm$0.18) $\times$10$^{14}$ & 2.33($\pm$0.36) $\times$10$^{14}$ \\  
& $N_\mathrm{wings}$ (cm$^{-2}$) & 1.12($\pm$0.01) $\times$10$^{14}$ & 8.31($\pm$0.10) $\times$10$^{13}$ & 6.51($\pm$0.09) $\times$10$^{13}$ & 1.40($\pm$0.01) $\times$10$^{14}$ & 4.07($\pm$0.13) $\times$10$^{13}$ & 4.34($\pm$0.09) $\times$10$^{13}$ & 4.33($\pm$0.12) $\times$10$^{13}$ & 2.62($\pm$0.02) $\times$10$^{14}$ & 8.23($\pm$0.11) $\times$10$^{13}$ & 1.03($\pm$0.01) $\times$10$^{14}$ \\  
& $N_\mathrm{wings}^\mathrm{corr}$ (cm$^{-2}$) & 2.81($\pm$0.01) $\times$10$^{14}$ & 8.29($\pm$0.10) $\times$10$^{13}$ & 6.51($\pm$0.09) $\times$10$^{13}$ & 2.95($\pm$0.01) $\times$10$^{14}$ & 4.07($\pm$0.13) $\times$10$^{13}$ & 7.76($\pm$0.09) $\times$10$^{13}$ & 4.3(3$\pm$0.12) $\times$10$^{13}$ & 4.95($\pm$0.02) $\times$10$^{14}$ & 8.23($\pm$0.11) $\times$10$^{13}$  & 2.94($\pm$0.01) $\times$10$^{14}$ \\  
\hline

\multirow{4}{*}{CS 3-2} & $N_\mathrm{tot}$ (cm$^{-2}$) & 2.04($\pm$0.06) $\times$10$^{14}$ & 1.31($\pm$0.08) $\times$10$^{14}$ & 6.34($\pm$0.87) $\times$10$^{13}$ & 2.31($\pm$0.17) $\times$10$^{14}$ & 5.22($\pm$0.27) $\times$10$^{13}$ & 1.22($\pm$0.07) $\times$10$^{14}$ & 3.35($\pm$0.37) $\times$10$^{13}$ & 2.94($\pm$0.14) $\times$10$^{14}$ & 6.13($\pm$0.44) $\times$10$^{13}$ & 7.54($\pm$0.84) $\times$10$^{13}$ \\  
& $N_\mathrm{tot}^\mathrm{corr}$ (cm$^{-2}$) & 3.25($\pm$0.62) $\times$10$^{13}$ & 4.76($\pm$0.81) $\times$10$^{13}$ & 4.61($\pm$0.87) $\times$10$^{13}$ & 8.32($\pm$1.70) $\times$10$^{13}$ & 1.35($\pm$0.27) $\times$10$^{13}$ & 4.84($\pm$0.74) $\times$10$^{13}$ & 2.50($\pm$0.37) $\times$10$^{13}$ & 7.01($\pm$1.40) $\times$10$^{13}$ & 1.97($\pm$0.44) $\times$10$^{13}$ & 5.49($\pm$0.84) $\times$10$^{13}$ \\ 
& $N_\mathrm{wings}$ (cm$^{-2}$) & 3.04($\pm$0.50) $\times$10$^{13}$ & 2.06($\pm$0.07) $\times$10$^{13}$ & 2.63($\pm$0.07) $\times$10$^{13}$ & 6.07($\pm$0.07) $\times$10$^{13}$ & 5.91($\pm$0.70) $\times$10$^{12}$ & 1.82($\pm$0.06) $\times$10$^{13}$ & 7.88($\pm$0.61) $\times$10$^{12}$ & 7.89($\pm$0.08) $\times$10$^{13}$ & 1.90($\pm$0.06) $\times$10$^{13}$ & 2.16($\pm$0.05) $\times$10$^{13}$\\  
& $N_\mathrm{wings}^\mathrm{corr}$ (cm$^{-2}$) & 8.33($\pm$0.50) $\times$10$^{13}$ & 4.12($\pm$0.07) $\times$10$^{13}$ & 2.63($\pm$0.07) $\times$10$^{13}$ & 1.14($\pm$0.01) $\times$10$^{14}$ & 1.69($\pm$0.07) $\times$10$^{13}$ & 1.82($\pm$0.06) $\times$10$^{13}$ & 7.88($\pm$0.61) $\times$10$^{12}$& 1.74($\pm$0.01) $\times$10$^{14}$ & 1.90($\pm$0.06) $\times$10$^{13}$  & 4.41($\pm$0.05) $\times$10$^{13}$ \\ 
\hline\hline
\end{tabular}
\end{sidewaystable*}

\begin{table}
\caption{Comparison of HCN column densities calculated from HCN 1-0 and H$^{13}$CN 1-0 observations}\label{table:HCNs}

\begin{tabular}{l | c c | c | c}
 \hline \hline
Source & $N$(HCN)$_{\tau_1}$ & $N$(HCN)$_\mathrm{\tau_2}$  & $N$(H$^{13}$CN)  & Ratios \\ 
& hyperfine & full profile & $\times68$ & \\
 & 10$^{14}$ (cm$^{-2}$) & 10$^{14}$ (cm$^{-2}$) &  10$^{14}$ (cm$^{-2}$) & \\
 \hline \hline
SMM1 & 4.8($\pm$0.8) & 1.7($\pm$0.3) & 6.0($\pm$0.1) & 1.3-3.5\\
SMM2 & 5.8($\pm$0.9) & 2.1($\pm$0.3) & 5.7($\pm$~0.1) & 1.0-2.7\\
SMM3 & 7.0($\pm$1.1) & 2.4($\pm$0.8) & 2.1($\pm$~0.1) & 0.3-0.9\\
SMM4 & 8.9($\pm$1.4) & 3.6($\pm$0.5) & 6.1($\pm$0.1) & 0.7-1.7\\
SMM5 & 2.3($\pm$~0.4) & 0.7($\pm$0.1) & 2.9($\pm$~0.1) & 1.3-4.1\\
SMM6 & 5.7($\pm$~0.9) & 2.2($\pm$0.3) & 3.3($\pm$~0.1) & 0.6-1.5\\ 
SMM8 & - & 1.1($\pm$0.2) & 0.5($\pm$~0.1) & 0.5\\
SMM9 & - & 3.4($\pm$0.5) & 9.0($\pm$~0.1) & 2.6\\
SMM10 & - & 1.2($\pm$0.2) & 3.9($\pm$~0.1) & 3.3\\
SMM12 & 7.5($\pm$~1.1) & 2.3($\pm$0.4) & 6.1($\pm$~0.1) & 0.8-2.7\\  
\hline
\end{tabular}
\tablefoot{$N$(HCN)$_{\tau_1}$ refers to HCN column densities corrected for the optical depths calculated using the ratio of F=2$\rightarrow$1 and F=1$\rightarrow$1 hyperfine components (see Table E.4). $N$(HCN)$_\mathrm{\tau_2}$ refers to HCN column densities calculated using optical depths assuming that H$^{13}$CN is optically thin (see Table E.3). The range of ratios of HCN column densities scaled from the H$^{13}$CN column densities and the HCN column densities corrected for optical depths are shown in the last column.}
\end{table}

\section{Line ratios and optical depths}
\begin{figure*} 
\begin{subfigure}{.5\textwidth} 
\centering
\includegraphics[width=1\linewidth]{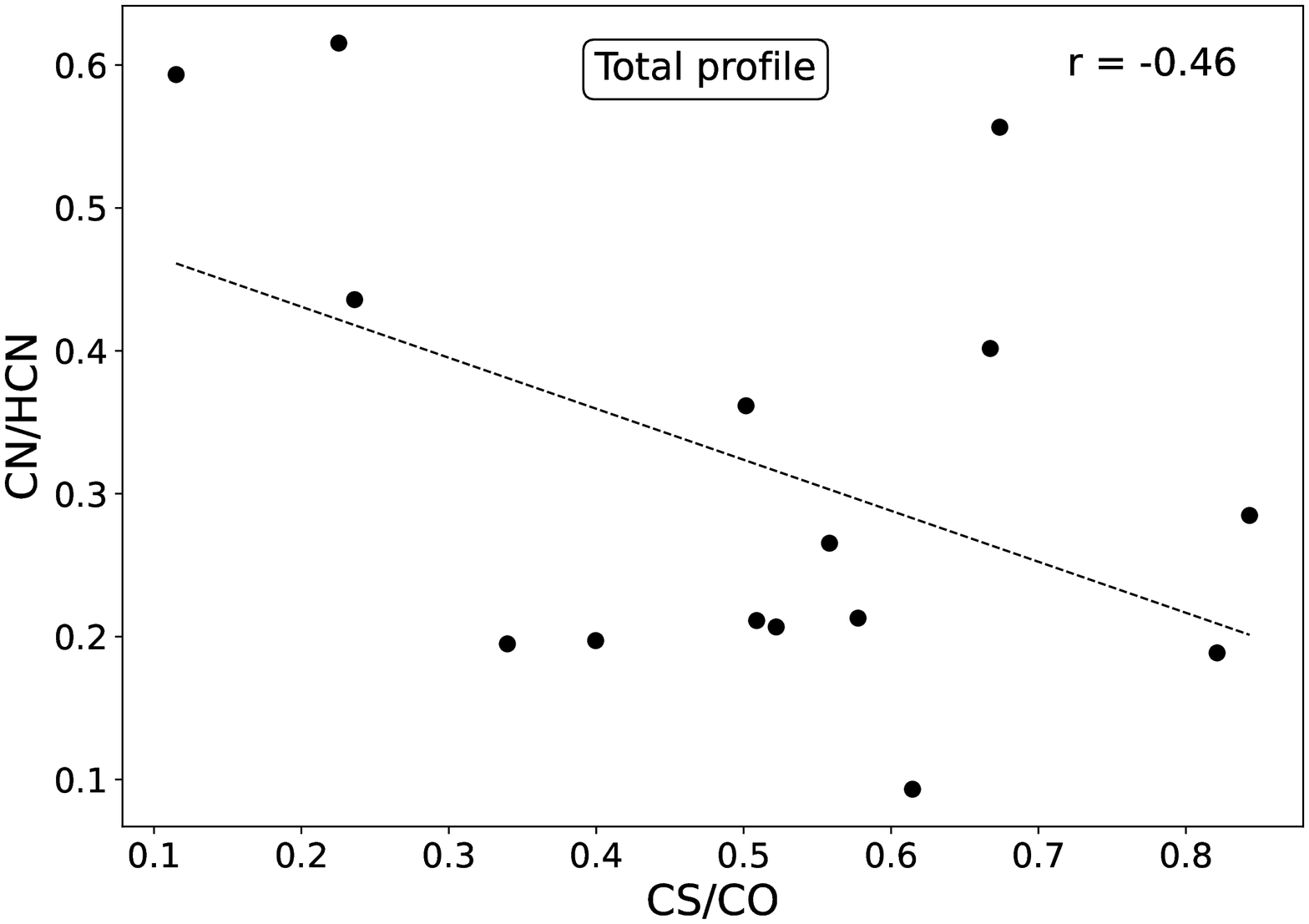} 
\end{subfigure}
\begin{subfigure}{.5\textwidth} 
\centering
\includegraphics[width=1\linewidth]{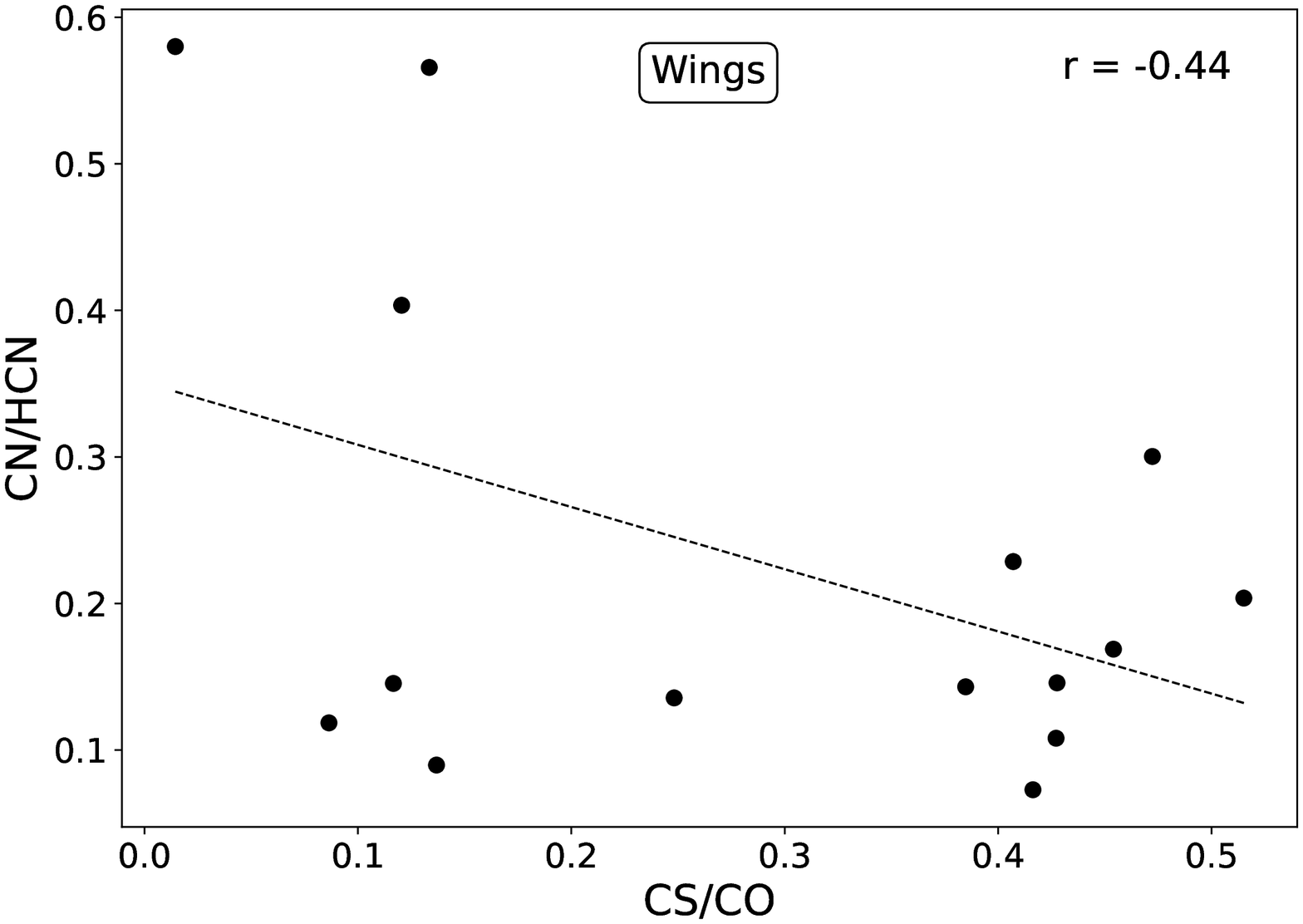} 
\end{subfigure}
\begin{subfigure}{.5\textwidth} 
\centering
\includegraphics[width=1\linewidth]{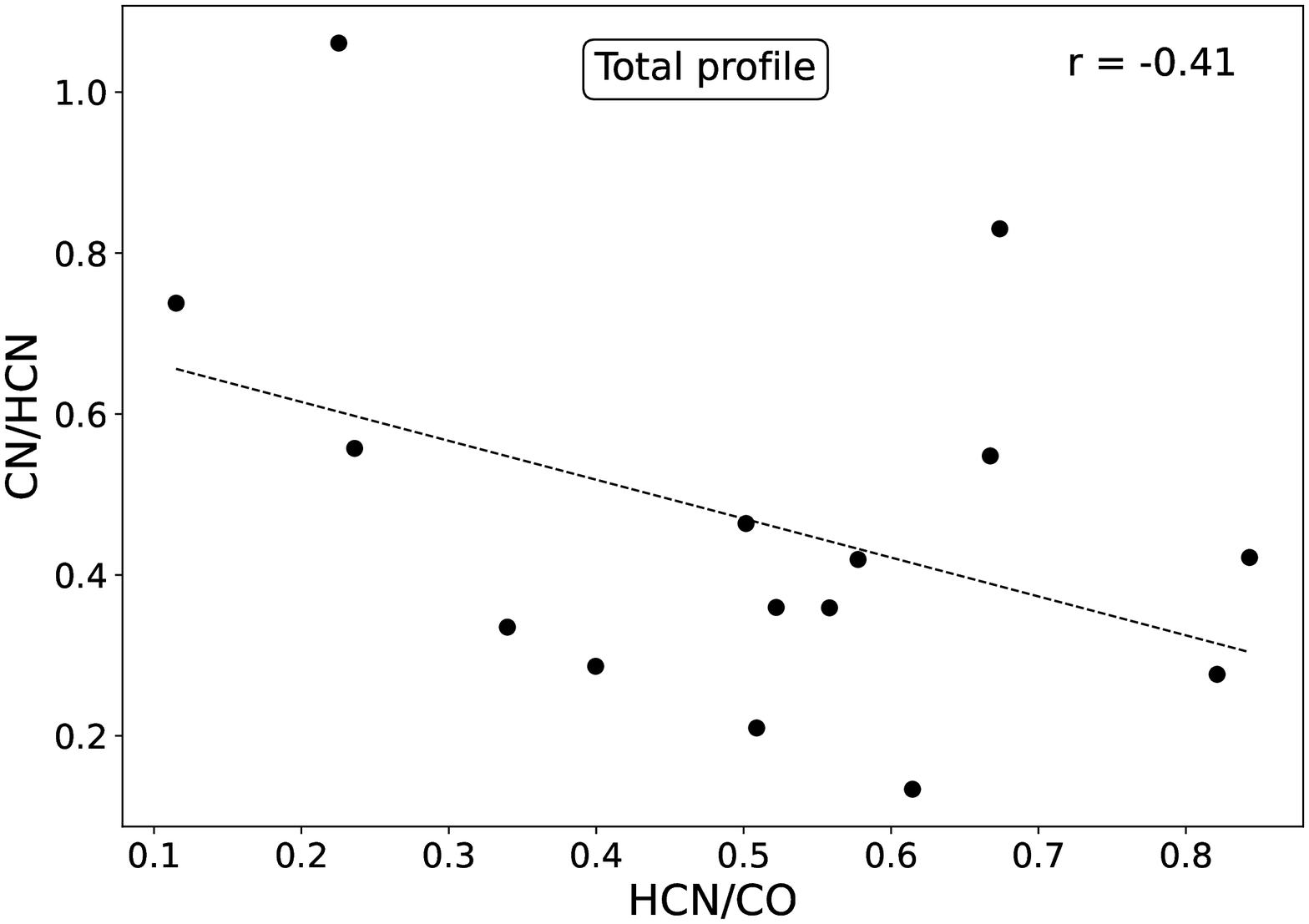} 
\end{subfigure}
\begin{subfigure}{.5\textwidth} 
\centering
\includegraphics[width=1\linewidth]{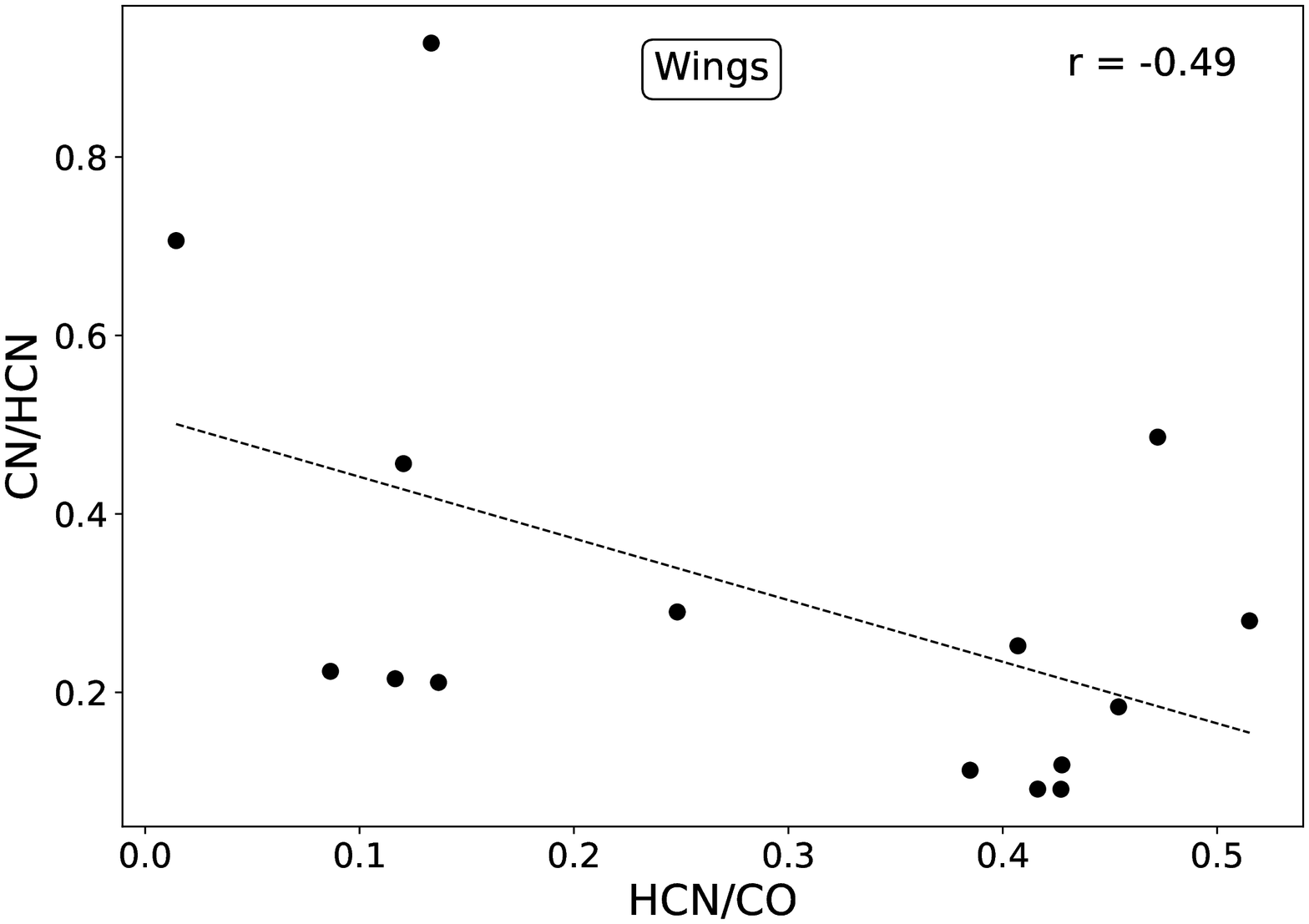} 
\end{subfigure}
\begin{subfigure}{.5\textwidth} 
\centering
\includegraphics[width=1\linewidth]{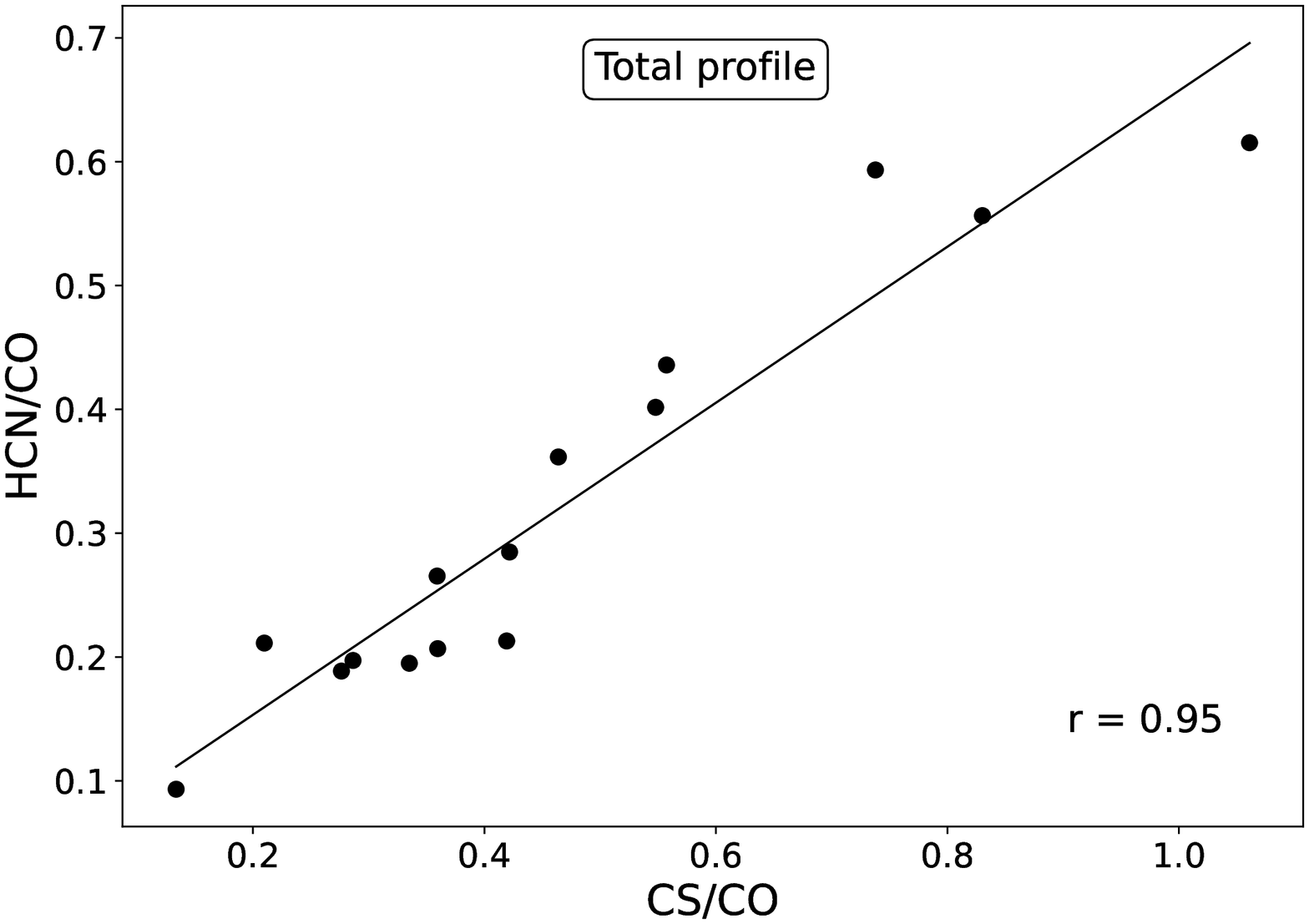} 
\end{subfigure}
\begin{subfigure}{.5\textwidth} 
\centering
\includegraphics[width=1\linewidth]{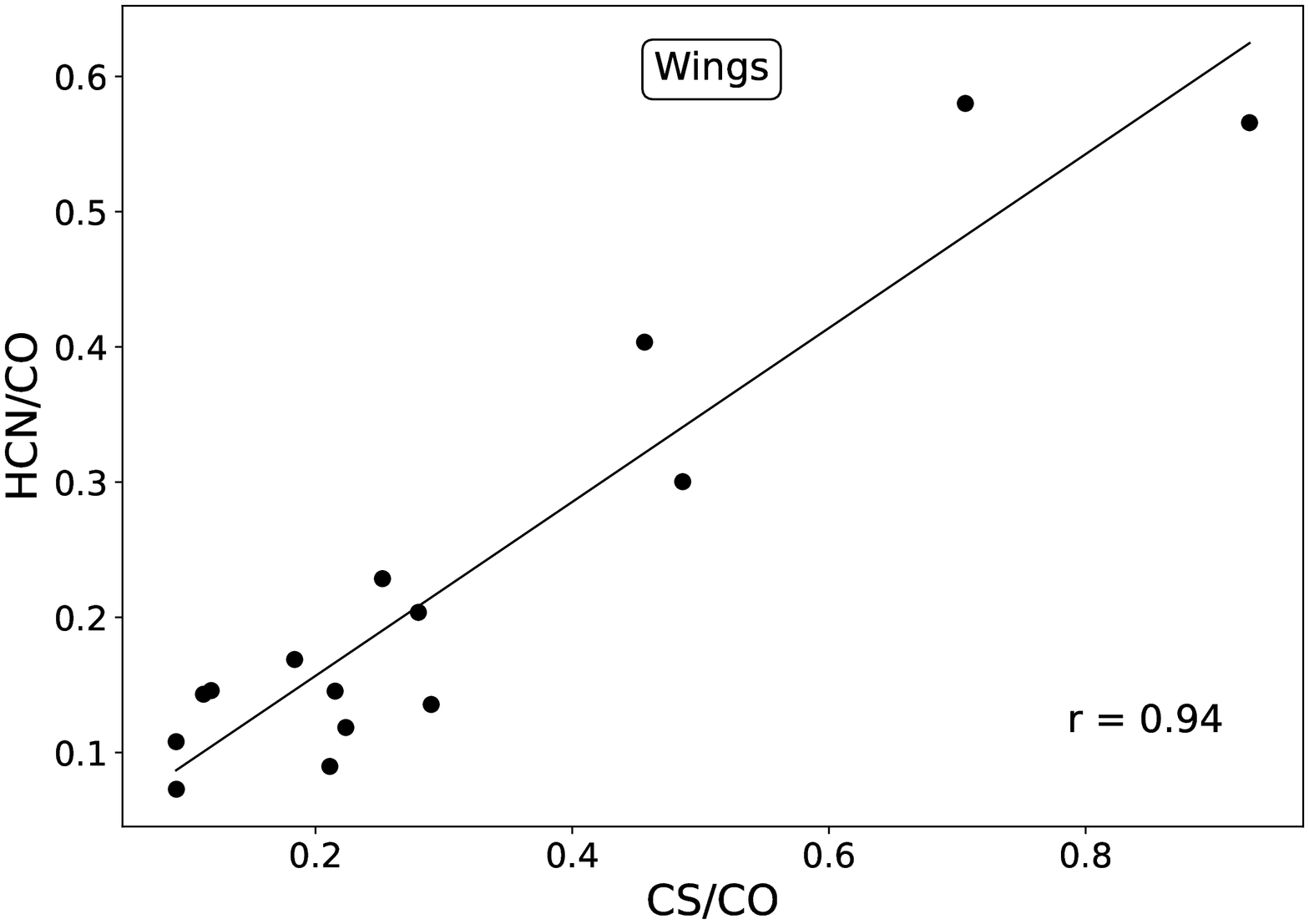} 
\end{subfigure}
\caption{\label{fluxes_correlation} Correlations of the molecular line ratios of the fully integrated line profile (left panel) and the line wings (right panel) derived in protostellar and outflow positions. The Pearson coefficients ($r$) are shown on the plots; the number of datapoints is 15 in all cases. Solid lines show the best linear fits obtained with a least-squares method for cases where the Pearson coefficient exceeds the 3$\sigma$ threshold. Dashed lines show the fits for cases where the Pearson coefficient corresponds to 1.5-1.8$\sigma$ levels.}
\end{figure*}

Table \ref{table:CN/HCN} shows line ratios of different species and Table \ref{table:CN/HCN-isotopolog} shows line ratios of different isotopologues of the same species. Figure~\ref{fluxes_correlation} shows a corresponding correlation plots including the Pearson coefficient, $r$.

Table \ref{table:taus} shows optical depths of HCN and CS lines calculated
 using their rare isotopologues. Table \ref{table:taus_hyperfine} shows optical depths calculated using the ratios of hyperfine-splitted components for HCN, H$^{13}$CN, and CN lines. 
 
 The Gaussian profiles were fitted to the hyperfine splitting components excluding wings with the PYSPECKIT package \footnote{https://pyspeckit.readthedocs.io/en/latest/}. The lines are blended at Ser SMM9, SMM10, and Outflow 4 positions, so the models cannot be fitted. Positive values of optical depths, $\tau>0$, are found when the following condition is met:
 \begin{equation} 
\label{eq5b}  \frac
{1-\exp(-\tau_{\mathrm{HCN}})}{\tau_{\mathrm{HCN}}} < 1 ,
\end{equation} 

Both methods show that HCN is optically thick ($\tau_\mathrm{HCN}$>1) in all positions. The median value of the optical depth for H$^{13}$CN is $\sim1$ and thus for some sources it is not optically thin. The exact values of $\tau_\mathrm{HCN}$ depend on the ratios which are used, which has been noted for other sources and referred to as hyperfine anomalies \citep{Lou12,Mul16}. CN is optically thin toward most positions (except Ser SMM1, SMM5, and SMM10), and CS is optically thick in all positions except Ser SMM3 and SMM10, and off-position 3.

\begin{table*} \caption{Molecular line ratios of the fully integrated line profile and the line wings}
\small
\label{table:CN/HCN}
\centering 
\begin{tabular}{l c c c c c c} 
\hline\hline 
Source & \multicolumn{2}{c}{CN / HCN} & \multicolumn{2}{c}{CS / CO} & \multicolumn{2}{c}{HCN / CO} \\
~ & Full & Line wings & Full & Line wings & Full & Line wings  \\
\hline 
\multicolumn{7}{c}{Protostellar positions} \\
\hline 
Ser SMM1 & 0.61$\pm$0.10 & 0.40$\pm$0.08 & 0.08$\pm$0.01 & 0.05$\pm$0.01 & 0.14$\pm$0.02 & 0.10$\pm$0.02 \\
Ser SMM2 & 0.65$\pm$0.10 & 0.53$\pm$0.15 & 0.38$\pm$0.06 & 0.23$\pm$0.05 & 0.58$\pm$0.09 & 0.35$\pm$0.07 \\
Ser SMM3 & 0.85$\pm$0.13 & 0.57$\pm$0.11 & 0.16$\pm$0.03 & 0.10$\pm$0.03 & 0.27$\pm$0.04 & 0.10$\pm$0.02  \\
Ser SMM4 & 0.47$\pm$0.08 & 0.40$\pm$0.08 & 0.16$\pm$0.03 & 0.12$\pm$0.03 & 0.23$\pm$0.03 & 0.13$\pm$0.03  \\
Ser SMM5 & 0.67$\pm$0.09 & 0.31$\pm$0.06 & 0.19$\pm$0.04 & 0.09$\pm$0.03 & 0.31$\pm$0.06 & 0.17$\pm$0.06 \\
Ser SMM6 & 0.81$\pm$0.13 & 0.55$\pm$0.11 & 0.29$\pm$0.04 & 0.19$\pm$0.03 & 0.44$\pm$0.07 & 0.14$\pm$0.03 \\
Ser SMM8 & 0.43$\pm$0.07 & 0.27$\pm$0.04 & - & - & - & -  \\
Ser SMM9 & 0.24$\pm$0.04 & 0.11$\pm$0.02 & 0.34$\pm$0.07 & 0.30$\pm$0.06 & 0.57$\pm$0.09 & 0.49$\pm$0.06 \\
Ser SMM10 & 0.41$\pm$0.06 & 0.15$\pm$0.03 & 0.14$\pm$0.03 & 0.09$\pm$0.03 & 0.29$\pm$0.04 & 0.23$\pm$0.04 \\
Ser SMM12 & 0.71$\pm$0.10 & 0.63$\pm$0.10 & 0.57$\pm$0.08 & 0.40$\pm$0.05 & 0.82$\pm$0.13 & 0.49$\pm$0.10 \\
\hline 
\multicolumn{7}{c}{Off-source positions} \\
\hline 
1 (-11.0,33.0) & 0.58$\pm$0.09 & 0.24$\pm$0.04 & 0.24$\pm$0.03 & 0.15$\pm$0.02 & 0.43$\pm$0.07 & 0.31$\pm$0.04\\
2 (114.9,-153.0) & 0.51$\pm$0.08 & 0.47$\pm$0.06 & 0.35$\pm$0.06 & 0.24$\pm$0.03 & 0.47$\pm$0.07 & 0.27$\pm$0.04 \\
3 (56.0,-112.0) & 0.13$\pm$0.02 & 0.03$\pm$0.01 & 0.54$\pm$0.09 & 0.56$\pm$0.16 & 0.74$\pm$0.13 & 0.72$\pm$0.20 \\
4 (-54.0,81.0) & 0.25$\pm$0.04 & 0.13$\pm$0.03 & 0.63$\pm$0.10 & 0.54$\pm$0.14 & 0.97$\pm$0.16 & 0.84$\pm$0.22 \\
5 (138.9,-162.6) & 0.56$\pm$0.09 & 0.49$\pm$0.09 & 0.26$\pm$0.04 & 0.18$\pm$0.03 & 0.36$\pm$0.06 & 0.21$\pm$0.03 \\ 
6 (-60.0,-100.0) & 0.35$\pm$0.08 & 0.16$\pm$0.10 & 0.38$\pm$0.04 & 0.16$\pm$0.04 & 0.59$\pm$0.06 & 0.49$\pm$0.05 \\ 
\hline \end{tabular} 
\tablefoot{Sources in the upper part of the table refer to the positions of protostars
within the IRAM maps (Figure 1 and 2), positions in the lower part refer to selected
 positions on the same maps. Coordinates of the selected positions refer (in arcsec) to the center of the map (J2000 RA 
18:29:49.6 and DEC 01:15:20.5).}
\end{table*}

\begin{table*} \caption{Molecular line ratios of the fully integrated line profile and the line wings using various isotopologues}
\small
\label{table:CN/HCN-isotopolog}
\centering 
\begin{tabular}{l c c c c } 
\hline\hline 
Source &  \multicolumn{2}{c}{HCN / H$^{13}$CN} & CS / C$^{34}$S & H$^{13}$CN 2-1/1-0\\
~ & Full & Line wings & Full & Full \\
\hline 
\multicolumn{5}{c}{Protostellar positions} \\
\hline 
Ser SMM1 & 6.51$\pm$1.18 & 12.45$\pm$4.02 & 4.66$\pm$2.01 & - \\
Ser SMM2 & 8.74$\pm$1.58 & 22.28$\pm$8.85 & 9.98$\pm$3.84 & - \\
Ser SMM3 & 25.99$\pm$5.08 & 42.90$\pm$21.95 & 16.50$\pm$8.23 & - \\
Ser SMM4 & 13.43$\pm$2.07 & 14.91$\pm$3.17 & 9.88$\pm$4.64 & 1.98$\pm$0.88 \\
Ser SMM5 & 5.26$\pm$1.41 & - & 7.95$\pm$4.14 & - \\
Ser SMM6 & 15.10$\pm$3.38 & 17.39$\pm$13.47 & 11.19$\pm$3.89 & - \\
Ser SMM8 & 49.39$\pm$10.63 & 117.2$\pm$88.52 & 17.57$\pm$14.39 & - \\
Ser SMM9 & 8.73$\pm$1.69 & 16.33$\pm$3.51 & 6.78$\pm$2.89 & 1.31$\pm$0.39 \\
Ser SMM10 & 7.00$\pm$1.33 & 31.11$\pm$25.76 & 9.12$\pm$5.97 & - \\
Ser SMM12 & 8.84$\pm$1.51 & 10.82$\pm$3.29 & 17.70$\pm$6.32 & - \\
\hline 
\multicolumn{5}{c}{Off-source positions} \\
\hline 
1 (-11.0,33.0) & 9.87$\pm$1.71 & - & 4.16$\pm$1.39 & - \\
2 (114.9,-153.0) & 13.48$\pm$2.18 & 27.12$\pm$12.16 & 10.71$\pm$3.53 & - \\
3 (56.0,-112.0) & 11.92$\pm$1.75 & 12.29$\pm$3.64 & 15.57$\pm$6.03 & - \\
4 (-54.0,81.0) & 8.38$\pm$1.41 & 12.12$\pm$2.78 & 8.90$\pm$3.18 & - \\
5 (138.9,-162.6) & 9.05$\pm$1.44 & 11.52$\pm$2.67 & 9.76$\pm$3.40 & - \\ 
6 (-60.0,-100.0) & - & - & - & - \\ 
\hline \end{tabular} 
\tablefoot{Sources in the upper part of the table refer to the positions of protostars
within the IRAM maps (Figure 1 and 2), positions in the lower part refer to selected
 positions on the same maps. Coordinates of the selected positions refer (in arcsec) to the center of the map (J2000 RA 
18:29:49.6 and DEC 01:15:20.5).}
\end{table*}

\begin{table*} \caption{Optical depths of line emission in HCN and CS for total profiles and line wings using rare isotopologues} 
\small
\label{table:taus}
\centering 
\begin{tabular}{l c c c c} 
\hline\hline 
Source & \multicolumn{2}{c}{$\tau_{\mathrm{HCN}}$} &  \multicolumn{2}{c}{$\tau_{\mathrm{CS}}$}  \\
~ & Full & Line wings &  Full & Line wings\\
  \\
\hline 
\multicolumn{5}{c}{Protostar positions} \\
\hline 
Ser SMM1 & 10.31 & 2.12 &   4.39 & 2.56 \\
Ser SMM2 & 7.60 & 0.63 &  1.69 & 1.53  \\
Ser SMM3 & 2.33 & -- &  0.47 & --  \\
Ser SMM4 & 4.96 & 1.61 &  1.71 & 1.42  \\
Ser SMM5 & 12.80 & -- &   2.35 & 2.66   \\
Ser SMM6 & 4.38 & 1.21 &   1.39 & 0.89  \\
Ser SMM9 & 7.72 & 1.37 & 2.88 & 1.86  \\
Ser SMM10 & 7.69 & -- &   1.94 & --   \\
Ser SMM12 & 9.61 & 2.56 &   0.32 & 1.59  \\
\hline 
\multicolumn{5}{c}{Outflow positions} \\
\hline 
1 (-11.00,33.00) & 6.80 & --  & 4.91 & 5.63  \\
2 (114.86,-153.00) & 4.94 & 0.31 &  1.49 & 1.55  \\
3 (56.00,-112.00) & 5.60 & 2.32  &0.59 & 0.46 \\
4 (-54.00,81.00) & 8.00 & 2.37  & 2.01 & 1.25  \\
5 (138.86,-162.60) & 7.40 & 2.53 & 1.74 & 1.77  \\ 
\hline \end{tabular} 
\tablefoot{The '--' shows sources with negative values of optical depths obtained from calculations.}
\end{table*}

\begin{table*} \caption{Optical depths of HCN, CN, and H$^{13}$CN calculated using hyperfine-splitted components}
\small
\label{table:taus_hyperfine}
\centering 
\begin{tabular}{l c c |c c | c c c} 
\hline\hline 
Source  & \multicolumn{2}{c}{$\tau_{\mathrm{HCN}}$} & \multicolumn{2}{c}{$\tau_{\mathrm{H^{13}CN}}$} &
 \multicolumn{3}{c}{$\tau_{\mathrm{CN}}$} \\
~  & F=2$\rightarrow$1 $\div$ & F=1$\rightarrow$1 $\div$ & F=2$\rightarrow$1 $\div$ & F=1$\rightarrow$1 $\div$ & F=5/2$\rightarrow$3/2 $\div$  & F=5/2$\rightarrow$3/2 $\div$ &  F=1/2$\rightarrow$1/2 $\div$  \\
  ~ & F=1$\rightarrow$1 & F=0$\rightarrow$1 & F=1$\rightarrow$1 & F=0$\rightarrow$1 & F=1/2$\rightarrow$1/2 & F=3/2$\rightarrow$1/2 & F=3/2$\rightarrow$1/2  \\
\hline 
\multicolumn{8}{c}{Protostar positions} \\
\hline 
Ser SMM1 & 2.75 & 0.48 & 1.18 & 0.67 & 1.31 & 1.06 & -- \\
Ser SMM2 & 2.65 & 1.46 & 1.15 & 1.25 & 0.77 & 0.81 & 0.03 \\
Ser SMM3 & 2.73 & 1.06 & 1.02 & 2.79 & 0.46 & 0.45 & --  \\
Ser SMM4 & 2.54 & 0.58 & 1.70 & 0.85 & 0.57 & 0.86 & 0.27  \\
Ser SMM5 & 2.77 & 4.97 & -- & 3.01 & 1.39 & 2.10 & 0.54  \\
Ser SMM6 &  2.56 & 0.84 & 1.81 & 0.71 & 0.58 & 0.36 & --  \\
Ser SMM9 & \dots & \dots & 1.45 & 1.27 & 0.73 & 0.83 & 0.09  \\
Ser SMM10 & \dots & \dots & 0.77 & 1.19 & 1.64 & 1.20 & --  \\
Ser SMM12 &  3.02 & 1.00 & 2.11 & 0.92 & 0.65 & 0.85 & 0.18  \\
\hline 
\multicolumn{8}{c}{Outflow positions} \\
\hline 
1 (-11.00,33.00) & 3.21 & -- & 0.83 & 0.86 & 0.76 & 1.02 & 0.23  \\
2 (114.86,-153.00) & 2.66 & 0.34 &2.76 & -- & 0.34 & 0.16 & --  \\
3 (56.00,-112.00) & 1.48 & 4.06 & 1.41 & 3.29 & -- & -- & --  \\
4 (-54.00,81.00) &\dots & \dots & 0.88 & 1.32  & 0.84 & 0.64 & --  \\
5 (138.86,-162.60) & 4.12 & 0.36 & 1.07 & 0.84 & 0.42 & 0.56 & 0.14  \\ 
\hline \end{tabular} 
\tablefoot{The optically thin LTE line intensity ratio of hyperfine-splitted components is 1:3:5, both for HCN 1-0 and H$^{13}$CN 1-0. Hyperfine-splitted components of HCN for Ser SMM9, SMM10, and off-position 4 are blended, and the optical depths cannot be determined. The '--' shows sources with negative values of optical depths obtained from calculations.}

\end{table*}

\section{Dominant reactions in CN, HCN chemistry}
\label{app:reactions}

Dominant processes in CN and HCN reactions modeled at a temperature of 50 K and 300 K
 are listed in Table~\ref{reactions_50} and Table~\ref{reactions_300} respectively. The
  processes are illustrated in the Nahoon model parameter space: main channels of HCN
   and CN destruction (Figure~\ref{dest_50} and Figure~\ref{CN_HCN_dest}, as well as HCN
    and CN production (Figure~\ref{prod_50} and Figure~\ref{prod_300}). Only the reactions
     contributing at least 30\% in total flux are taken into consideration. Dominant
      reactions are more dependent on the strength of the UV radiation than on the
       hydrogen density, although in same cases the hydrogen density parameter plays
        role (See Tab.~\ref{reactions_weight}). Tab.~\ref{reactions_weight} provides detailed information illustrated in Figure~\ref{dest_50}, Figure~\ref{CN_HCN_dest}, Figure~\ref{prod_50} and Figure~\ref{prod_300}. The weights of dominant reactions at each point of the calculated parameter space are presented as the percentage of the total flux. Reaction networks for UV radiation
         in weak (G$_0$ = $10^{-4} - 10^{-1}$), intermediate (G$_0$ = $10^{-1} - 10^{1}$)
          and strong (G$_0$ = $10^{1} - 10^{6}$) regime are presented in
           Figure~\ref{reactions_small_mediumG0} and Figure~\ref{reactions_largeG0}.
            Not only the dominant reactions, but also the main route of the reactants
             production are illustrated.

\begin{figure} 
\centering \includegraphics[width=10cm]{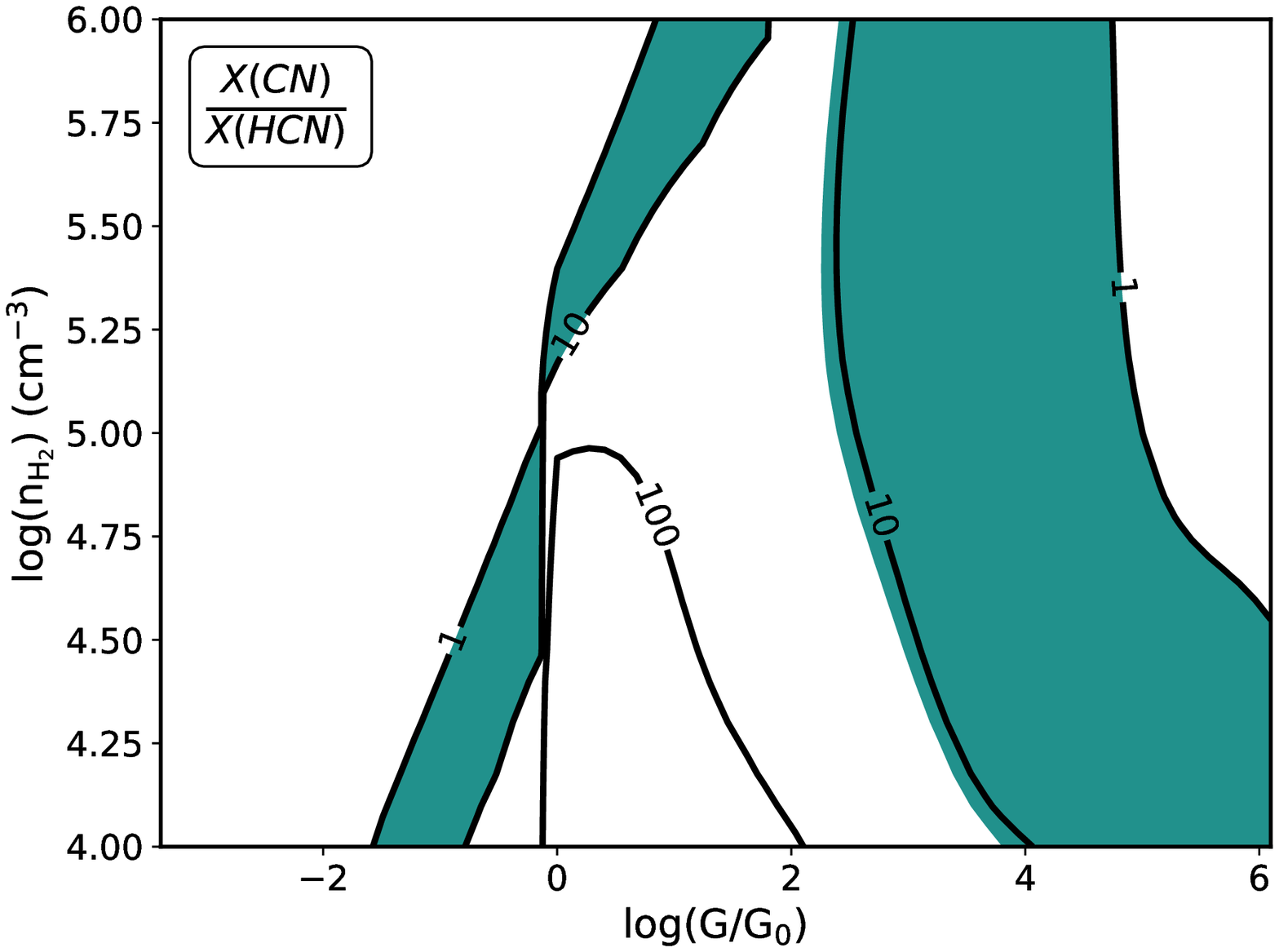} 
\caption{\label{G0_300} The column density ratio of CN to HCN from Nahoon for a 
range of hydrogen densities and UV field strengths assuming $T = 300$ K. 
The corresponding ratios from observations and radiative transfer models are shown in green.} 
\end{figure}
\begin{table*} 
\caption{Dominant processes in CN and HCN chemistry at 300 K} 
\centering 
\label{reactions_300} 

\begin{tabular}{c c c c} 
\hline\hline 
Molecule & Weak UV fields & Medium UV fields & Strong UV fields \\ 
& (G$_0$ = $10^{-3} - 10^{-1})$ & (G$_0$ = $10^{-1} - 10^{1})$ & (G$_0$ = $10^{1} - 10^{6})$ \\ 
\hline 
\multirow{8}{*}{CN} & \multicolumn{3}{c}{\textbf{Destruction}}\\ 
&  CN + H$_2$ $\rightarrow$ H + HCN & O + CN $\rightarrow$ N + CO & CN + h$\nu$ $\rightarrow$ C + N\\ 
& & CN + H$_2$ $\rightarrow$ H + HCN & CN + H$_2$ $\rightarrow$ H + HCN \\
& & CN + h$\nu$ $\rightarrow$ C + N & \\
\vspace{2.5 pt} &\multicolumn{3}{c}{\textbf{Production}}\\ 
& HCN + h$\nu$ $\rightarrow$ H + CN & CNC$^+$ + e$^-$ $\rightarrow$ C + CN & H + CN$^+$ $\rightarrow$ CN + H$^+$ \\
& CNC$^+$ + e$^-$ $\rightarrow$ C + CN & H + CN$^+$ $\rightarrow$ CN + H$^+$ & N + CH $\rightarrow$ H + CN\\
& HCNH$^+$ + e$^-$ $\rightarrow$ H + H+ CN & HCN + h$\nu$ $\rightarrow$ H + CN & HCN + h$\nu$ $\rightarrow$ H + CN \\
\hline
\multirow{9}{*}{HCN} & \multicolumn{3}{c}{\textbf{Destruction}}\\ 
& HCN + h$\nu$ $\rightarrow$ H + CN & HCN + C$^+$ $\rightarrow$ H + CNC$^+$ & HCN + h$\nu$ $\rightarrow$ H + CN\\
&HCN + C$^+$ $\rightarrow$ H + CNC$^+$ & HCN + h$\nu$ $\rightarrow$ H + CN & HCN + C$^+$ $\rightarrow$ H + CNC$^+$ \\
&HCN + H$_3$O$^+$ $\rightarrow$ H$_2$O + HCNH$^+$ & & \\
\vspace{2.5 pt} 
&\multicolumn{3}{c}{\textbf{Production}}\\ 
&CN + H$_2$ $\rightarrow$ H + HCN & H + CCN $\rightarrow$ C + HCN & HCNH$^+$ + e$^-$ $\rightarrow$ H + HCN\\
&H + CCN $\rightarrow$ C + HCN & CN + H$_2$ $\rightarrow$ H + HCN & H + CCN $\rightarrow$ C + HCN\\
& & HCNH$^+$ + e$^-$ $\rightarrow$ H + HCN & N + CH$_2$ $\rightarrow$ H + HCN\\ 
\hline \end{tabular} 
\end{table*}
\longtab[2]{
\label{reactions_weight} 
\begin{longtable}{c | c c c c | c c c c} 
\caption{Weights of dominant processes in CN and HCN chemistry}\\
\hline\hline 
&\multicolumn{4}{c}{\textbf{T = 50 K}} & \multicolumn{4}{c}{\textbf{T = 300 K}}\\
\textbf{n$_{\mathrm{H_2}}$ (cm$^{-3}$)} & \multicolumn{4}{c}{\textbf{UV fields (G$_0$)}} & \multicolumn{4}{c}{\textbf{UV fields (G$_0$)}}\\ 
& 10$^{-2}$ & 10$^{0}$ & 10$^{2}$ & 10$^{4}$ & 10$^{-2}$ & 10$^{0}$ & 10$^{2}$ & 10$^{4}$ \\ \hline
\endfirsthead
\caption{continued.}\\
\hline\hline 
&\multicolumn{4}{c}{\textbf{T = 50 K}} & \multicolumn{4}{c}{\textbf{T = 300 K}}\\
\textbf{n$_{\mathrm{H_2}}$ (cm$^{-3}$)} & \multicolumn{4}{c}{\textbf{UV fields (G$_0$)}} & \multicolumn{4}{c}{\textbf{UV fields (G$_0$)}}\\ 
& 10$^{-2}$ & 10$^{0}$ & 10$^{2}$ & 10$^{4}$ & 10$^{-2}$ & 10$^{0}$ & 10$^{2}$ & 10$^{4}$ \\ \hline
\endhead

\endfoot
\endlastfoot

&\multicolumn{8}{c}{\textbf{CN + h$\nu$ $\rightarrow$ C + N}}\\ 
$10^4$ & 0\% & 66\% & 100\% & 100\% & 0\% & 62\% & 100\% & 100\% \\
$10^5$ & 0\% & 16\% & 98\% & 100\% & 0\% & 0\% & 98\% & 100\%  \\
$10^6$ & 0\% & 0\% & 80\% & 100\% & 0\% & 0\% & 75\% & 99\% \\
& \multicolumn{8}{c}{\textbf{O + CN $\rightarrow$ N + CO}}\\ 
$10^4$ & 71\% & 27\% & 0\% & 0\%  & 21\% & 25\% & 0\% & 0\% \\
$10^5$ & 70\% & 69\% & 0\% & 0\%  & 18\% & 50\% & 0\% & 0\%  \\
$10^6$ & 68\% & 81\% & 0\% & 0\% & 0\% & 25\% & 10\% & 0\% \\ 
& \multicolumn{8}{c}{\textbf{CN + N $\rightarrow$ C + N$_2$}}  \\ 
$10^4$ & 29\% & 0\% & 0\% & 0\% &  & &  &  \\
$10^5$ & 30\% & 0\% & 0\% & 0\% &  &  &  & \\
$10^6$ & 31\% & 0\% & 0\% & 0\% &  &  &  & \\ 
& \multicolumn{8}{c}{\textbf{CN + H$_2$ $\rightarrow$ H + HCN}} \\ 
$10^4$ &  &  & & & 60\% & 0\% & 0\% & 0\% \\
$10^5$ &  & &  &  & 64\% & 0\% & 0\% & 0\%\\
$10^6$ &  & &  & & 88\% & 50\% & 0\% & 0\%\\ \hline
& \multicolumn{8}{c}{\textbf{HCN + h$\nu$ $\rightarrow$ H + CN}}\\ 
$10^4$ & 0\% & 0\% & 83\% & 100\% & 47\% & 0\% & 93\% & 97\%\\
$10^5$ & 12\% & 0\% & 33\% & 97\%  & 69\% & 0\% & 59\% & 94\%  \\
$10^6$ & 35\% & 0\% & 0\% & 74\% & 79\% & 75\% & 0\% & 63\% \\
&\multicolumn{8}{c}{\textbf{HCN + C$^+$ $\rightarrow$ H + CNC$^+$}}\\ 
$10^4$ & 79\% & 91\% & 0\% & 0\% & 52\% & 83\% & 0\% & 0\% \\
$10^5$ & 74\% & 98\% & 66\% & 0\% & 23\% & 98\% & 41\% & 0\% \\
$10^6$ & 30\% & 99\% & 95\% & 26\% & 0\% & 25\% & 87\% & 37\% \\
& \multicolumn{8}{c}{\textbf{HCN + H$^+$ $\rightarrow$ H + HNC$^+$}}\\ 
$10^4$ & 13\% & 0\% & 0\% & 0\% &  &  & &  \\
$10^5$ & 0\% & 0\% & 0\% & 0\% & & & & \\
$10^6$ & 0\% & 0\% & 0\% & 0\% &  & & & \\ 
& \multicolumn{8}{c}{\textbf{HCN + H$_3$O$^+$ $\rightarrow$ H$_2$O + HCNH$^+$}}\\ 
$10^4$ &  &  & & & 0\% & 0\% & 0\% & 0\% \\
$10^5$ &  & &  &  & 0\% & 0\% & 0\% & 0\%\\
$10^6$ &  & & &  & 20\% & 0\% & 0\% & 0\%\\ \hline
&\multicolumn{8}{c}{\textbf{N + C$_2$ $\rightarrow$ C + CN}}\\ 
$10^4$ & 14\% & 0\% & 0\% & 0\% &  &  & &   \\
$10^5$ & 7\% & 45\% & 0\% & 0\% & & &  & \\
$10^6$ & 0\% & 57\% & 47\% & 0\% & &  & &  \\
&\multicolumn{8}{c}{\textbf{HCN + h$\nu$ $\rightarrow$ H + CN}}\\ 
$10^4$ & &  &  && 39\% & 0\% & 0\% & 0\%  \\
$10^5$ & &  &  & & 56\% & 0\% & 9\% & 15\% \\
$10^6$ & &  &  &  & 79\% & 61\% & 0\% & 21\% \\
&\multicolumn{8}{c}{\textbf{N + CH $\rightarrow$ H + CN}}\\ 
$10^4$ & 52\% & 0\% & 0\% & 0\% & 0\% & 0\% & 0\% & 0\% \\
$10^5$ & 31\% & 15\% & 33\% & 20\% & 0\% & 0\% & 9\% & 18\% \\
$10^6$ & 8\% & 0\% & 24\% & 56\% & 0\% & 0\% & 10\% & 37\%  \\
&\multicolumn{8}{c}{\textbf{H + CN$^+$ $\rightarrow$ CN + H$^+$}}\\ 
$10^4$ & 0\% & 69\% & 58\% & 33\% & 0\% & 88\% & 85\% & 69\% \\
$10^5$ & 0\% & 13\% & 18\% & 0\% & 0\% & 27\% & 44\% & 32\% \\
$10^6$ & 0\% & 0\% & 0\% & 0\% & 0\% & 0\% & 0\% & 0\% \\
&\multicolumn{8}{c}{\textbf{CNC$^+$ + e$^-$ $\rightarrow$ C + CN}}\\ 
$10^4$ & 21\% & 0\% & 0\% & 0\% & 42\% & 0\% & 0\% & 0\% \\
$10^5$ & 36\% & 0\% & 10\% & 0\% & 18\% & 36\% & 1\% & 0\% \\
$10^6$ & 31\% & 21\% & 0\% & 8\% & 0\% & 19\% & 43\% & 16\% \\
&\multicolumn{8}{c}{\textbf{HCN$^+$ + e$^-$ $\rightarrow$ H + CN}}\\ 
$10^4$ & 0\% & 13\% & 25\% & 56\% & &  &  &  \\
$10^5$ & 0\% & 0\% & 19\% & 34\% & &  &  &   \\
$10^6$ & 0\% & 0\% & 0\% & 0\% &  &  &  & \\  \hline
&\multicolumn{8}{c}{\textbf{CN + H$_2$ $\rightarrow$ H + HCN}}\\ 
$10^4$ &  & & & & 72\% & 45\% & 0\% & 0\%  \\
$10^5$ &  &  &  & & 80\% & 49\% & 0\% & 0\%  \\
$10^6$ &&  &  & & 87\% & 60\% & 20\% & 0\% \\ 
&\multicolumn{8}{c}{\textbf{N + CH$_2$ $\rightarrow$ H + HCN}}\\ 
$10^4$ & 55\% & 0\% & 0\% & 0\% & 0\% & 0\% & 0\% & 0\% \\
$10^5$ & 59\% & 0\% & 0\% & 13\% & 0\% & 0\% & 0\% & 12\% \\
$10^6$ & 50\% & 0\% & 0\% & 0\% & 0\% & 0\% & 0\% & 0\%\\
&\multicolumn{8}{c}{\textbf{H + CCN $\rightarrow$ C + HCN}}\\ 
$10^4$ & 21\% & 0\% & 0\% & 0\% & 21\% & 0\% & 0\% & 0\% \\
$10^5$ & 0\% & 32\% & 23\% & 0\% & 0\% & 24\% & 17\% & 0\% \\
$10^6$ & 0\% & 67\% & 71\% & 46\% & 0\% & 35\% & 67\% & 60\% \\
&\multicolumn{8}{c}{\textbf{HCNH$^+$ + e$^-$ $\rightarrow$ H + HCN}}\\ 
$10^4$ & 0\% & 76\% & 81\% & 86\% & 0\% & 46\% & 90\% & 88\% \\
$10^5$ & 0\% & 49\% & 55\% & 73\% & 0\% & 21\% & 66\% & 69\%\\
$10^6$ & 0\% & 13\% & 11\% & 23\% & 0\% & 0\% & 0\% & 21\% \\
\hline 
\end{longtable}}

\begin{figure} 
\begin{subfigure}{.45\textwidth} 
\includegraphics[width=10cm]{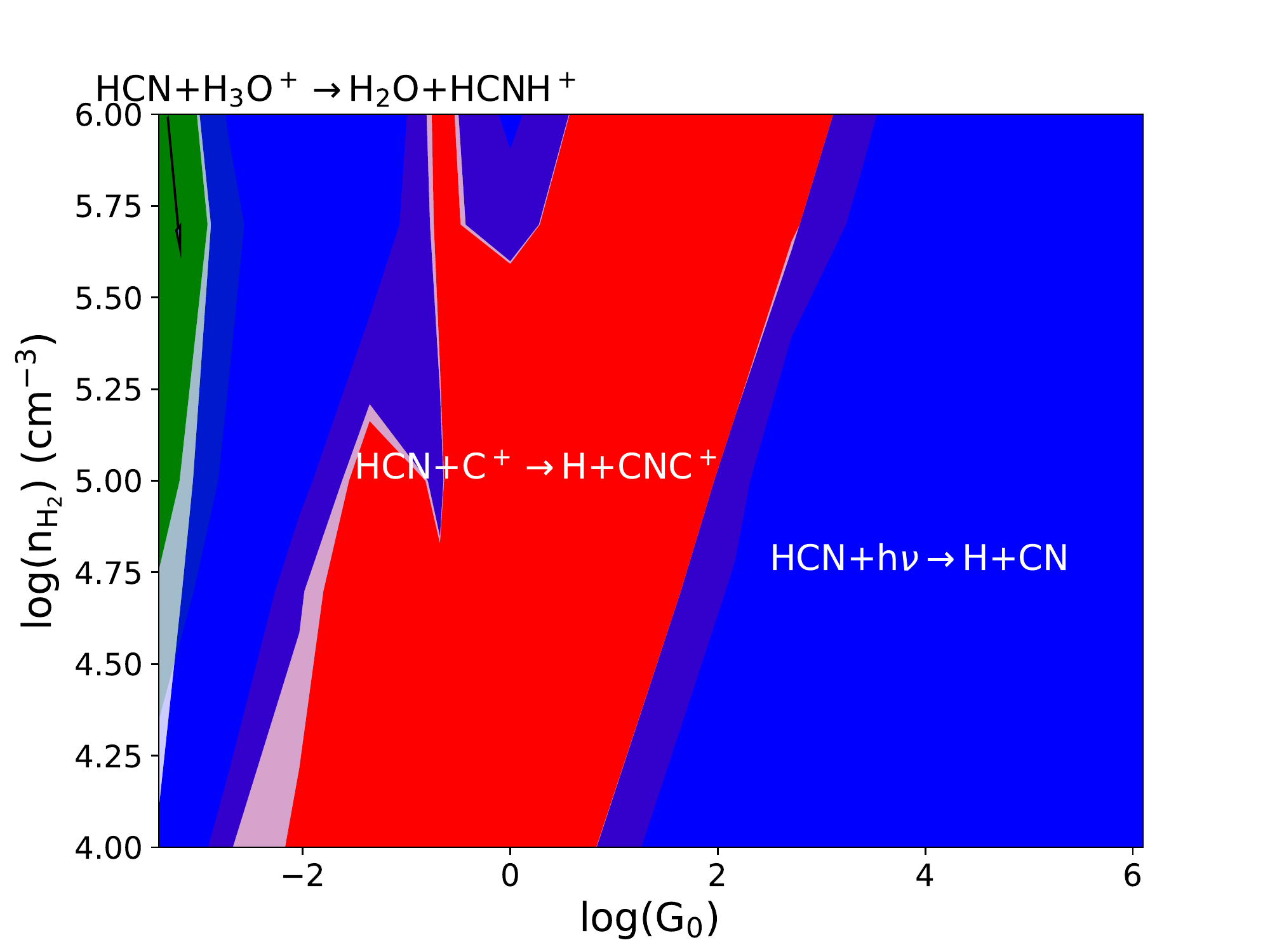} 
\end{subfigure}
\begin{subfigure}{.45\textwidth} 
\includegraphics[width=10cm]{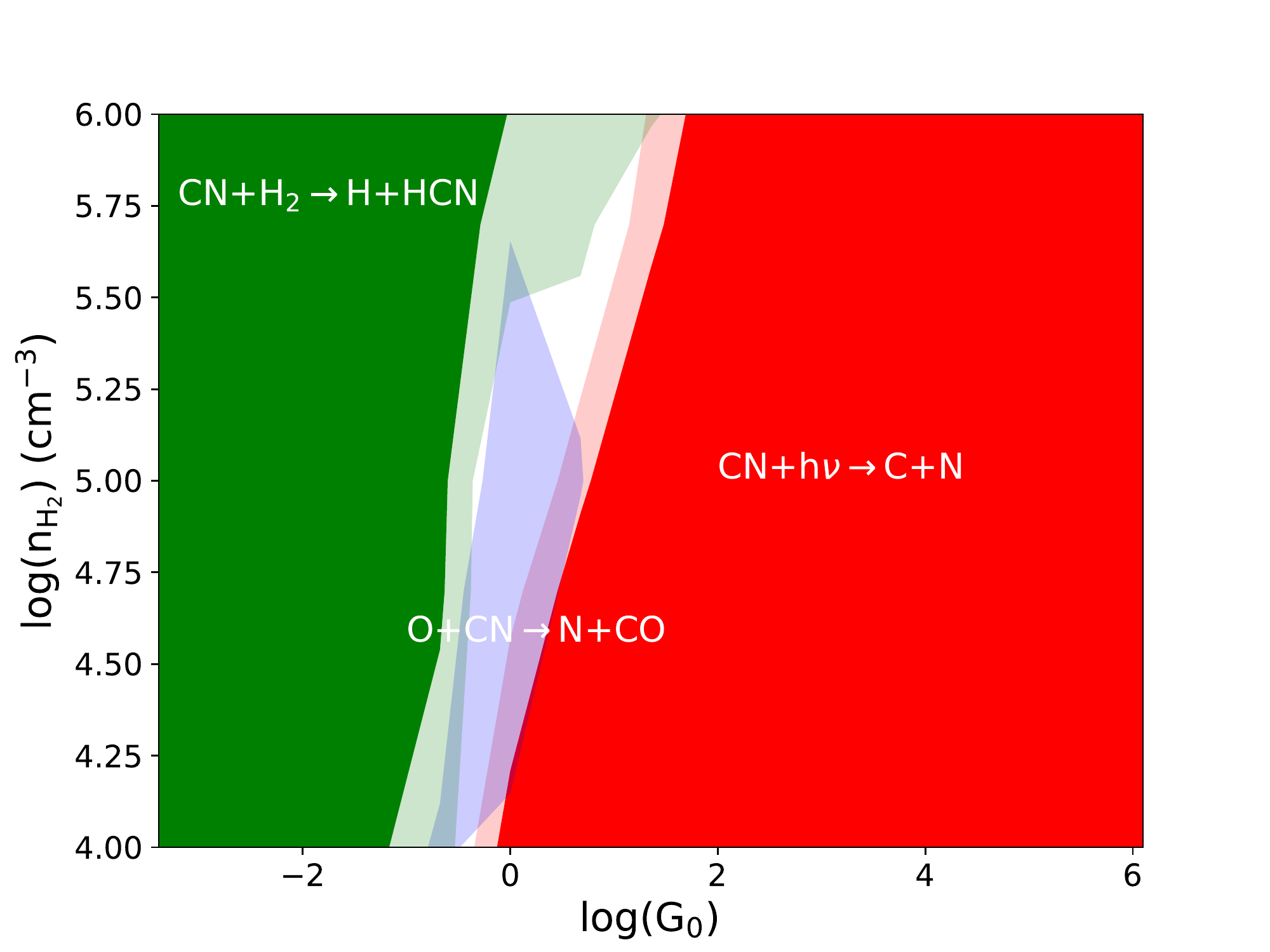} 
\end{subfigure}
\caption{Dominant reactions
of HCN (upper panel) and CN (bottom panel) destruction assuming $T~=~300$ K.
 Reactions contributed at least 50$\%$ of total flux are marked with full colors.
Transparent colors correspond to 30$\%$-50$\%$ contribution. HCN + h$\nu$
 $\rightarrow$ H + CN reaction is shown in blue, 
HCN + C$^+$ $\rightarrow$ H + CNC$^+$ in red, 
and HCN + H$_3$O$^+$ $\rightarrow$ H$_2$O + HCNH$^+$ in green. 
\mbox{CN + h$\nu$ $\rightarrow$ C + N} reaction is marked with red color,
 \mbox{O + CN $\rightarrow$ N + CO} with blue color and \mbox{CN + H$_2$ $\rightarrow$ H + HCN} with green color.} 
\label{CN_HCN_dest} 
\end{figure}

\begin{figure} 
\begin{subfigure}{.45\textwidth} 
\includegraphics[width=10cm]{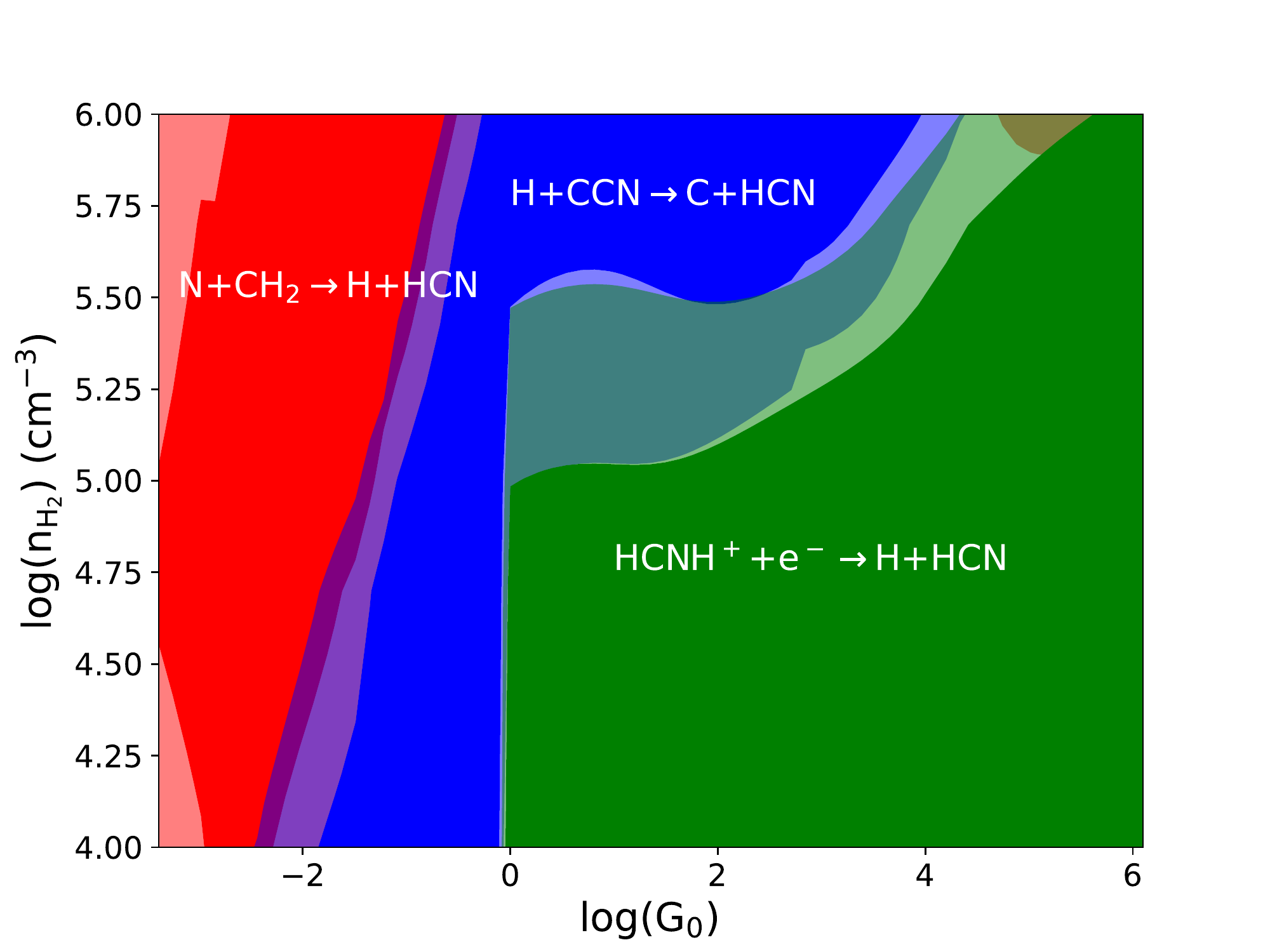} 
\end{subfigure}
\begin{subfigure}{.45\textwidth} 
\includegraphics[width=10cm]{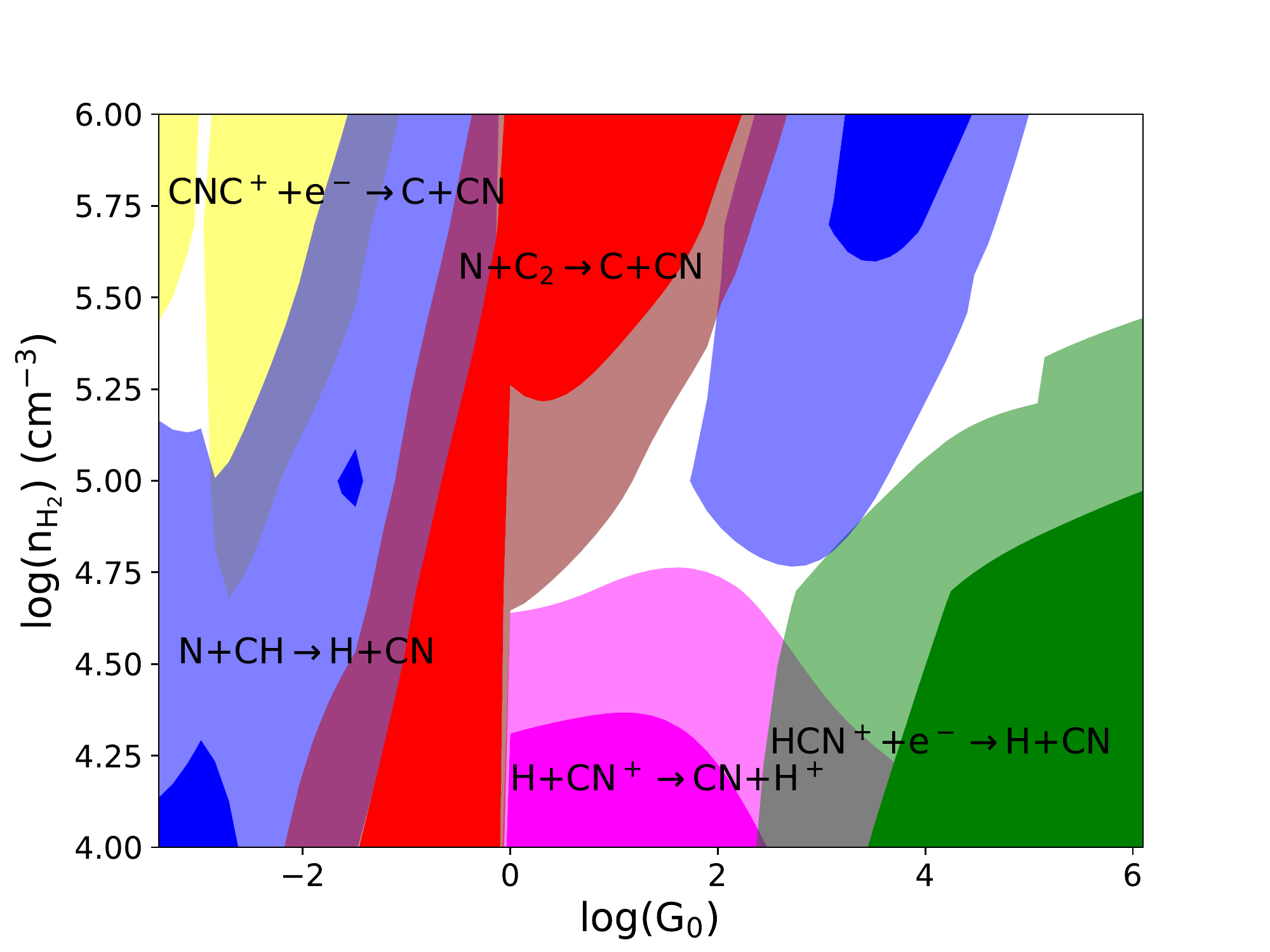} 
\end{subfigure}
\caption{Similar to Figure~\ref{CN_HCN_dest} but for HCN (upper panel) and CN (bottom panel)
 production assuming $T~=~50$~K. \mbox{HCNH$^+$ + e$^-$ $\rightarrow$ H + HCN} reaction
  is marked with green color, \mbox{H + CCN $\rightarrow$ C + HCN} with blue color and
   \mbox{N + CH$_2$ $\rightarrow$ H + HCN} with red color. 
   \mbox{HCN$^+$ + e$^-$ $\rightarrow$ H + CN} reaction is marked with green color,
    \mbox{N + C$_2$ $\rightarrow$ C + CN} with red color, \mbox{CNC$^+$ + e$^-$ $\rightarrow$ C + CN}
     with yellow color, \mbox{N + CH $\rightarrow$ H + CN} with blue color and
      \mbox{H + CN$^+$ $\rightarrow$ CN + H$^+$} with magenta. }
\label{prod_50} 
\end{figure}

\begin{figure} 
\begin{subfigure}{.45\textwidth} 
\includegraphics[width=10cm]{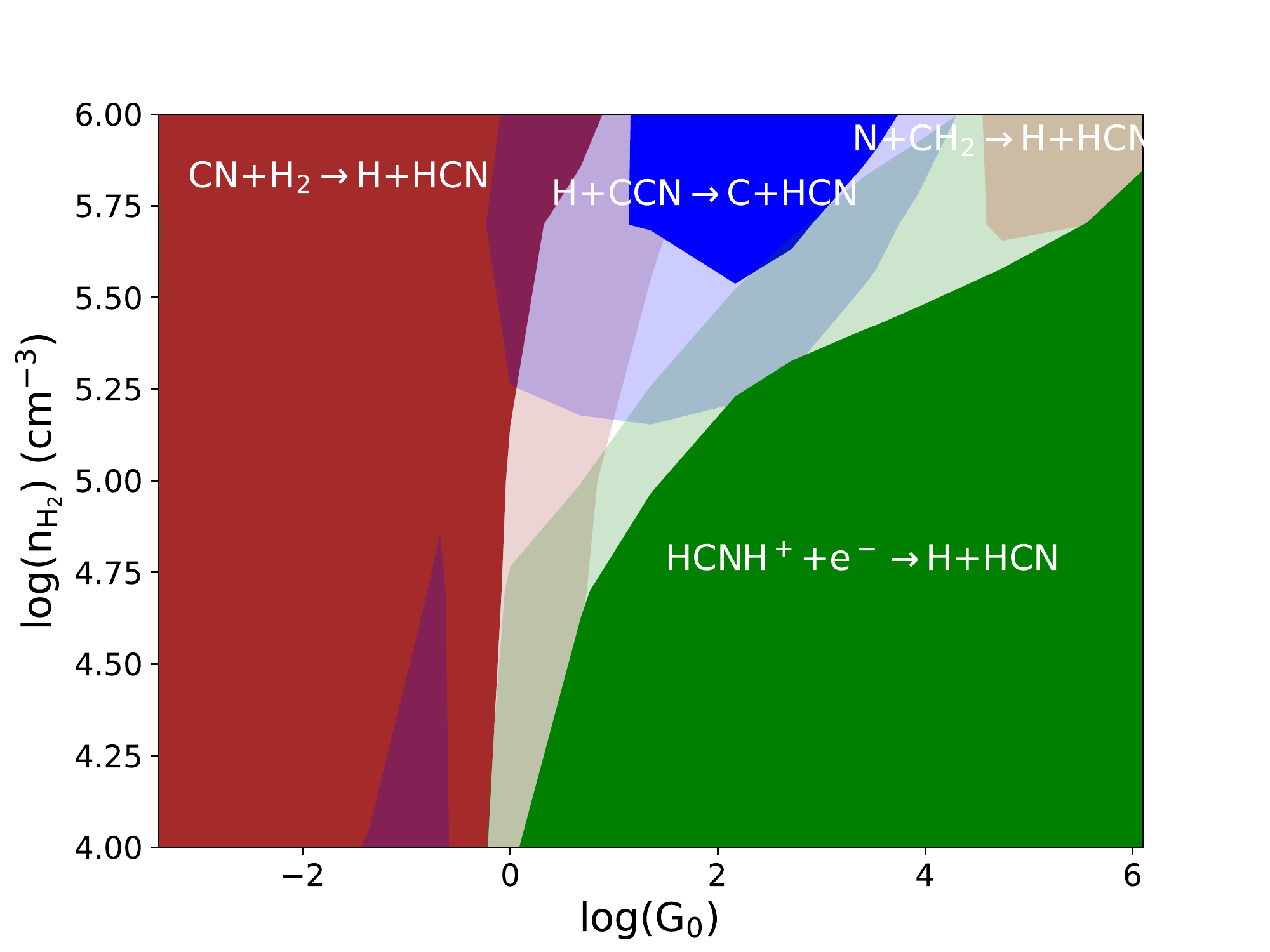} 
\end{subfigure}
\begin{subfigure}{.45\textwidth} 
\includegraphics[width=10cm]{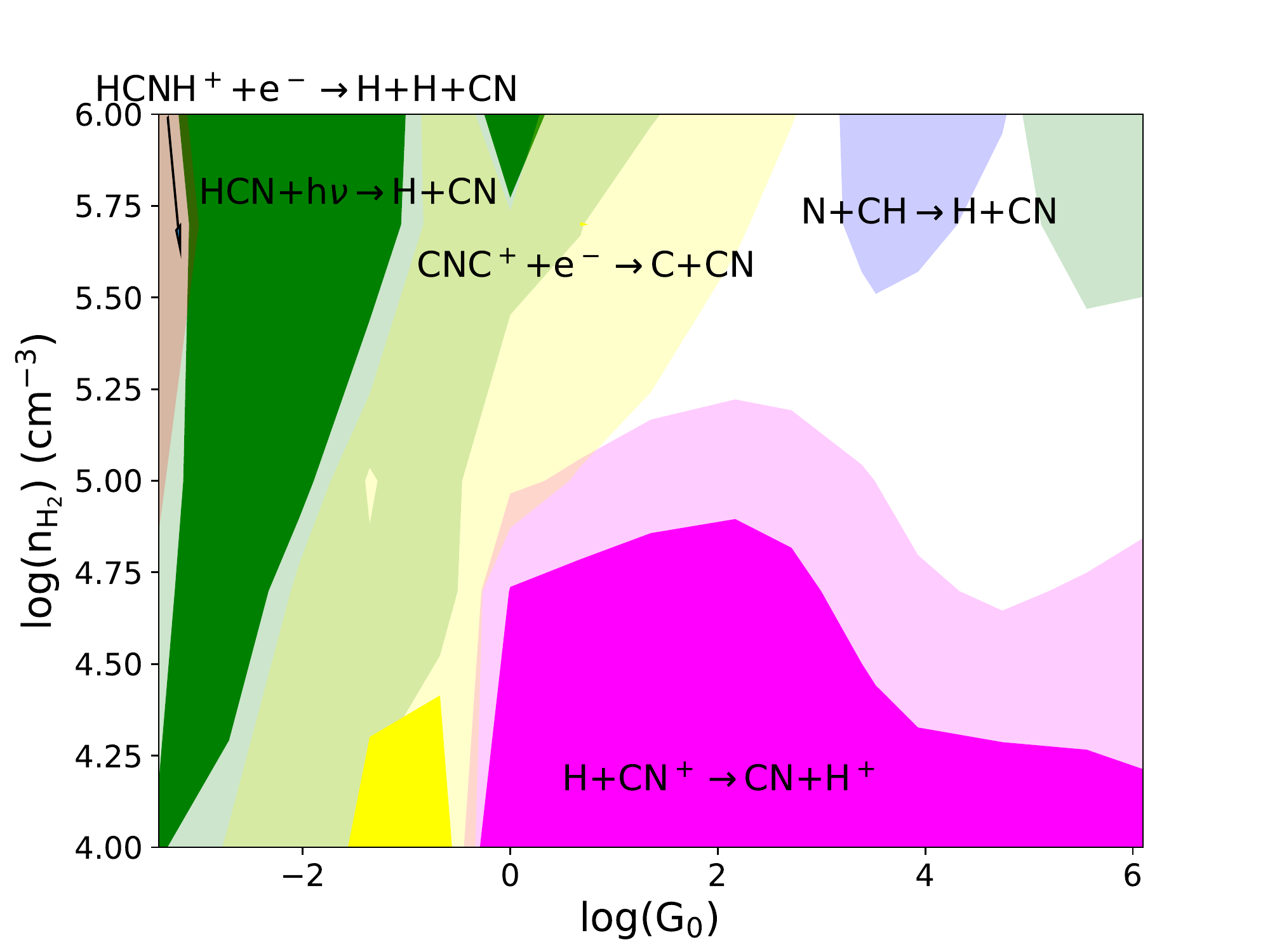} 
\end{subfigure}
\caption{Similar to Figure~\ref{CN_HCN_dest} but for HCN (upper panel) and CN (bottom panel)
 production assuming $T~=~300$~K. \mbox{HCNH$^+$ + e$^-$ $\rightarrow$ H + HCN}
  reaction is marked with green color, \mbox{H + CCN $\rightarrow$ C + HCN} with blue
   color, \mbox{CN + H$_2$ $\rightarrow$ H + HCN} with red color and
    \mbox{N + CH$_2$ $\rightarrow$ H + HCN} with orange. \mbox{HCN + h$\nu$ $\rightarrow$ H + CN}
     reaction is marked with green color, \mbox{CNC$^+$ + e$^-$ $\rightarrow$ C + CN}
      with yellow color, \mbox{N + CH $\rightarrow$ H + CN} with blue color and
       \mbox{H + CN$^+$ $\rightarrow$ CN + H$^+$} with magenta.} 
\label{prod_300}
\end{figure}

\begin{figure} 
\begin{subfigure}{.49\textwidth} 
\includegraphics[width=10cm]{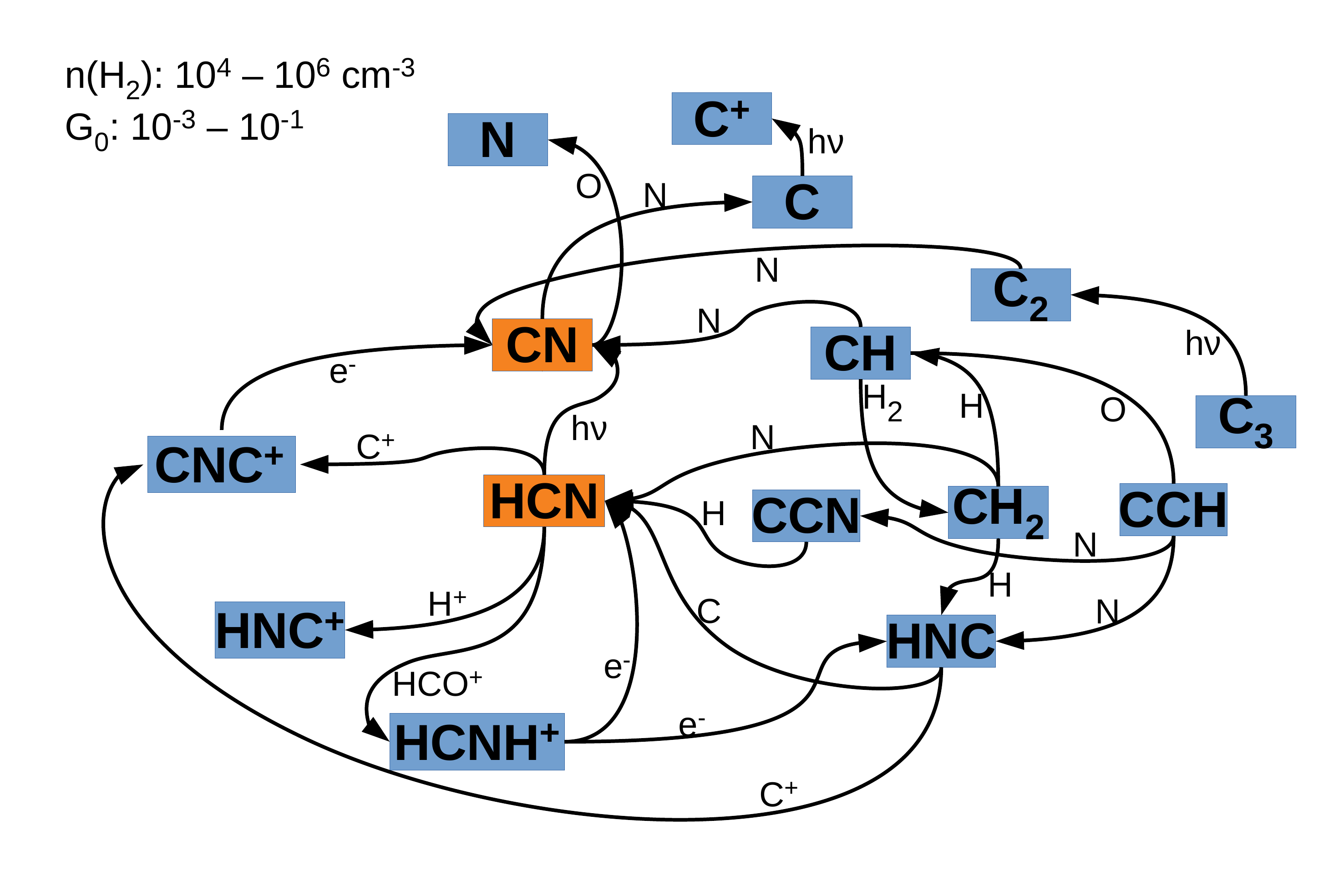} 
\end{subfigure}
\begin{subfigure}{.4\textwidth} 
\includegraphics[width=9cm]{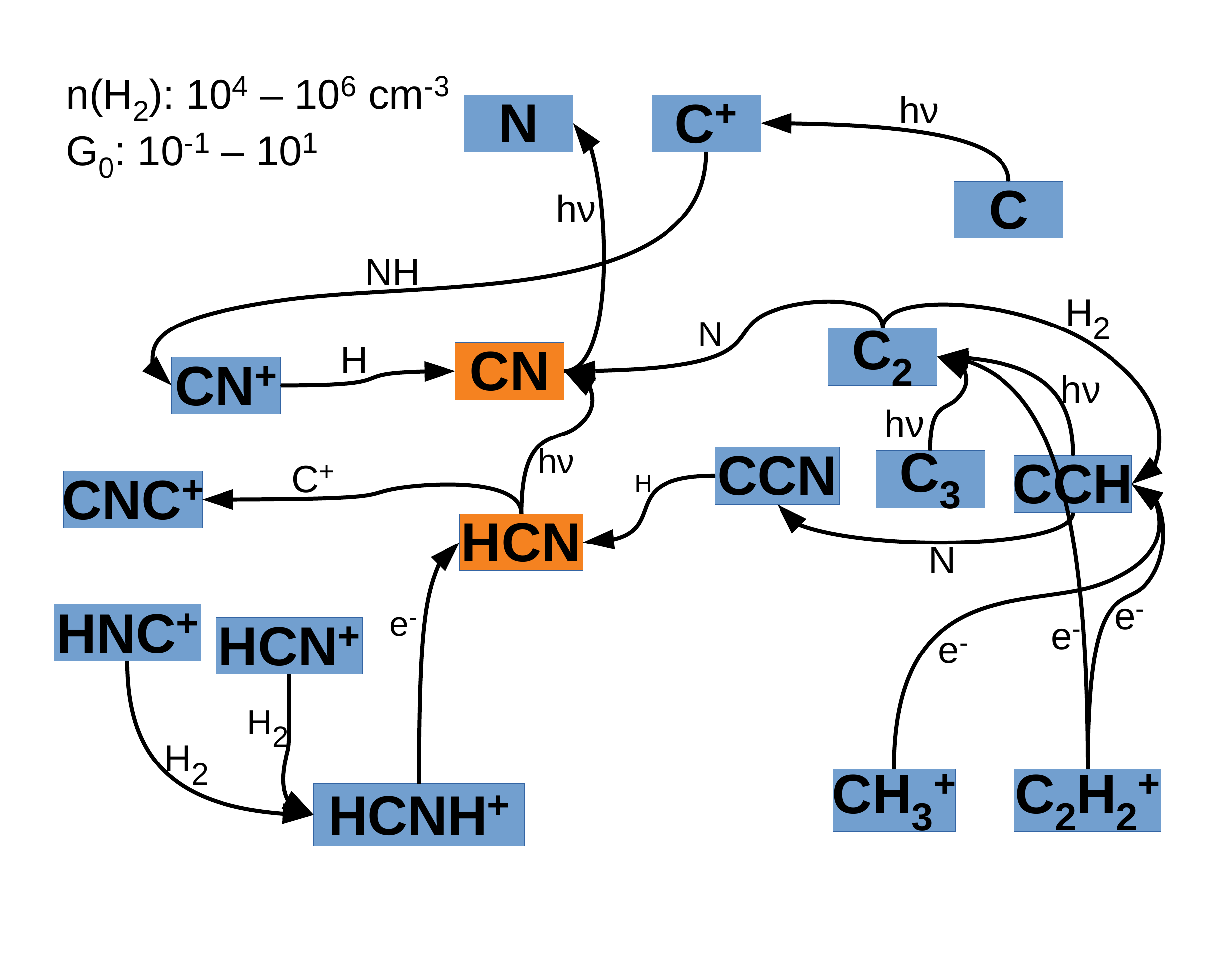} 
\end{subfigure}
\caption{Reactions network for weakly and medium UV irradiated gas. The dominant reactions are listed in Table~\ref{reactions_50}.}
\label{reactions_small_mediumG0}
\end{figure}

\end{appendix}

\end{document}